\documentclass{aa}
\usepackage{natbib}         
\usepackage{graphicx}
\usepackage{txfonts}
\usepackage{longtable}
\usepackage{xcolor}
\usepackage[flushleft]{threeparttable}
\bibpunct{(}{)}{;}{a}{}{,} 

\newcommand{\ind}[1]{_{\mathrm{#1}}}
\newcommand{\diff}{\mathrm{d}}

\newcommand\Dnu{\Delta\nu}

\newcommand\Dnup{\Delta\nu\ind{p}}
\newcommand\DPi{\Delta\Pi\ind{1}}

\newcommand\dnurotcore{\delta\nu\ind{rot,core}}

\newcommand\nup{\nu\ind{p}}

\newcommand\numax{\nu\ind{max}}
\newcommand\nmax{n\ind{max}}

\newcommand\Dtaum{\Delta\tau\ind{m}}

\newcommand\np{{n\ind{p}}}

\newcommand\epsp{\varepsilon\ind{p}}
\newcommand\epsg{\varepsilon\ind{g}}
\newcommand\N{\mathcal{N}}

\newcommand\Rsol{{R\ind{\odot}}}

\begin{document}

\author{%
 C. Gehan\inst{1},
 B. Mosser\inst{2},
 E. Michel\inst{2},
M. S. Cunha\inst{1,3}}

\institute{Instituto de Astrofísica e Ciências do Espaço, Universidade do Porto, CAUP, Rua das Estrelas, PT4150-762 Porto, Portugal ; \texttt{charlotte.gehan@astro.up.pt}
\and LESIA, Observatoire de Paris, Université PSL, CNRS, Sorbonne Université, Université de Paris, 5 place Jules Janssen, 92195 Meudon, France
\and School of Physics and Astronomy, University of Birmingham, Birmingham, B15 2TT, United Kingdom}


\abstract{Measuring stellar inclinations is fundamental to understanding planetary formation and dynamics as well as the physical conditions during star formation. Oscillation spectra of red giant stars exhibit mixed modes that have both a gravity component from the radiative interior and a pressure component from the convective envelope. Gravity-dominated (g-m) mixed modes split by rotation are well separated inside frequency spectra, allowing accurate measurement of stellar inclinations.}
{We aim to develop an automated and general approach to measuring stellar inclinations that can be applied to any solar-type pulsator for which oscillation modes are identified. We also aim to validate this approach using red giant branch stars observed by \textit{Kepler}.}
{Stellar inclination impacts the visibility of oscillation modes with azimuthal orders $m=\lbrace -1, 0, +1 \rbrace$. We used the mean height-to-background ratio of dipole mixed modes with different azimuthal orders to measure stellar inclinations. We recovered the underlying statistical distribution of inclinations in an unbiased way using a probability density function for the stellar inclination angle.}
{We derive stellar inclination measurements for 1139 stars on the red giant branch for which \cite{Gehan_2018} identified the azimuthal order of dipole g-m mixed modes. Raw measured inclinations exhibit strong deviation with respect to isotropy which is expected for random inclinations over the sky. When taking uncertainties into account, the reconstructed distribution of inclinations actually follows the expected isotropic distribution of the rotational axis.}
{This work highlights the biases that affect inclination measurements and provides a way to infer their underlying statistical distribution. When a star is seen either pole on or equator on, measurements are challenging and result in a biased distribution. Correcting biases that appear in low- and high-inclination regimes allows us to recover the underlying inclination distribution.}


\title{Automated approach to measure stellar inclinations: validation through large-scale measurements on the red giant branch}
\authorrunning{C. Gehan et al.}
\titlerunning{Automated approach to measure stellar inclinations: validation on the red giant branch}
\maketitle

\section{Introduction}\label{introduction}

The ultra-high-precision photometry space mission \textit{Kepler} has recorded very long observation runs, providing us with seismic data of unprecedented quality \citep{Gilliland}. Red giants represent an ideal laboratory to study the physical mechanisms governing deep stellar interiors because their oscillation spectra exhibit mixed modes. These modes result from a coupling between pressure (p) waves in the convective envelope and gravity (g) waves in the radiative interior and allow us to probe the highly condensed core of red giants \citep{Scuflaire}.

Dipole mixed modes present the strongest coupling between p and g waves, as the evanescent region that separates p-mode and g-mode cavities is thinner compared to mixed modes with higher angular degrees. Dipole mixed modes are therefore the most suitable mixed modes with which to probe the core of red giants \citep{Deheuvels_2017}, and have already been used to derive large-scale measurements of several parameters characterising the red giant interior:
\begin{itemize}
\item The period spacing between consecutive pure g modes, $\DPi$, provides information on the size of the radiative core \citep{Montalban} and was measured for almost 5000 stars both on the red giant branch (RGB) and in the red clump \citep{Vrard}.
\item The mean core rotational splitting, $\dnurotcore$, is a proxy of the mean core rotation rate and was measured for almost 900 stars on the RGB \citep{Gehan_2018} and 200 stars in the red clump \citep{Mosser_2012c}.
\item The coupling factor, $q$, characterises the strength of the coupling in the evanescent region between p and g waves and was measured for about 5000 RGB and red clump stars \citep{Mosser_2017, Pincon_2020}.
\item The gravity offset, $\epsg$, provides information on the stratification occurring in the radiative region and was measured for almost 400 stars both on the RGB and in the red clump \citep{Mosser_2018, Pincon_2019}.
\end{itemize}

Additionally, dipole mixed modes can also be used to derive high-precision measurements of the stellar inclination angle $i$, which is the angle between the stellar rotation axis and the line of sight that take values between 0$^\circ$ and 90$^\circ$. Characterising stellar inclinations is important when studying both planetary dynamics and star formation.

On the one hand, in the case of planet host stars, measurements of $i$ are required to constrain the angle between the stellar spin axis and the planetary orbit axis, namely the obliquity $\psi$ \citep[e.g.][]{Winn, Chaplin, Huber, Campante}, through \citep{Fabrycky}
\begin{equation}\label{eqt-obliquity}
\cos \psi = \sin i \, \sin i\ind{p} \, \cos \lambda + \cos i \, \cos i\ind{p},
\end{equation}
where $i\ind{p}$ is the inclination angle of the planetary orbit and $\lambda$ is the sky-projected spin–orbit angle.
The angle $i\ind{p}$ can be constrained from the transit light curve \citep[see][]{Huber_2018}. The angle $\lambda$ can be measured through high-resolution spectroscopic observations of transiting planetary systems, using the so-called Rossiter-McLaughlin effect \citep{Holt, McLaughlin, Rossiter, Queloz, Ohta}.
The measurement of $i$ is thus a prerequisite to the measurement of the real spin-orbit angle $\psi$, \citep{Kamiaka}, which can lead to a better understanding of the formation and evolution of exoplanetary systems. Measuring obliquities helps in particular to constrain the mechanism driving the post-formation migration of hot Jupiters closer to their host star, which is still raising questions \citep{Lin, Chatterjee, Fabrycky_2007}. Indeed, observations indicate that many hot Jupiters show a wide range of obliquities \citep{Johnson, Winn_2010} that can result from  planet-star tidal interactions: the stronger the tidal interaction between the star and the planet, the lower the obliquity \citep{Albrecht_2012}. However, other scenarios may also result in such high obliquities, which may in this case be observed for different types of planetary systems and not only for hot Jupiters. In this context, measuring stellar obliquities in multiplanet systems is key to testing whether or not spin-orbit misalignment is common \citep{Albrecht_2013}. If these systems mostly present low obliquities, then the high obliquities measured for hot Jupiters might be due to planet migration mechanisms. On the contrary, a distribution of obliquities that appears to be similar to that of hot Jupiters would indicate that high obliquities result from formation processes that are common to planetary systems in general, and thus not specific to hot Jupiter migration. Asteroseismology is highly valuable in this context. Indeed, seismic inclinations do not depend on planet size and can therefore be measured for multiplanet systems with small planets, where Rossiter-McLaughlin measurements are usually not possible \citep[see][]{Huber_2018}. Asteroseismology has already provided an example of spin-orbit misalignment in a multiplanet system \citep{Huber_2013}.
\newline

On the other hand, the statistical distribution of stellar inclinations provides clues as to the physical conditions during star formation \citep[e.g.][]{McKee}. Theoretical models suggest that the angular momentum of forming stars is dominated by small-scale turbulence \citep[e.g.][]{Rey-Raposo}. If some stellar-spin alignment occurs during star formation, one might expect it to be erased by turbulence in the absence of important interactions between forming stars. For a given set of stars with randomly oriented rotation axes, inclination angles follow a distribution in $\sin(i)$, which we refer to as an isotropic inclination distribution in this study.
 For a given stellar population, an isotropic inclination distribution indicates that turbulence during star formation was strong enough to prevent any stellar spin alignment. On the contrary, a non-isotropic distribution highlights that stellar spins show preferential orientations, and that strong interactions with the progenitor cloud(s) occurred during star formation. (Corsaro et al. \citeyear{Corsaro}; C17 hereafter) reported stellar spin alignment towards low inclinations inside two open clusters, however the findings of \cite{Mosser_2018} contradict this result. This point is therefore under debate, and inclination measurements can help to discriminate among star formation mechanisms inside open clusters.
\newline

The majority of transiting exoplanets are detected around F, G, and K solar-type stars on the main sequence, but measuring $i$ is difficult in this evolutionary stage because oscillations have low amplitudes and modes are sometimes blended \citep[see][]{Appourchaux}. In this context, \cite{Kamiaka} carried out a systematic verification of the accuracy of the seismic measurement of the stellar inclination angle for main sequence stars using 3000 simulated power spectra and found that reliable seismic inclination measurements are possible for $20^\circ \lesssim i \lesssim 80^\circ$.

However, accurate measurements of $i$ are much easier to obtain for evolved stars presenting mixed modes: g-dominated (g-m) mixed modes have lifetimes of the order of years and probe regions that rotate faster than the envelope \citep{Beck_2012, Deheuvels_2012, Benomar, Deheuvels, Deheuvels_2015, Mosser_2017}. Therefore, rotational splittings are much larger than mode line widths and dipole modes split by rotation are well separated. (Kuszlewicz et al. \citeyear{Kuszlewicz}; K19 hereafter) recently developed a Bayesian hierarchical method to extract inclination angles from red giants and successfully applied it to artificial oscillation spectra. They also derived inclination measurements for about 100 red giant stars. As \cite{Kamiaka} did not address red giants and K19 measured inclinations for
only a limited number of red giants, it is worthwhile obtaining large-scale measurements of the stellar inclination angle for evolved stars.

Nevertheless, it is difficult to detect exoplanetary transits around evolved stars because the associated depth in the light curve varies as the squared ratio between planet and star radii, $\left(R\ind{p}/R\ind{\star}\right)^2$ \citep{Heller}. The \textit{Kepler} mission is optimised to detect Earth-size planets in the habitable zone of solar-like stars, and can therefore detect a depth of 0.01 \% in relative stellar brightness fluctuations, which corresponds to an Earth-sized planet transiting a Sun-like star with $R\ind{p} \sim R\ind{\star}/100$. Red giants have a mean radius on the order of $R\ind{\star} \sim 10 \, \Rsol$, and therefore the transit of an Earth-sized planet leads to a depth of only $10^{-6}$ in relative brightness. If we consider a Jupiter-like planet with a radius $R\ind{p} \sim 10 \, R\ind{\oplus}$, where $R\ind{\oplus}$ is the Earth's radius, the transit in front of a red giant corresponds to a depth of 0.01 \% in relative brightness, which is the same order of magnitude as an Earth-sized planet transiting a Sun-like star. The detection of hot Jupiters transiting red giant stars is therefore possible, in contrast to transiting Earth-size planets.

About 100 exoplanets are already known to orbit red giants and subgiants \citep{Johnson_2010, Johnson_2011, Hirano, Lillo-Box_2014, Jones_2015, Lee_2015, Quinn_2015, Lillo-Box_2016, Hrudkova}. The detection of a much larger number of planets around evolved stars is crucial to constrain theoretical models of planet engulfment by the expanding host star \citep{Lillo-Box_2016}, and to understand the effect of stellar evolution on the orbital and physical properties of planetary systems \citep{Jones_2015}. In this context, measuring stellar inclination angles on a large scale for evolved stars is a major stepping stone in understanding planetary formation, evolution, and death.

The method presented in \cite{Gehan_2018} identifies rotational splittings of g-m modes for stars on the RGB. It is therefore possible to measure the stellar inclination angle using the ratio of the power spectral densities (PSDs) of dipole mixed modes with the same mixed-mode order but different azimuthal orders \citep{Gizon}. In this work we present a general and automated method to derive seismic measurements of the stellar inclination angle that can be applied to any solar-type pulsator for which oscillation modes are identified. We check the consistency of our approach on a large sample of red giant stars, from which we should recover an isotropic inclination distribution.
In Sect. \ref{measurements} we provide details of the way in which we derive stellar inclinations and their associated uncertainties. In Sect. \ref{results} we apply the method to the RGB stars of the \textit{Kepler} public catalogue studied by \cite{Gehan_2018} and infer the underlying statistical distribution of the measured inclinations. In Sects. \ref{Kuszlewicz} and \ref{Corsaro} we check the consistency of our results with those obtained by K19 and C17. Section \ref{planets} focuses on a discussion of the inclinations measured for RGB stars with a planet candidate(s) or a confirmed planet(s). We present our conclusions in Sect. 7.

\section{Deriving stellar inclinations}\label{measurements}

Measuring stellar inclinations using frequency oscillation spectra is not straightforward and requires several steps.

\subsection{Disentangling mixed modes with different azimuthal orders}

\begin{figure}
\centering
\includegraphics[width=10cm]{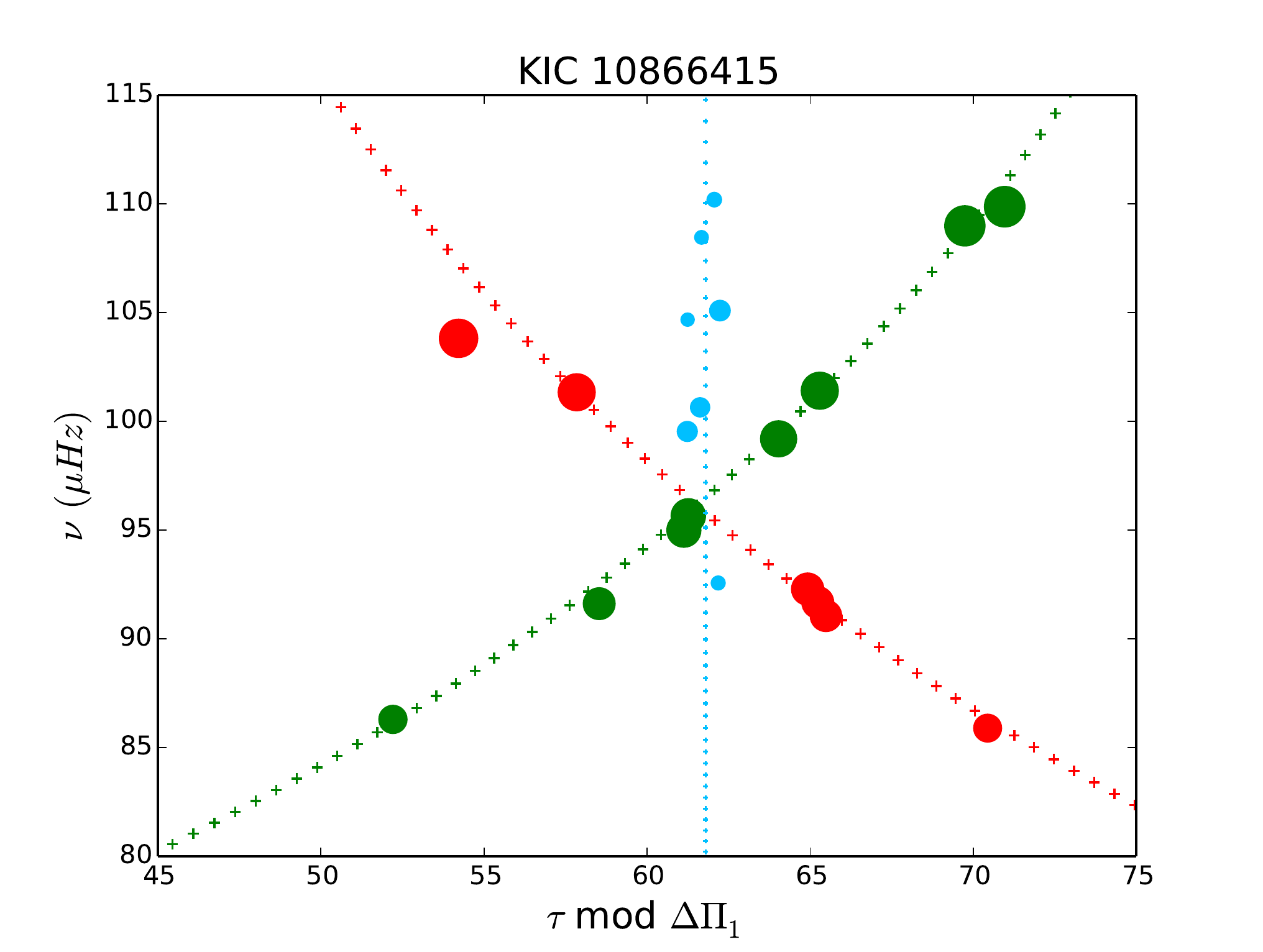}
\caption{Stretched-period échelle diagram for the RGB star KIC 10866415 with a high inclination angle. Colours indicate the azimuthal order: the $m=\lbrace-1,0,+1\rbrace$ rotational components are represented in green, light blue, and red, respectively. Observed modes are represented by dots and the symbol size varies as a function of the measured power spectral density. Rotational components with $m=\pm \, 1$  are identified in an automatic way through a correlation of the observed spectrum with a synthetic one constructed using Eqs.~(\ref{eqt-spacing-rot}) and (\ref{eqt-N}),  represented here by crosses \citep{Gehan_2018}. The location of the $m=0$ rotational component is identified considering that the $ m=0 $ ridge is median with respect to $m = \pm \, 1$ ridges. As there are six modes identified in the $m = +1$ ridge and eight modes in the $m = -1$ ridge, this gives a mean number of visible modes $N\ind{\pm \, 1}=7$. We thus consider the first seven modes with the highest height-to-background ratio belonging to the $m = 0$ ridge.}
\label{fig-echelle-2-components}
\end{figure}

The identification of dipole mixed modes with different azimuthal orders relies on oscillation spectra in stretched period, where radial and quadrupole modes have been removed using the red giant universal oscillation pattern which characterises pressure modes. \citep{Mosser_2011}. This stretching is done using the function
\begin{equation}\label{eqt-zeta}
    \zeta = \left[1 + \frac{\nu^2}{q} \frac{\DPi}{\Dnup} \frac{1}{\frac{1}{q^2} \sin^2 \left(\pi \frac{\nu - \nup}{\Dnup}\right) + \cos^2 \left(\pi \frac{\nu - \nup}{\Dnup}\right)}\right]^{-1}
,\end{equation}
which characterises the contribution of the core and envelope to the inertia of dipole modes \citep{Goupil, Mosser_2015, Hekker, Cunha_2019}. The $\zeta$ function as expressed in Eq.~(\ref{eqt-zeta}) is continuous with frequency $\nu$ and depends on:
\begin{itemize}
\item the coupling parameter $q$ between g and p modes \citep{Mosser_2017b};
\item the asymptotic large separation,
\begin{equation}
\Dnup = \Dnu \left( 1 + \alpha (\np - \nmax) \right),
\end{equation}
where $\nmax = \numax / \Dnu - \epsp$ is the non-integer order at the frequency $\numax$ of maximum oscillation signal, with $\epsp$ the phase shift of pure pressure modes, and where $\Dnup$ increases with the pressure radial order $\np$ \citep{Mosser_2013};
\item \ and the pure dipole pressure mode frequencies, $\nup$, computed using the universal pattern \citep{Mosser_2011}.
\end{itemize}
Stretched-period spectra are then obtained using the differential equation \citep{Mosser_2015}
\begin{equation}\label{eqt-asymp}
\diff \tau = \frac{1}{\zeta} \frac{\diff \nu}{\nu^2}.
\end{equation}
Dipole mixed modes are expected to be evenly spaced in stretched period with a spacing equal to $\DPi$,  as in pure dipole g modes. However, p-dominated (p-m) dipole mixed modes present large line widths and consequently appear as several modes scattered close to the expected unique p-m mode. This spread is exacerbated when stretching frequency spectra, as the correction induced by $\zeta$ is most important for p modes \citep{Gehan_2018}. As p-m dipole mixed modes scramble the expected regular pattern in stretched period, we remove them from stretched-period spectra using once again the red giant universal pattern \citep{Mosser_2011}. We are now left with g-m dipole modes only, which should in theory present an equal $\DPi$ spacing in stretched period. However, rotation associated to the g-mode cavity disturbs this regular pattern, meaning that the stretched period spacing between modes with the same azimuthal order is \citep{Mosser_2015}
\begin{equation}\label{eqt-spacing-rot}
 \Dtaum \simeq \DPi \left( 1 + 2 \, m \, \zeta \, \frac{\dnurotcore}{\nu} \right).
\end{equation}
In this equation, $\zeta$ can be estimated for g-m mixed modes as $\N / (\N+1)$, where $\N$ is the mixed-mode density representing the number of g modes per $\Dnu$-wide frequency range and is defined as
\begin{equation}\label{eqt-N}
\N = {\Dnu \over \DPi \, \numax^2}.
\end{equation}
Mixed modes with different azimuthal orders can now be easily disentangled using stretched period échelle diagrams \citep{Mosser_2015, Gehan_2018, Mosser_2018}, where mixed modes with the same azimuthal order draw ridges called rotational components (Fig.~\ref{fig-echelle-2-components}). Peaks associated to the g-m modes that have a maximum PSD significantly above the background are kept (see Sect.2.2 of \cite{Gehan_2018}), and their maximum PSD is used in the following. The number of visible ridges depends on the stellar inclination angle, which modifies the visibility of the different rotational components: when the inclination is low, only the central rotational component associated to $m=0$ is visible, for intermediate $i$ values all the three ridges with $m=\lbrace -1, 0, +1 \rbrace$ are visible, and in the case of high $i$ values, only the $m=\pm \, 1$ components are visible. Each ridge with a given azimuthal order is identified through a correlation of the observed spectrum with a synthetic one built upon asymptotic seismic parameters using Eqs.~(\ref{eqt-spacing-rot}) and (\ref{eqt-N}), as described by \cite{Gehan_2018}. The azimuthal order of each g-m mode is thus identified, making it possible to measure the stellar inclination angle.

\subsection{Measurement of the stellar inclination angle}

We now detail how we measure the stellar inclination angle $i$ . We write $p\ind{-1}(\nu)$, $p\ind{0}(\nu),$ and $p\ind{1}(\nu),$ the PSD of individual g-m modes with azimuthal orders $m=\lbrace-1,0,+1\rbrace$, respectively. The PSD is related to the stellar inclination by \citep{Gizon}
\begin{equation}\label{eqt-p-m}
\begin{aligned}
p\ind{0}(\nu) &\propto \cos^2(i)\\
p\ind{\pm \, 1}(\nu) &\propto \frac{1}{2} \sin^2(i).
\end{aligned}
\end{equation}
The background signal close to $\numax$ can be approximated by
\begin{equation}\label{eqt-background}
B(\nu) = \beta \left(\frac{\nu}{\numax}\right)^\gamma,
\end{equation}
where $\beta$ and $\gamma$ are parameters that take real values \citep{Mosser_2012b}. This expression is valid in the frequency range where oscillations are detected, because it provides the precise measurement of $\numax$ \citep[Fig. 8 of][]{Pinsonneault_2018}.

We then subtract the background from the PSD of each mode and normalise the result by $B(\nu)$. We thus consider the background-free and normalised PSD for each mode of azimuthal order $m$, which we denote
\begin{equation}\label{eqt-P-over-B}
\mathcal{P}\ind{m}(\nu) = \frac{p\ind{m}(\nu)}{B(\nu)} - 1.
\end{equation}

As the height-to-background ratio (HBR) of modes is supposed to follow a Gaussian envelope centred on $\numax$, we then weight the contribution of modes with same azimuthal order by a Gaussian function denoted $f\ind{G}(\nu)$. This step maximises the contribution of modes in the close vicinity of $\numax$, which are expected to have the highest HBR, and minimises the contribution of modes in the tails of the Gaussian envelope, which are expected to have the lowest HBR. The standard deviation of the Gaussian function can be written \citep{Mosser_2012b}
\begin{equation}\label{eqt-sigma}
\sigma = \frac{\delta\nu\ind{env}}{2 \sqrt{2 \, \ln{2}}},
\end{equation}
where $\delta\nu\ind{env}$ is a frequency interval chosen to be slightly larger than the frequency range of observed dipole modes.
We write the weighting of $\mathcal{P}\ind{m}(\nu)$ (Eq.~(\ref{eqt-P-over-B})) by the Gaussian function $f\ind{G}(\nu)$ as
\begin{equation}\label{eqt-P-m-ini}
P\ind{m}(\nu) = \mathcal{P}\ind{m}(\nu) \, f\ind{G}(\nu).
\end{equation}
We then consider the mean value of $P\ind{m}(\nu)$ in Eq.~(\ref{eqt-P-m-ini}) over all $n$ modes with a given azimuthal order such as
\begin{equation}\label{eqt-P-m}
\left < P\ind{m} \right > = \frac{\sum^n\ind{i=1} P\ind{m,i}(\nu)}{\sum^n\ind{i=1} f\ind{G,i}(\nu)}.
\end{equation}
We can finally measure $i$ from Eq.~(\ref{eqt-p-m}) through
\begin{equation}\label{eqt-i-3}
\tan(i) = \sqrt{\frac{\left < P\ind{1} \right > + \left < P\ind{-1} \right >}{\left < P\ind{0} \right >}}.
\end{equation}

\begin{figure}
\centering
\includegraphics[width=10cm]{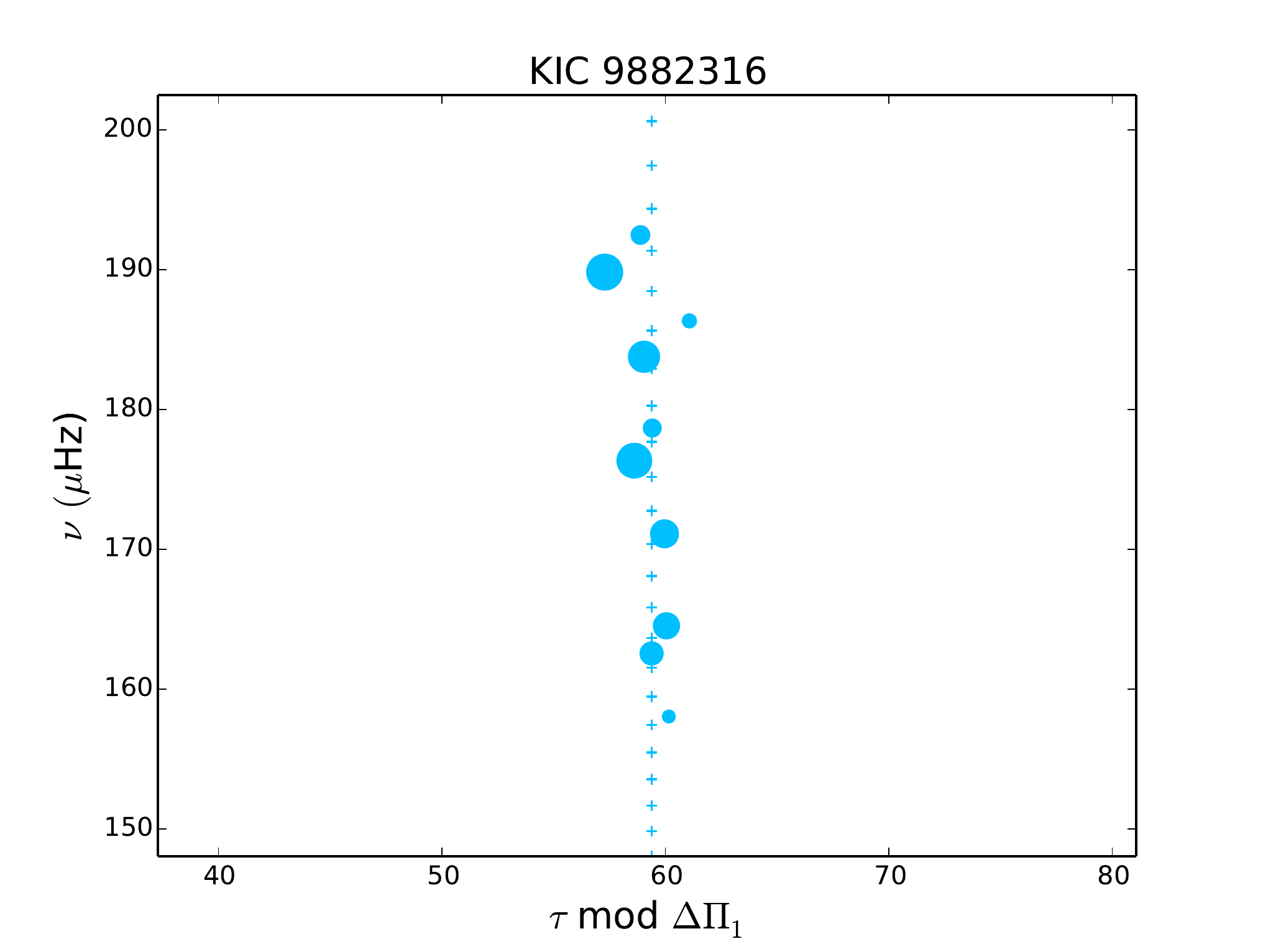}
\caption{Same as Fig.~\ref{fig-echelle-2-components} but for the RGB star KIC 9882316 with a low inclination angle.}
\label{fig-echelle-1-component}
\end{figure}

\subsubsection{Intermediate inclinations}\label{section-int-i}
When all the three rotational components associated to $m=\lbrace-1,0,+1\rbrace$ are visible, we can directly derive a measurement of the inclination angle using Eq.~(\ref{eqt-i-3}).

\subsubsection{Low inclinations}\label{low-inclinations}
Measuring low inclinations is particularly tricky, and we can only derive a rough estimate of $i$.
 Indeed, when the stellar inclination is low, the PSD of mixed modes associated to $m=\pm \, 1$ is so low that these modes are lost in the background and are no longer visible (Fig.~\ref{fig-echelle-1-component}). We do not know where we can find the missing $m=\pm \, 1$ components in the signal; we can only make the assumption that non-visible modes have a HBR below a given threshold $x$, corresponding to $\mathcal{P}\ind{\pm \, 1}(\nu) = x - 1$ in Eq.~(\ref{eqt-P-over-B}). In the absence of direct information on the HBR of the $m = \pm \, 1$ components, we suppose that they are symmetric and have similar $x$ values.
Using Eq.~(\ref{eqt-P-m}), we thus have an upper limit $i\ind{max}$ for low stellar inclinations, such as
\begin{equation}\label{eqt-i-1}
\tan(i) < \sqrt{\frac{2 \, \left < P\ind{\pm \, 1} \right >}{\left < P\ind{0} \right >}} = \tan(i\ind{max}).
\end{equation}

\begin{figure}
\centering
\includegraphics[width=9.2 cm]{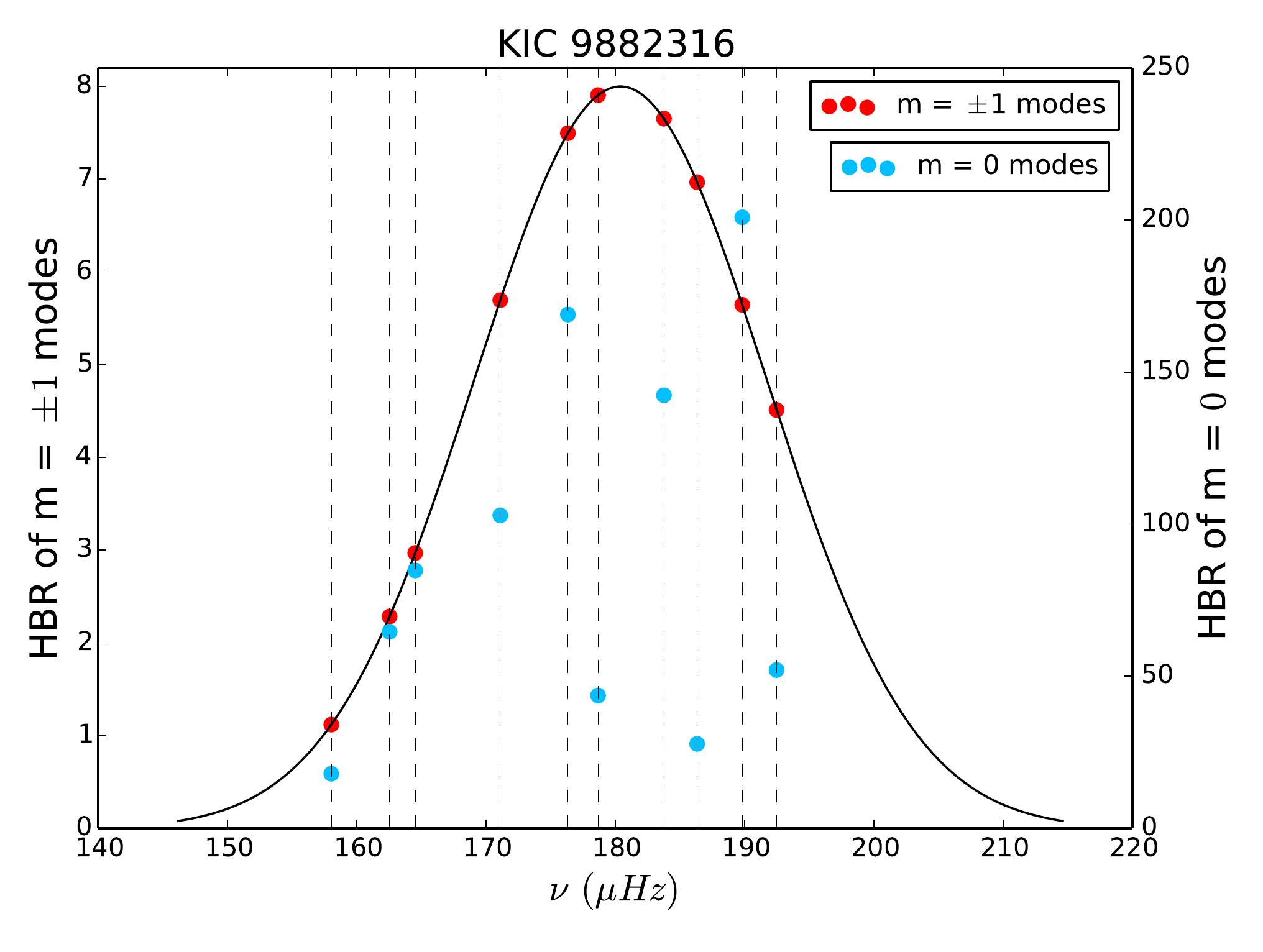}
\caption{HBR of g-m modes for KIC 9882316, for which only the central $m=0$ rotational component is visible. Observed modes with $m=0$ are in light blue (characterised by the right ordinate axis). Missing modes with $m=\pm \, 1$ are in red (characterised by the left ordinate axis), and their HBR follows a Gaussian envelope (in black) referred to as $x$ in the core text, which is normalised to the value of 8. Vertical dashed lines identify frequencies of observed $m=0$ modes. The number of missing $m=\pm \, 1$ modes has been chosen to be the same as visible $m=0$ modes.}
\label{fig-HBR-1-component}
\end{figure}

\paragraph{Estimating the PSD of the missing $m = \pm \, 1$ modes}\label{PSD-estimate}

\begin{figure}
\centering
\includegraphics[width=10cm]{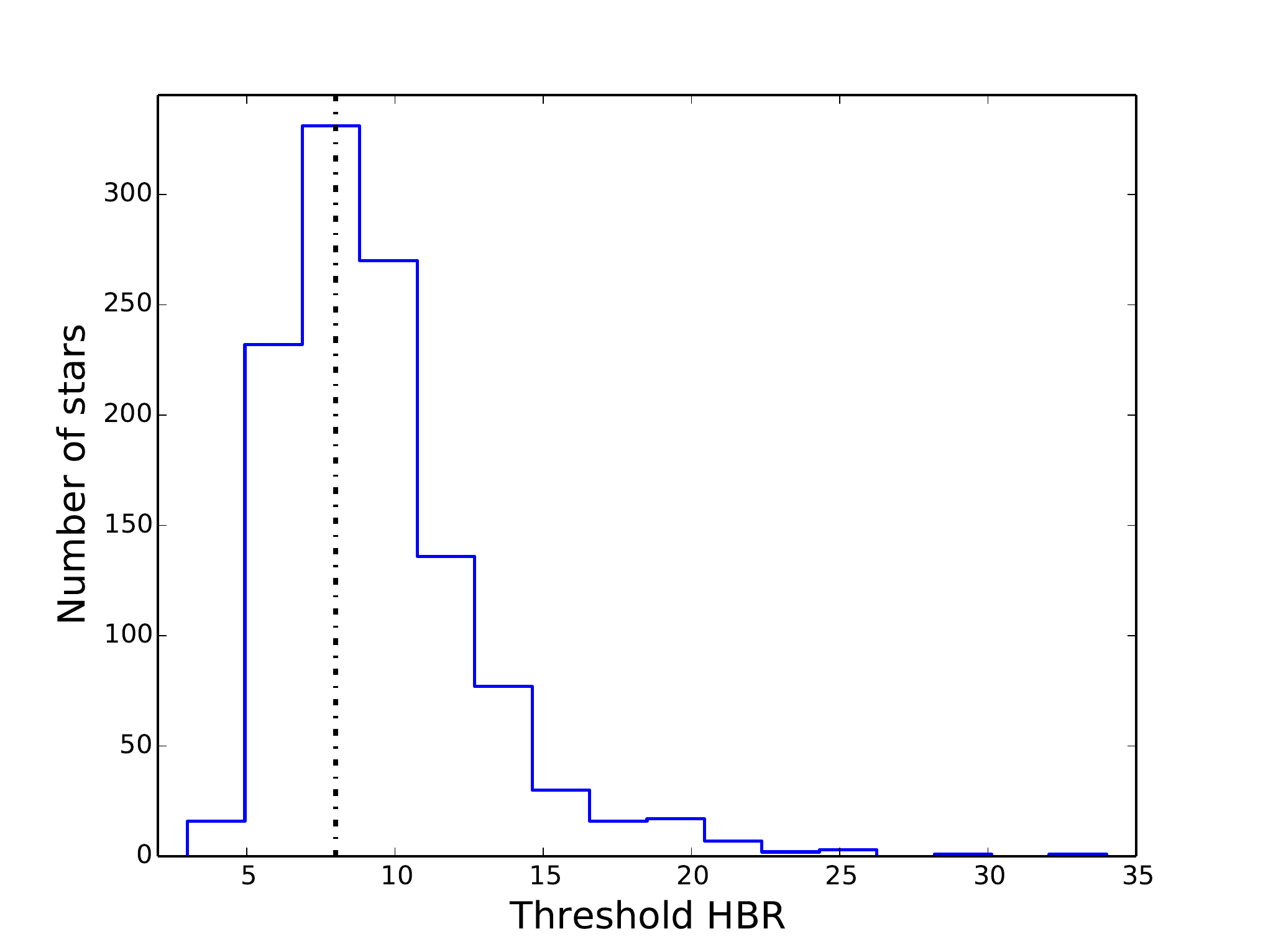}
\caption{Histogram of the threshold HBR above which g-m modes were considered as significant and were kept to initiate the fit of rotational components in the study of \cite{Gehan_2018}, for our sample of 1139 RGB stars. The vertical dot-dashed line indicates a threshold HBR of 8.}
\label{fig-HBR-threshold}
\end{figure}

We have to provide an estimate of the maximum possible PSD of the missing $m = \pm \, 1$ modes in order to derive an upper limit on $i$. As in the case of stars presenting two and three rotational components, we consider that the threshold $x$ follows a Gaussian with a standard deviation computed as in Eq.~(\ref{eqt-sigma}), in order to follow the Gaussian shape of the power spectrum around $\numax$ (Fig.~\ref{fig-HBR-1-component}). In order to chose the maximum value taken by $x$ around $\numax$, we considered the threshold HBR above which g-m modes were considered as significant and were kept to initiate the fit of rotational components in the study of \cite{Gehan_2018}. Among the 1139 RGB stars for which \cite{Gehan_2018} have identified rotational components, we note that the distribution of the threshold HBR peaks around 8, and that the number of stars in our sample sharply decreases below a threshold HBR of 8 (Fig.~\ref{fig-HBR-threshold}). This indicates that rotational components can only be poorly identified through the method developed by \cite{Gehan_2018} when the threshold HBR drops below 8. We also notice that only three stars in our sample have a threshold HBR that is below 4 (Fig.~\ref{fig-HBR-threshold}), which is compatible with the minimum HBR of 4 that is found by K19 to reliably extract the inclination angle from their artificial spectra.

According to the distribution of the HBR shown in Fig.~\ref{fig-HBR-threshold}, we chose a maximum value of $x\ind{max}=8$ for the Gaussian that models the HBR of the missing $m=\pm \, 1$ components (Fig.~\ref{fig-HBR-1-component}). For cases where the mean HBR is below 8 for modes belonging to the $m=0$ component, we took this mean HBR value as a maximum value for $x\ind{max}$. In order to compute $\left < P\ind{\pm \, 1} \right >$ in Eq.~(\ref{eqt-i-1}), we estimated the maximum PSD of the missing $m=\pm \, 1$ modes by interpolating the Gaussian threshold $x$ to the frequencies of observed $m=0$ modes so that we have the same number of $m=0$ and $m=\pm \, 1$ modes (Fig.~\ref{fig-HBR-1-component}). We then computed $\left < P\ind{\pm \, 1} \right >$ using Eq.~(\ref{eqt-P-m}).

The threshold $x\ind{max}$ is linked to the total observation duration $T\ind{obs}$ such as \citep{Mosser_2018}
\begin{equation}\label{eqt-x}
x\ind{max} \simeq \ln \left(\frac{T\ind{obs} \, \Delta\ind{\nu}}{p}\right),
\end{equation}
where $\Delta\ind{\nu}$ is the width of the frequency range where a mode is expected and $p$ is the rejection probability of the null hypothesis, that is, the hypothesis that background noise alone is enough to explain the presence of the observed peaks. Red giants observed by \textit{Kepler} have $T\ind{obs} = 4$ years, and we have $\Delta\ind{\nu} = 0.1 \, \mu$Hz in the case of g-m modes. The maximum value $x\ind{max}=8$ considered in this work corresponds to a rejection probability of the null-hypothesis at the 0.4 \% level.

\begin{figure}
\centering
\includegraphics[width=10cm]{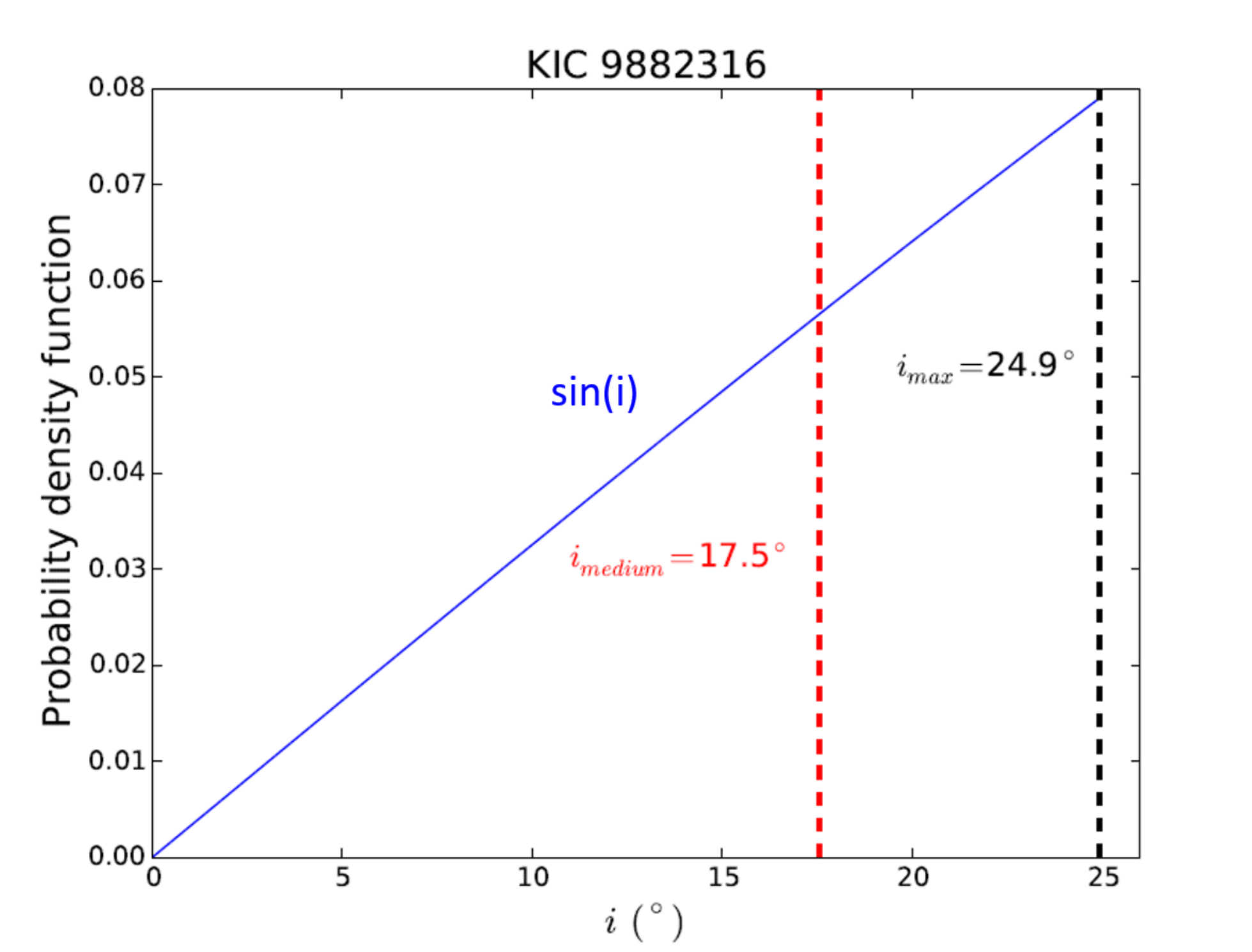}
\caption{Probability density function in $\sin(i)$ between $0^\circ$ and $i\ind{max}$ for KIC 9882316 (in blue). Black dashed lines indicate the upper limit on the inclination derived from Eq.~(\ref{eqt-i-1}), here $i\ind{max} = 24.9 ^\circ$. Red dashed lines indicate the estimated inclination corresponding to a PDF value of 0.5, here $i\ind{medium} = 17.5 ^\circ$.}
\label{fig-pdf-1-component}
\end{figure}

\begin{figure*}
\centering
\includegraphics[width=13.8cm]{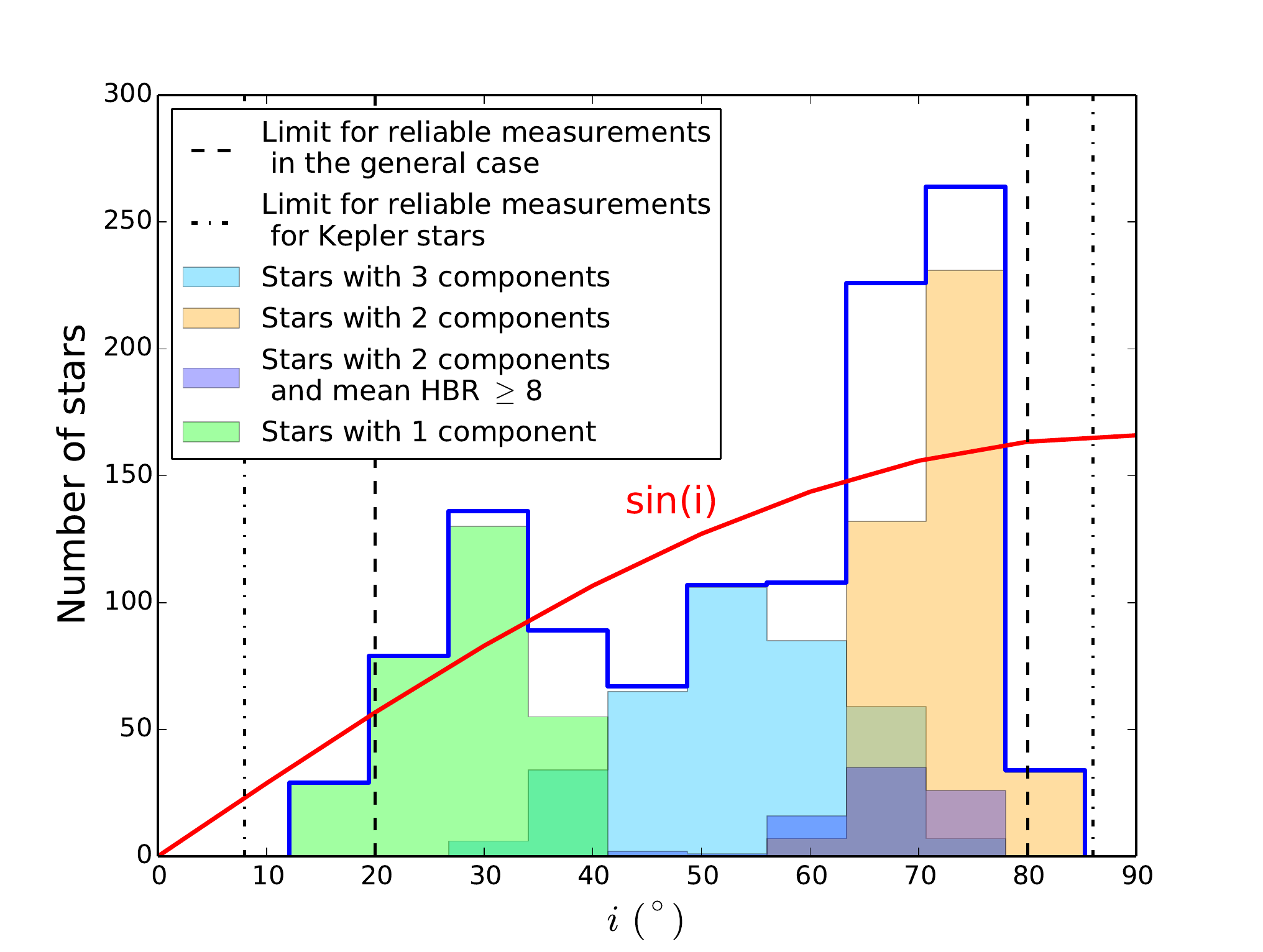}
\caption{Distribution of raw stellar inclinations for 1139 RGB stars analysed by \cite{Gehan_2018} (in blue). The red line represents an isotropic distribution in $\sin(i)$ derived from a fit to the histogram in blue. Inclinations measured for stars presenting three rotational components are in light blue, while inclination estimates for stars presenting one (two) rotational components are in green (orange). Inclination measurements derived for stars presenting two rotational components and a mean HBR equal to or above 8 for the $m=0$ component are in violet. Vertical dashed lines represent the limits of 20$^\circ$ and 80$^\circ$ between which \cite{Kamiaka} consider seismic inclination measurements to be reliable in the general case. Vertical dot-dashed lines represent the limits of 8$^\circ$ and 86$^\circ$ between which \cite{Kamiaka} show that seismic inclination measurements can be reliable for \textit{Kepler} stars.}
\label{fig-i-distribution}
\end{figure*}

\paragraph{Inclination estimate}\label{PSD-estimate}

When only the $m=0$ rotational component is visible, we can only have an estimate of the inclination angle, which can take any value between $0^\circ$ and $i\ind{max}$ given by Eq.~(\ref{eqt-i-1}). In order to provide an estimate of $i$, rather than just the upper limit $i\ind{max}$, we assume for a given star that the possible inclination values are isotropically distributed between $0^\circ$ and $i\ind{max}$. Under this hypothesis, the probability of having an inclination between $i$ and $i + \diff i$, where $\diff i$ is an infinitesimal inclination angle, is proportional to the spherical area covered by $\diff i$, and therefore to the corresponding specific solid angle $\diff \Omega = 2 \pi \sin(i) \, \diff i$. We therefore consider that the inclination follows a probability density function (PDF) in $\sin(i)$ between $0^\circ$ and $i\ind{max}$ (Fig.~\ref{fig-pdf-1-component}). We then derive an estimate of the mean inclination by selecting the $i$ value corresponding to a PDF value of 0.5 ($i\ind{medium}$ in Fig.~\ref{fig-pdf-1-component}).

\subsubsection{High inclinations}\label{high-inclinations}

When the inclination is high and only the $m = \pm \, 1$ components are visible, we can rely on the existing signal to find the missing $m=0$ component. The location of the $m=0$ component is derived considering that the $m=0$ ridge is median with respect to $m = \pm \, 1$ ridges. We consider, as in the fitting procedure of the rotational components used in \cite{Gehan_2018}, that peaks belong to the $m=0$ rotational component if they are at a distance of no more than $\DPi/80$ from the mid-point between the $m=-1$ and $m=+1$ components in the stretched-period spectrum. We then compute $N\ind{\pm \, 1}$ , the mean number of identified modes with azimuthal orders $m = \pm \, 1$, and we keep the $N\ind{\pm \, 1}$ modes belonging to the $m=0$ ridge that have the highest HBR estimated using Eq.~(\ref{eqt-P-over-B}); see Fig.~\ref{fig-echelle-2-components}. This step allows us to compute $\left < P\ind{0} \right >$ in Eq.~(\ref{eqt-P-m}). 
We then distinguish between two different configurations, depending on the mean HBR $\left < P\ind{0} \right >$ of the retrieved $m=0$ component. If $\left < P\ind{0} \right >$ is sufficiently high, then we have clear oscillation modes and we can derive a measurement of the inclination through Eq.~(\ref{eqt-i-3}). On the contrary, if $\left < P\ind{0} \right >$ is below a given threshold, we cannot know if the visible signal is due to oscillation modes or to stochastic variations of the background noise and we can only derive an estimate of the inclination instead of a firm measurement.
We consider the threshold value $\left < P\ind{0} \right > = 8$ to separate stars with two components for which we obtain an inclination measurement from stars with a $m=0$ component hidden in the background noise, for which we can only derive an estimate of $i$.

\begin{figure}
\centering
\includegraphics[width=10cm]{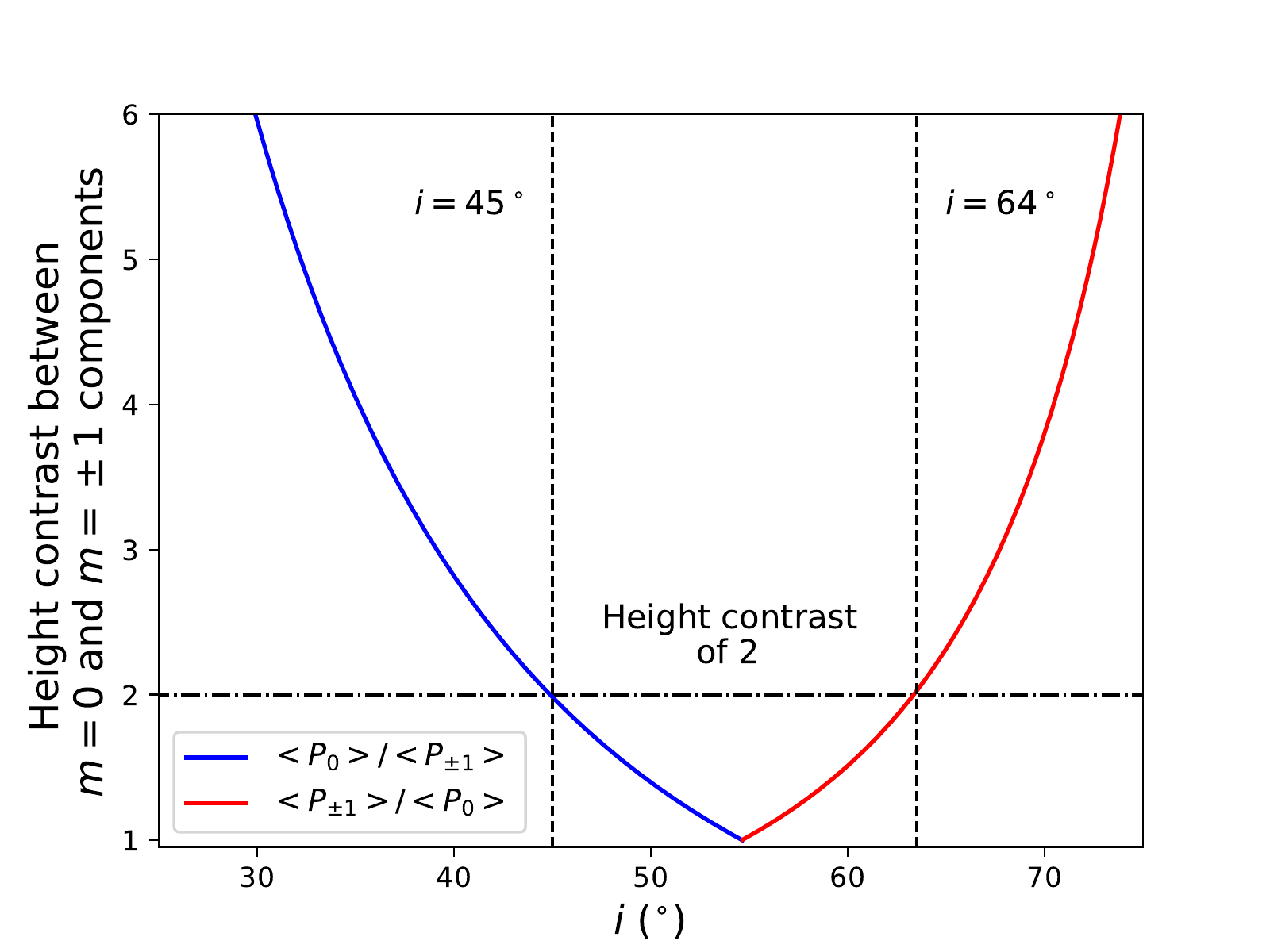}
\caption{Evolution with inclination of the height contrast between rotational components with azimuthal orders 0 and $\pm \, 1$. The blue line represents the contrast $\left < P\ind{0} \right > / \left < P\ind{\pm \, 1} \right >$ computed from Eq.~(\ref{eqt-P-m}), while the red line represents the contrast $\left < P\ind{\pm \, 1} \right > / \left < P\ind{0} \right >$. Vertical dashed lines indicate inclinations of 45 and 64$^\circ$, while the horizontal dot-dashed line indicates a height contrast of 2.}
\label{fig-HBR-contrast}
\end{figure}

\subsection{Estimate of the uncertainties}

Uncertainties on the measured stellar inclinations mainly arise from the stochastic nature of oscillations that is a source of spread in the PSD of observed modes, and therefore in their HBR, impacting the precision of the inclination measurement. Uncertainties are derived from Eq.~(\ref{eqt-i-3}). We can write
\begin{equation}\label{eqt-dtan-i}
\frac{\diff \tan^2(i)}{\tan^2(i)} = \frac{\diff P\ind{+1}}{\left < P\ind{+1} \right >} + \frac{\diff P\ind{-1}}{\left < P\ind{-1} \right >} + \frac{\diff P\ind{0}}{\left < P\ind{0} \right >},
\end{equation}
where $P\ind{m}(\nu)$ is defined by Eq.~(\ref{eqt-P-m-ini}) and $\left < P\ind{m} \right >$ is defined by Eq.~(\ref{eqt-P-m}).
Moreover, we have
\begin{equation}\label{eqt-dtan-i-bis}
\frac{\diff \tan^2(i)}{\tan^2(i)} = 2 \, \frac{\diff \tan(i)}{\tan(i)} = 2 \, \left ( \frac{1 + \tan^2(i)}{\tan(i)} \right ) \diff i = \frac{4}{\sin(2i)} \, \diff i.
\end{equation}
The equality of Eqs.~(\ref{eqt-dtan-i}) and (\ref{eqt-dtan-i-bis}) gives
\begin{equation}\label{eqt-di}
\diff i = \frac{1}{4} \sin(2i) \left (\frac{\diff P\ind{+1}}{\left < P\ind{+1} \right >} + \frac{\diff P\ind{-1}}{\left < P\ind{-1} \right >} + \frac{\diff P\ind{0}}{\left < P\ind{0} \right >} \right ).
\end{equation}
We finally obtain the uncertainty $\sigma\ind{i}$ on the measured inclination through
\begin{equation}\label{eqt-sigma-i}
\sigma \ind{i} = \frac{1}{4} \left | \sin(2i) \right | \sqrt{\sum_{m=-1}^{m=+1} \left ( \frac{\Delta P\ind{m}}{\left < P\ind{m}\right >} \right) ^2},
\end{equation}
where $\Delta P\ind{m}$ is the standard deviation of $P\ind{m}(\nu)$ used in Eq.~(\ref{eqt-P-m-ini}).

Equation~(\ref{eqt-sigma-i}) provides 1-$\sigma$ uncertainties. Error bars are therefore symmetric around the measured inclination value, and we have $\sigma \ind{i} = \sigma \ind{i,-} = \sigma \ind{i,+}$. However error bars should not be symmetric when the inclination is low or high and we should have different lower and upper error bars $\sigma\ind{i,-}$ and $\sigma\ind{i,+}$, respectively. For these two extreme cases, uncertainties differ from 1-$\sigma$ uncertainties.

\begin{figure*}
\centering
\includegraphics[width=13.8cm]{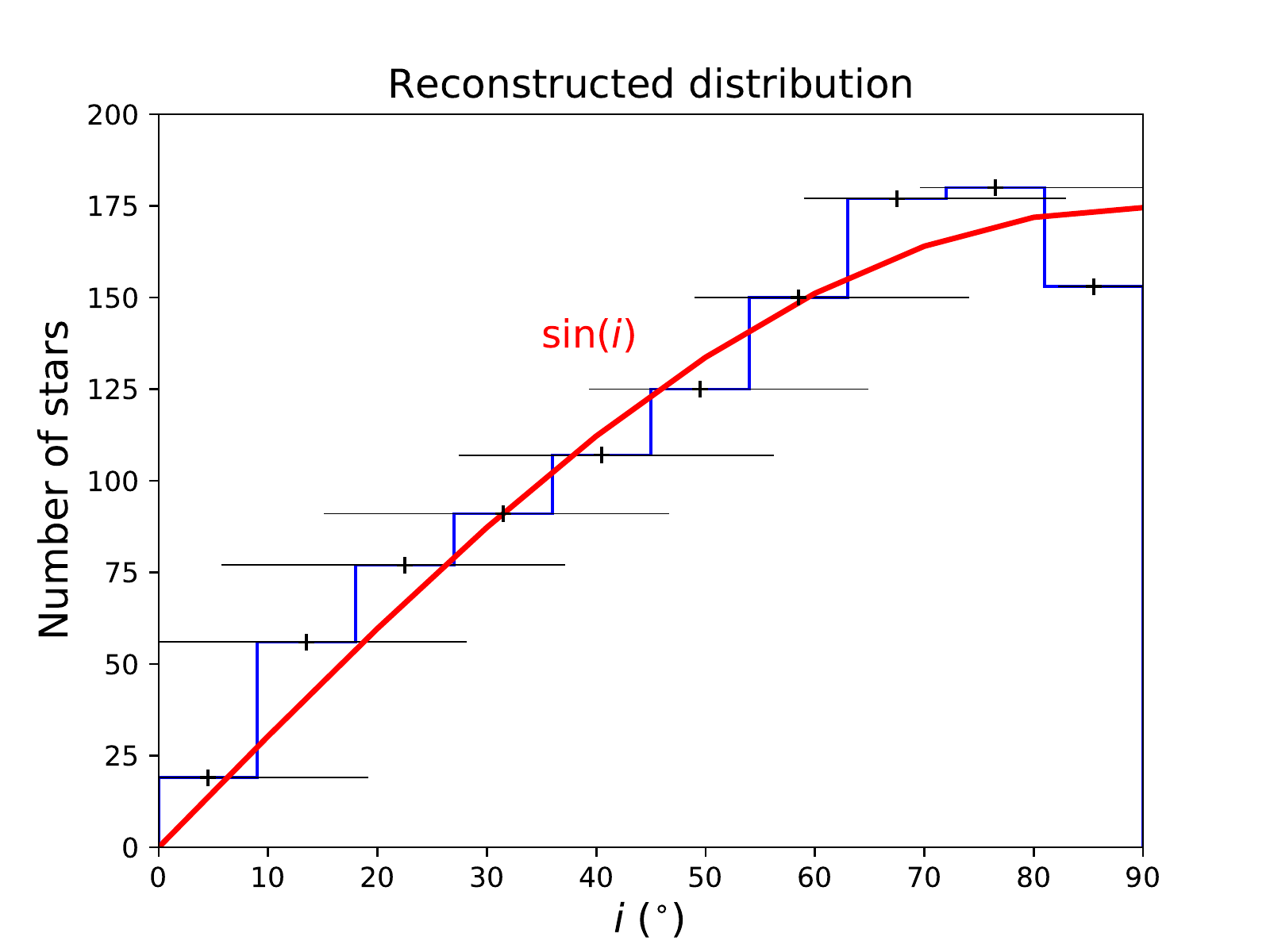}
\caption{Reconstructed distribution of stellar inclinations for our sample of 1139 RGB stars. We assumed an isotropic PDF in $\sin(i)$ for all stars presenting one rotational component as well as for stars presenting two rotational components and a $m=0$ component lost in the background. We used a Gaussian PDF for stars presenting three rotational components as well as for stars with two rotational components and a mean HBR equal to or above 8 for the retrieved $m=0$ component. The red line represents an isotropic distribution in $\sin(i)$ derived from a fit to the histogram. Mean error bars are represented in black for each inclination bin.}
\label{fig-i-pdf}
\end{figure*}

\begin{figure}
\centering
\includegraphics[width=10cm]{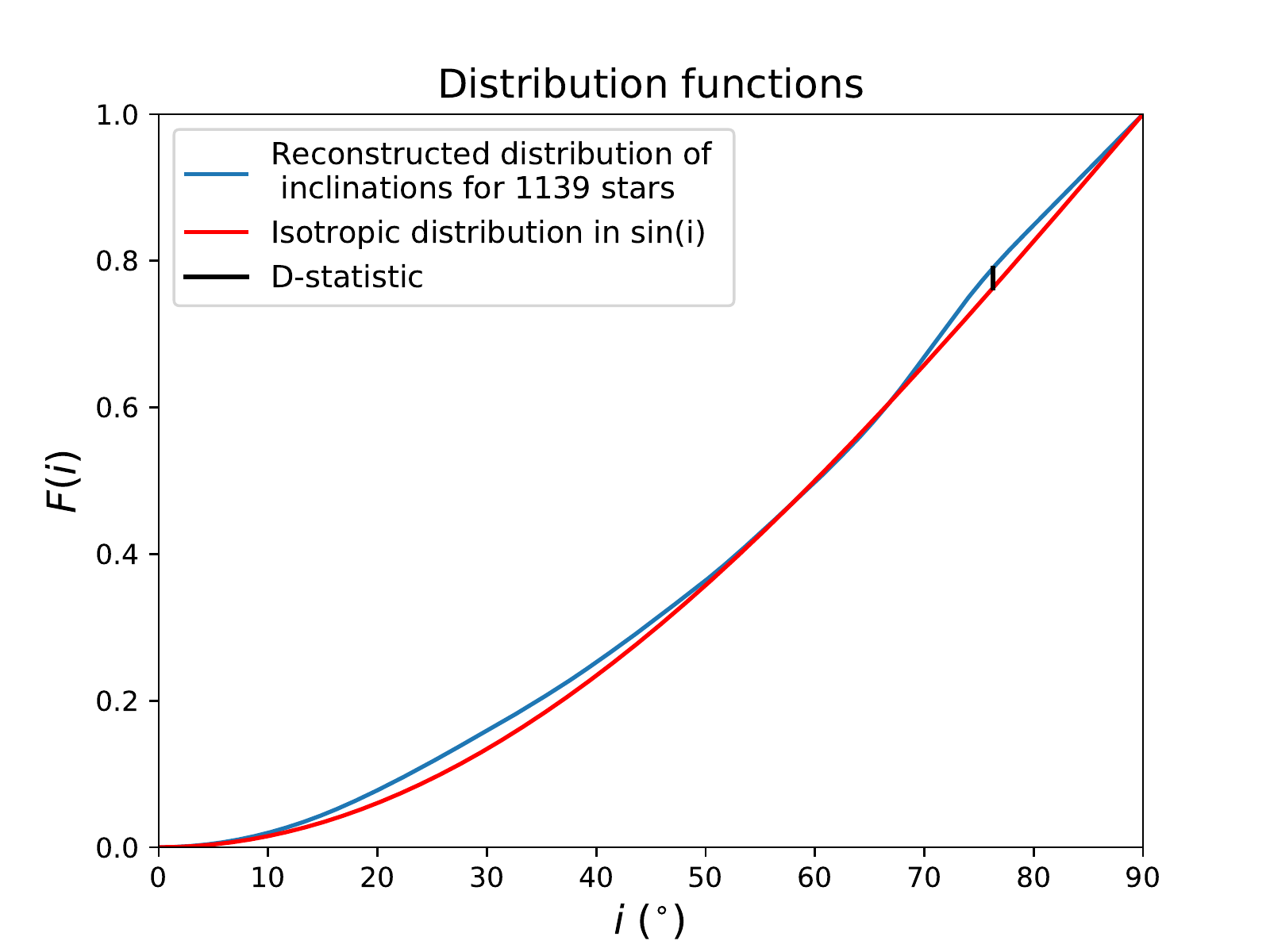}
\caption{Cumulative distribution functions of our sample of 1139 stars (in blue) computed from the inclination distribution in Fig.~\ref{fig-i-pdf}, and of a purely isotropic distribution in $\sin(i)$ (in red). The D-statistic is highlighted in black.}
\label{fig-distribution-function-us}
\end{figure}

\subsubsection{Low inclinations}

When the $m = \pm \, 1$ rotational components are missing, the lower error bar $\sigma\ind{i,-}$ is computed assuming that the lowest possible inclination is 0$^\circ$. The upper error bar $\sigma\ind{i,+}$ on the $i\ind{max}$ value derived from Eq.~(\ref{eqt-i-1}) corresponds to a 1-$\sigma$ uncertainty and is computed using Eq.~(\ref{eqt-sigma-i}). The upper error bar on $i\ind{medium}$ is computed assuming that the highest possible inclination is $i\ind{max} + \sigma\ind{i,+}$.

\subsubsection{High inclinations}\label{high-i}

When only the two rotational components with azimuthal orders $m = \pm \, 1$ are visible, there are two possible configurations (Sect. \ref{high-inclinations}) impacting the way uncertainties are derived.

For stars with a mean HBR of the m=0 rotational component $\left < P\ind{0} \right > \geq 8$, we have a firm measurement of the inclination (Sect. \ref{high-inclinations}) and we have symmetric error bars corresponding to 1-$\sigma$ uncertainties computed using Eq.~(\ref{eqt-sigma-i}).

The other configuration corresponds to stars with a mean HBR $\left < P\ind{0} \right > < 8$, for which the inclination can in principle be as high as 90$^\circ$. However, as stressed by \cite{Kamiaka}, it is in practice impossible to measure $i = 90^\circ$, as this would require an infinite HBR corresponding to a strict absence of background noise from granulation. Therefore, we cannot access the inclination regime close to 90$^\circ$. Hence, stars that are in this configuration in our sample should have upper error bars allowing  the value $i = 90^\circ$ to be encompassed as a possible extreme value for the inclination. The upper error bar $\sigma\ind{i,+}$ for these stars is thus computed assuming that the highest possible inclination is $i = 90^\circ$. The lower error bar $\sigma\ind{i,-}$ corresponds to a 1-$\sigma$ uncertainty and  is computed using Eq.~(\ref{eqt-sigma-i}).

\section{Results}\label{results}

We analysed 1139 stars on the RGB\footnote{Results are publicly available on ADS} for which \cite{Gehan_2018} identified rotational components; we refer to this paper for the details of the data analysis. The oscillation spectra and the stretched-period échelle diagrams of some of these stars are shown in Appendix~\ref{appendix-spectra}, with the azimuthal order of mixed modes identified. We analyse the distribution of inclinations and test its compatibility with isotropy.

\subsection{Raw stellar inclinations}\label{raw-i}

When looking at stars randomly selected over a large fraction of the sky, we expect stellar inclinations to be randomly distributed following a $\sin(i)$ distribution. Stars analysed here have been considered following increasing KIC numbers. This is not supposed to induce any selection bias and inclinations derived in this study should be isotropically distributed.

The raw distribution of stellar inclinations of our sample appears to deviate significantly from an isotropic distribution (Fig.~\ref{fig-i-distribution}). This is expected because measuring $i=0$ or 90$^\circ$ is impossible in practice as explained in Sect.~\ref{high-i}, resulting in a dramatic deficit of values in these extreme inclination regimes. We obtain firm inclination measurements between 27$^\circ$ and 78$^\circ$ for stars presenting three rotational components and stars presenting two rotational components for which the mean HBR of the retrieved $m=0$ component is equal to or above 8 (Fig.~\ref{fig-i-distribution}). This is consistent with the conclusion of \cite{Kamiaka}, namely that reliable seismic inclination measurements are possible in the general case for $20^\circ \lesssim i \lesssim 80^\circ$. For $i < 27^\circ$, we can only derive an overestimated value of the inclination. The same phenomenon occurs at high inclination, where we can only derive an underestimated value of the inclination for $i > 77^\circ$. We have no inclination estimate below 12$^\circ$ and above 85$^\circ$. This is consistent with Fig. 3 of \cite{Kamiaka}, which  shows that reliable inclination measurements are possible between $8^\circ \lesssim i \lesssim 86^\circ$ for stars with a high HBR of 30 and an observation duration of 4 years. Our results for \textit{Kepler} red giants confirm that we cannot derive inclination estimates outside of these limit values.

We note that the inclination distribution for stars presenting three rotational components (light blue in Fig.~\ref{fig-i-distribution}) peaks around 55$^\circ$ and decreases below 45$^\circ$ and above 64$^\circ$. This is expected because heights of rotational components with different azimuthal orders are equal at 55$^\circ$, giving a minimal height contrast between components (Fig.~\ref{fig-HBR-contrast}). We checked that the values $i =45^\circ$ and $64^\circ$ correspond to a height ratio between components with different $m$ values that reaches values above approximately 2 (Fig.~\ref{fig-HBR-contrast}). This highlights the fact that the efficiency of the detection of all the $m$ components decreases when the height contrast between ridges with different azimuthal orders surpasses a certain threshold. For $i$ values above 64$^\circ$, the non-detected $m=0$ component can be retrieved at the mid-point between $m=-1$ and $m=+1$ components (Sect.~\ref{high-i} and Fig.~\ref{fig-echelle-2-components}), allowing us to avoid a drop in measured inclinations in this regime (violet in Fig.~\ref{fig-i-distribution}). However, we miss the  $m=\pm \, 1$ components for stars with $i$ values below 45$^\circ$. These stars are therefore lost among the population of stars showing only one rotational component, leading to the dip observed in Fig.~\ref{fig-i-distribution} between 40$^\circ$ and 50$^\circ$.

\subsection{Unbiased inclination distribution}\label{unbiased-i}

We cannot obtain a firm measurement of the inclination for stars presenting one rotational component or for stars with two components and a low mean HBR for the retrieved $m=0$ component, but only a range of possible values. We take individual uncertainties into account to test the inclination distribution in a statistical sense.

In the particular cases where only one ($m=0$) or only two ($m=\pm \, 1$) rotational components are hidden in the background noise, we assume that the inclination is isotropically distributed between the minimum possible value $i\ind{min}$ and the maximum possible value $i\ind{max}$ for each star. We thus assumed a PDF in $\sin(i)$ for these stars (Fig.~\ref{fig-pdf-1-component}). Additionally, we assumed a Gaussian PDF for stars presenting three rotational components as well as for stars with two rotational components and a mean HBR equal to or above 8 for the retrieved $m=0$ component. The Gaussian PDF is centred on the measured inclination value and has a standard deviation equal to $\sigma\ind{i} / (2 \sqrt{2 \ln 2})$ (Eq.~\ref{eqt-sigma}). By integrating the PDF of all these stars, we retrieve a global distribution of inclinations for which the statistic is free from the observational limitations hampering the measurement of low and high $i$ values.

We obtain a reconstructed global distribution that is roughly isotropic (Fig.~\ref{fig-i-pdf}). This is expected for the large sample of stars analysed in this study and emphasises that our method is not affected by bias coming from sample selection effects. The inclination estimates obtained between $12$ and $40^\circ$ in Fig.~\ref{fig-i-distribution} have been redistributed towards lower inclinations, and the estimates derived between $62$ and $85^\circ$ have been redistributed towards higher inclinations.
\newline

K19 highlighted that it is easier to identify modes when the star presents one or two rotational components compared to three components, as the power distribution between modes with different azimuthal orders leads to higher mode visibilities in these two configurations. This can lead to an observational bias when checking the inclination distribution, with a possible excess of stars with low and high inclinations compared to stars with intermediate $i$ values. Figure~\ref{fig-i-pdf} demonstrates that our method allows us to identify triplets as efficiently as singlets and doublets, as the recovered inclination distribution is compatible with isotropy. Hence, modes are still easily identified in the case of stars presenting three rotational components, even if they have lower relative heights compared to stars with one or two components.

Our results confirm the necessity to take uncertainties on the estimated angle into account through PDFs in order to recover the underlying inclination distribution, correcting the biased distribution obtained in Fig.~\ref{fig-i-distribution} and obtaining the reconstructed isotropic one in Fig.~\ref{fig-i-pdf}.
The fact that we recover an inclination distribution compatible with an isotropic trend when randomly analysing a large number of stars validates the approach developed in this work, which can be directly applied to any solar-type pulsator for which the azimuthal order of oscillation modes is identified.

\begin{figure}
\centering
\includegraphics[width=10cm]{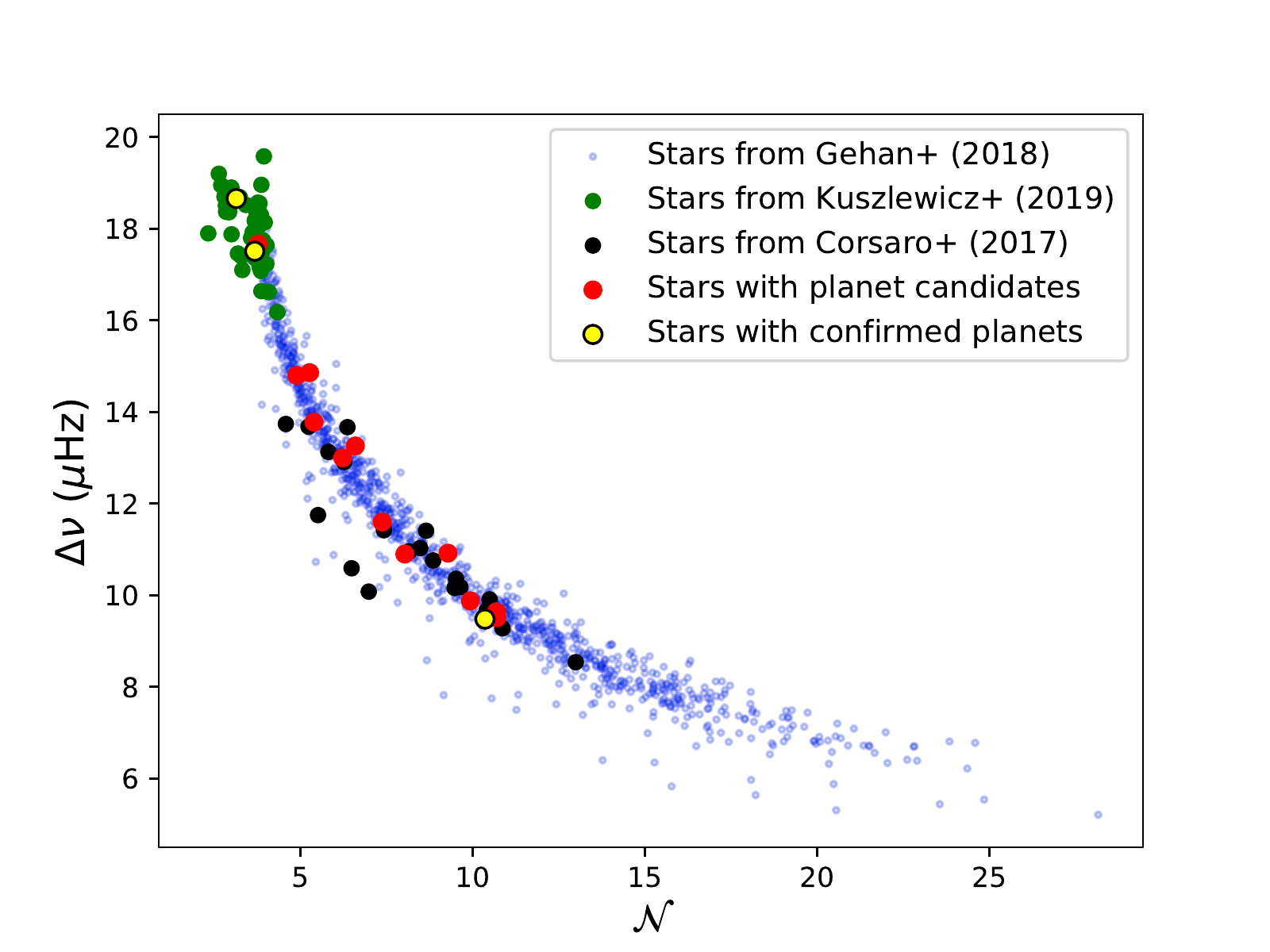}
\caption{Large separation as a function of the mixed-mode density. Stars analysed in Sect. \ref{results} are in blue, stars in common with \cite{Kuszlewicz} are in green, stars in common with \cite{Corsaro} are in black, stars with planet candidate(s) are in red, and stars with confirmed planet(s) are in yellow.}
\label{fig-Dnu-N}
\end{figure}

\subsection{Statistical analysis}\label{K-S-us}

In order to quantify the extent to which the reconstructed distribution of inclinations in Fig.~\ref{fig-i-pdf} is compatible with isotropy, we performed a Kolmogorov-Smirnov (K-S) test \citep{Kolmogorov, Smirnov} as used by K19. The K-S test consists in comparing the cumulative distribution functions (CDFs) $F\ind{1}(x)$ and $F\ind{2}(x)$ of two samples of size $n\ind{1}$ and $n\ind{2}$ by computing the maximum deviation between them at fixed inclination angle. In this work, we use a one-sample K-S test because we compare the CDF of our observational sample of size $n = 1139$ with a theoretical isotropic one \citep{Kolmogorov}. We computed the maximum deviation (D-statistic) between the CDF of our sample and the CDF of a purely isotropic distribution and found $D = 2.72 \, \%$ (Fig.~\ref{fig-distribution-function-us}).

The H$\ind{0}$ hypothesis we considered is that our observational sample comes from a purely isotropic distribution of inclinations \citep{Santos}.
For samples of size $n > 35$, the H$\ind{0}$ hypothesis is rejected if \citep{Smirnov}
\begin{equation}\label{eqt-D-bis}
D > \frac{c(\alpha)}{\sqrt{n}} = D\ind{H\ind{0}},
\end{equation}
where $c(\alpha)$ is a constant coming from the one-sample K-S table \citep{Smirnov} corresponding to a probability $\alpha$ of incorrectly rejecting the H$\ind{0}$ hypothesis.
We consider values of $\alpha$ from 1 to 10 \% to check what evidence there may be to reject H$\ind{0}$. We note that $\alpha=10$ \% provides only weak evidence for the rejection of the null hypothesis, and therefore there is no justification to consider values larger than that. For $\alpha=10$ \%, the K-S table gives $c(\alpha) = 1.22$, from which we obtain $D\ind{H\ind{0}} = 4.83$ \%. This value is above $D = 2.72$ \% computed for our sample, which indicates that the $H\ind{0}$ hypothesis is not rejected at the 10 \% level, and that no significant deviation from isotropy is detected.
We emphasise that the non-rejection of the $H\ind{0}$ hypothesis does not allow us to consider that the $H\ind{0}$ hypothesis is verified.
However, the non-rejection of the $H\ind{0}$ hypothesis indicates that the distribution of inclinations derived in this study is fully compatible with isotropy, as expected.

K19 and C17 determined the inclination of some \textit{Kepler} RGB and red clump stars through a Bayesian analysis of oscillation spectra. Below we explore the compatibility of these results with ours.

\begin{figure}
\centering
\includegraphics[width=10cm]{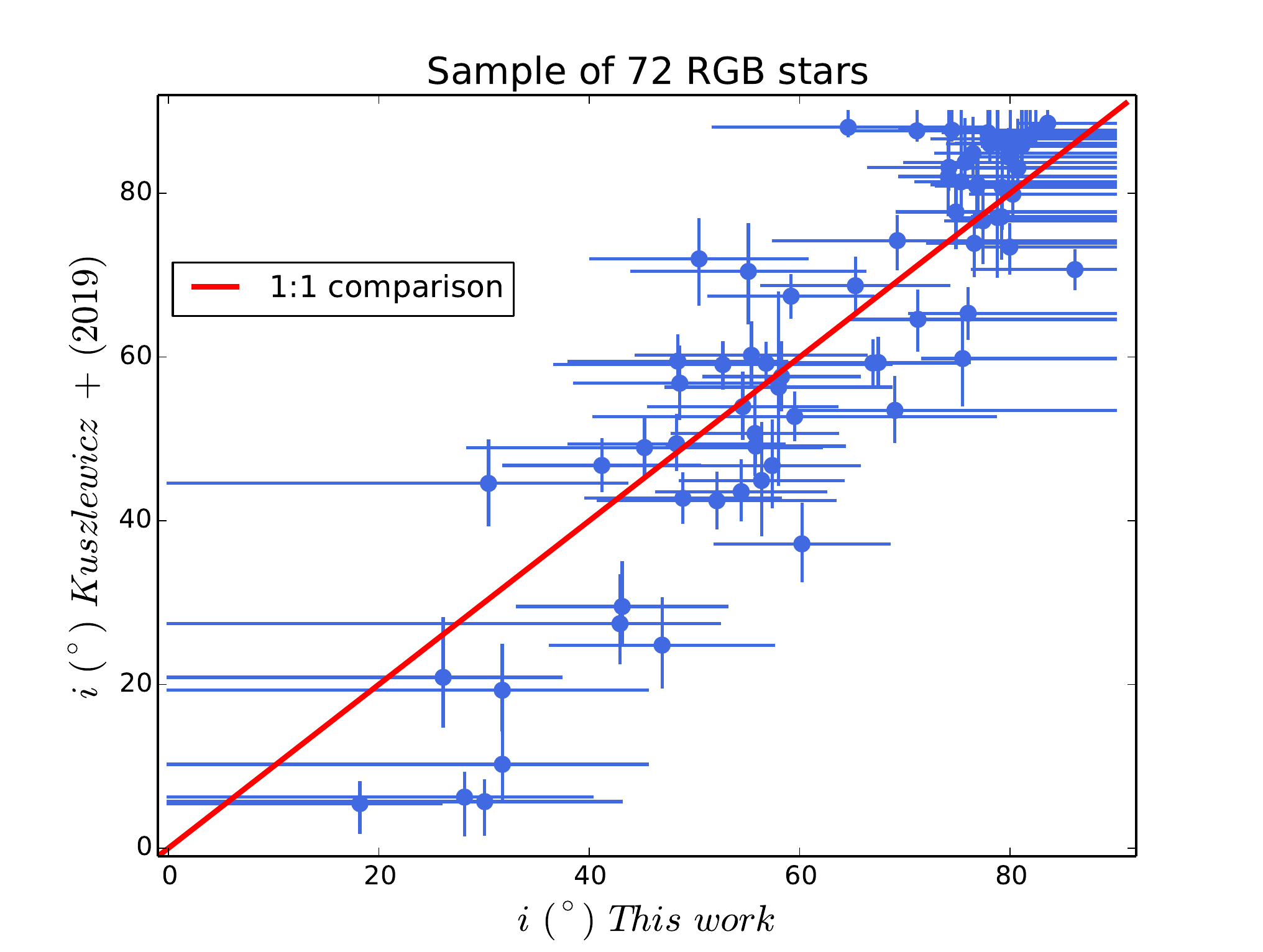}
\caption{Star-by-star comparison between \cite{Kuszlewicz} measurements and ours. The red line represents the 1:1 comparison.}
\label{fig-Kuszlewicz-star-by-star}
\end{figure}

\section{Comparison with \cite{Kuszlewicz} results}\label{Kuszlewicz}

K19 derived inclination measurements for 89 RGB stars. These latter authors selected stars with high $\numax$ values, which are therefore mostly above the confusion limit between rotational splittings and mixed-mode frequency spacings \citep{Mosser_2012c}, for which the identification of mixed modes is straightforward without using stretched oscillation spectra \citep{Mosser_2015} as in \cite{Gehan_2018}. Using the mixed-mode density defined in Eq.~(\ref{eqt-N}) as a proxy of stellar evolution on the RGB \citep{Gehan_2018}, we note that these stars lie on the lower RGB, with $\N$ values below 5 (in green in Fig.~\ref{fig-Dnu-N}). The 1139 stars analysed in this study have higher $\N$ values on average, namely as high as 28 (blue in Fig.~\ref{fig-Dnu-N}). Indeed, using stretched periods allows us to identify rotational multiplets even when they overlap, and therefore we can work at much lower $\numax$ than K19 and we are not limited to the lower RGB only. We were able to derive inclination measurements for 72 stars analysed by K19 (Table~\ref{table:Kuszlewicz}). We were not able to draw conclusions for 17 stars for which g-m modes have low HBR.

\begin{figure}
\centering
\includegraphics[width=10cm]{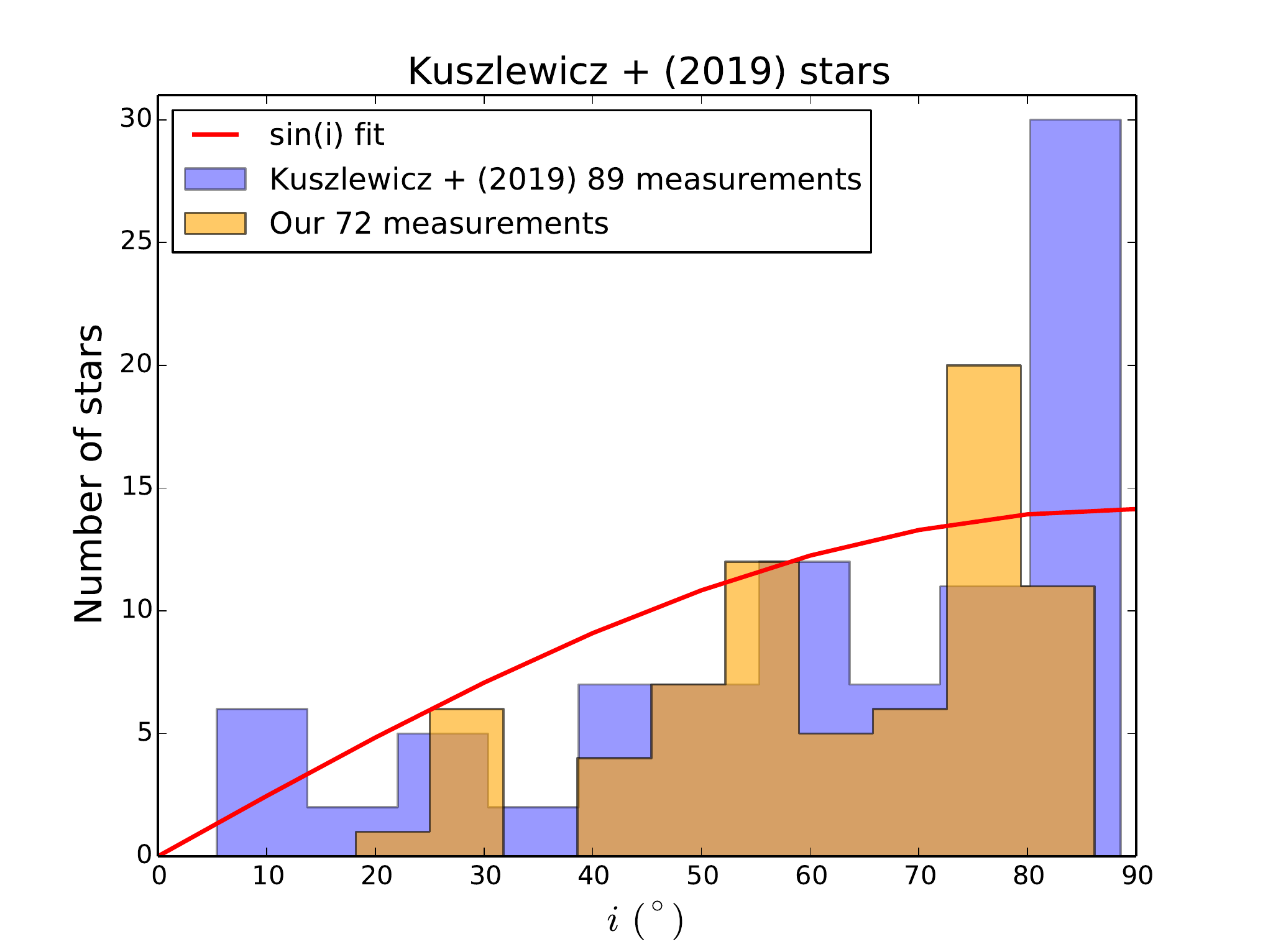}
\caption{Distribution of measured inclinations for the sample studied by \cite{Kuszlewicz}. The distribution resulting from \cite{Kuszlewicz} measurements for 89 stars is in blue, and the distribution resulting from our measurements for 72 stars is in orange. The red line represents an isotropic distribution in $\sin(i)$ deriving from a fit to the histogram in blue.}
\label{fig-Kuszlewicz-histograms}
\end{figure}

\subsection{Raw stellar inclinations}\label{raw-Kuszlewicz}

Our results globally agree with those of K19 for the 72 stars analysed (Fig.~\ref{fig-Kuszlewicz-star-by-star}). The oscillation spectra and the stretched-period échelle diagrams of some of these stars are shown in Appendix~\ref{appendix-spectra}, with the azimuthal order of mixed modes identified. The small differences we find for some stars can be explained by the fact that K19 included p-m modes to derive an estimate of the inclination, and not only g-m modes as in our study. We note that K19 used a Bayesian hierarchical method that provides inclination measurements derived from the combination of local estimates of $i$ instead of coming from a global analysis using all modes with the same azimuthal order together, as we did in this study. While \cite{Kamiaka} highlighted the fact that fitting oscillation modes individually can lead to biases in the inclination measurement, K19 used synthetic oscillation spectra to verify  that using hierarchical inference makes it possible to reliably extract $i$, even when using a local analysis.

\begin{figure}
\centering
\includegraphics[width=10cm]{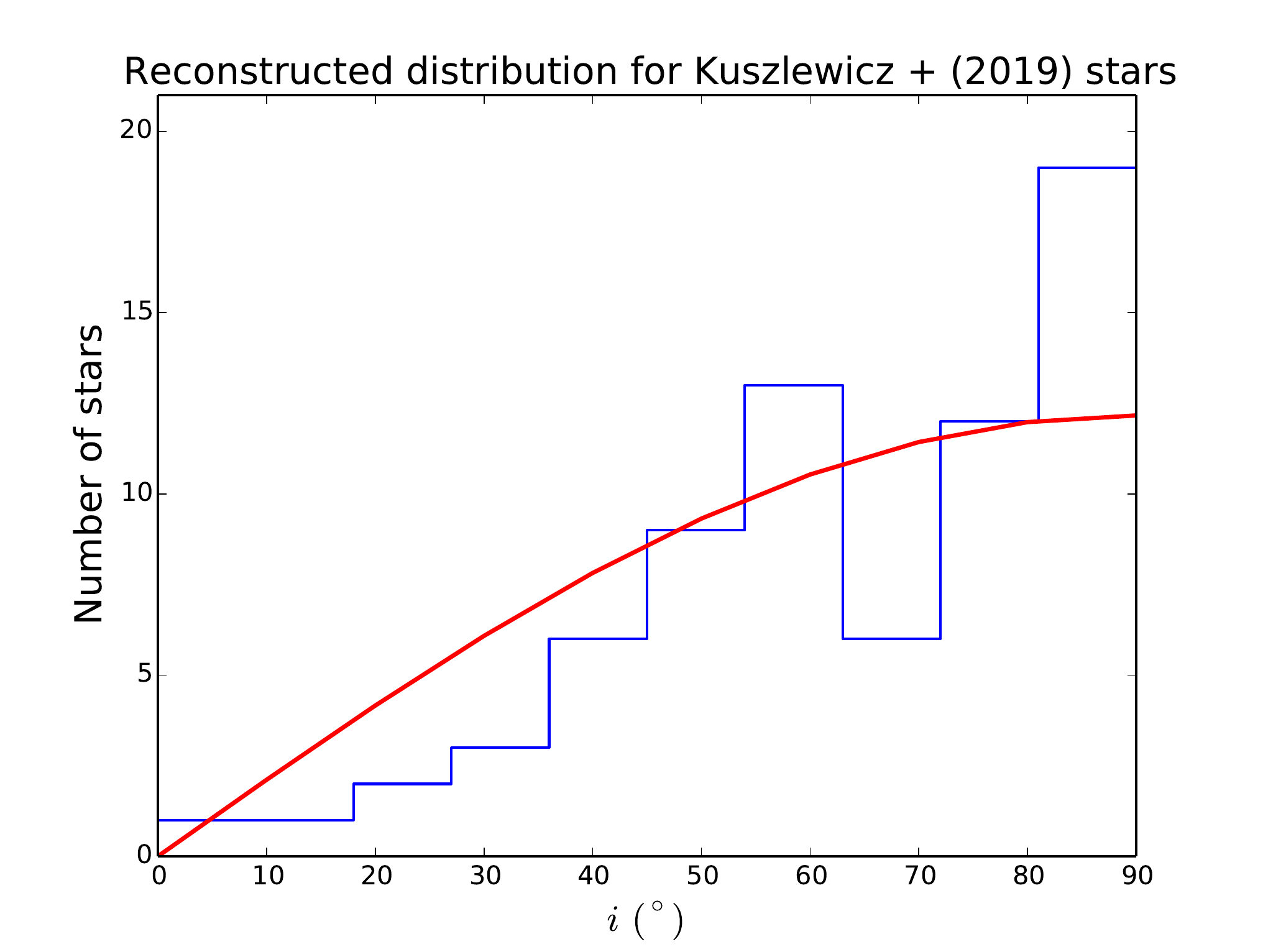}
\caption{Reconstructed distribution of stellar inclinations such as Fig.~\ref{fig-i-pdf} for the 72 stars we have in common with \cite{Kuszlewicz}. The red line represents an isotropic distribution in $\sin(i)$ derived from a fit to the histogram.}
\label{fig-Kuszlewicz-reconstructed}
\end{figure}

\begin{figure}
\centering
\includegraphics[width=10cm]{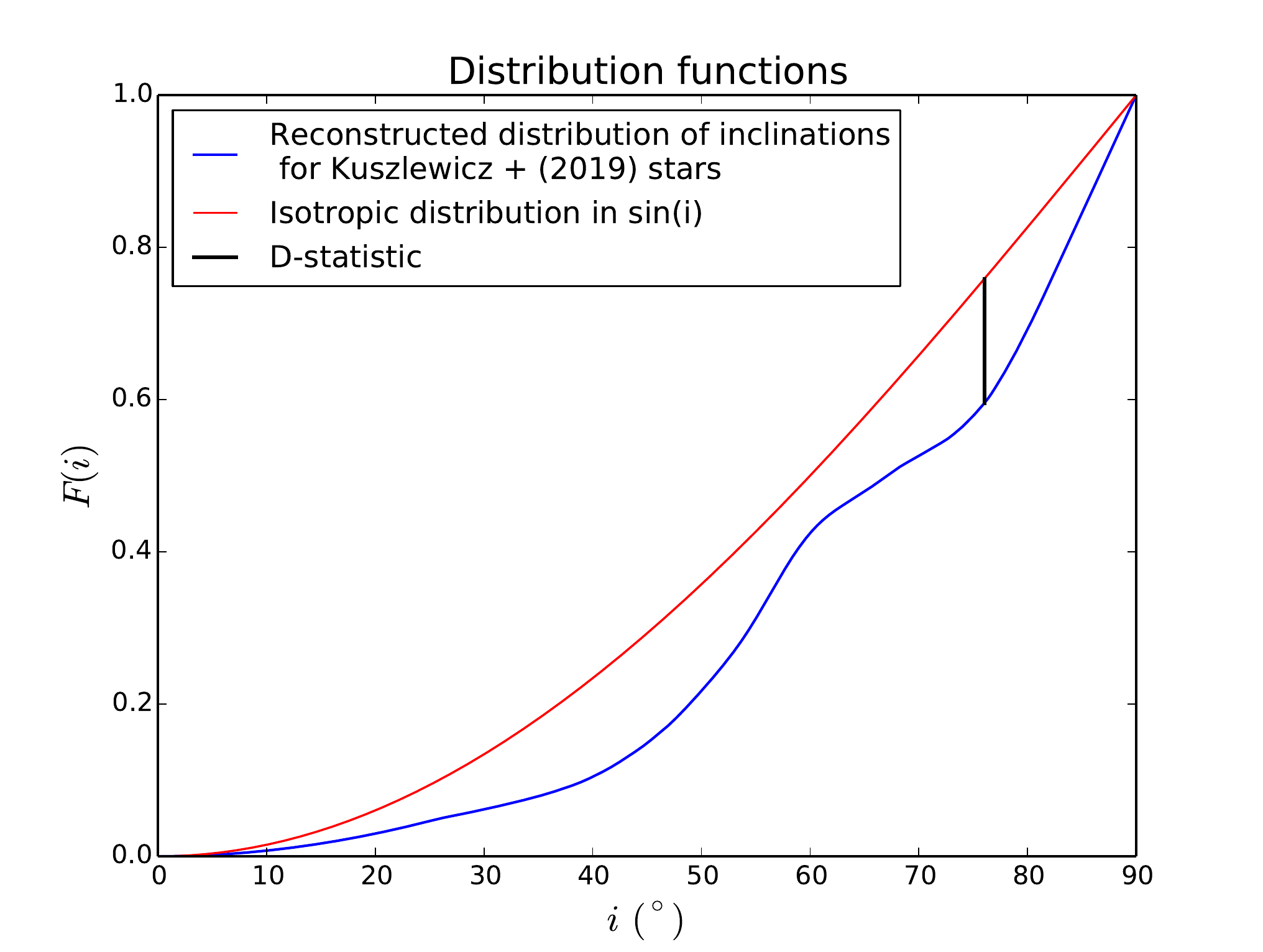}
\caption{Cumulative distribution functions of the sample of 72 stars from \cite{Kuszlewicz} (blue) computed from the inclination distribution in Fig.~\ref{fig-Kuszlewicz-reconstructed}, and of a purely isotropic distribution in $\sin(i)$ (red). The D-statistic is highlighted in black.}
\label{fig-Kuszlewicz-distribution}
\end{figure}

We also note that we have much larger uncertainties than those derived by K19 (Fig.~\ref{fig-Kuszlewicz-star-by-star}). K19 determined uncertainties as the 68.3 \% highest posterior density interval, which corresponds to formal 1-$\sigma$ uncertainties when the posterior distribution follows a normal law. Our uncertainties are also 1-$\sigma$, except with lower error bars for stars presenting only one rotational component and upper error bars for some stars presenting two rotational components. K19 used a Bayesian approach based on an isotropic prior for the inclination. They assumed that individual angles are all drawn from an isotropic underlying distribution, allowing the inclination angle of individual modes to differ slightly under the assumption that these differences are small. By contrast, we adopt a frequentist approach with no prior on the possible values of individual inclinations. We derive our uncertainties considering the spread in PSD for all modes with the same azimuthal order. These variations in PSD for a given azimuthal order can be significant along the spectrum due to the stochastic nature of oscillations. Therefore, it is not surprising that we end up with larger uncertainties compared to K19.

We obtain a very similar distribution of inclinations  to that found by K19, the main difference being that our distribution is squeezed with no extreme low and high inclination values (Fig.~\ref{fig-Kuszlewicz-histograms}).

\subsection{Unbiased inclination distribution}\label{unbiased-Kuszlewicz}

The reconstructed histogram of inclinations using PDFs for each of the 72 stars from the K19 sample (Fig.~\ref{fig-Kuszlewicz-reconstructed}) is not isotropic, in contrast to the reconstructed distribution we obtain for 1139 stars (Fig.~\ref{fig-i-pdf}). It is not too surprising that we do not recover a clean isotropic distribution because the sample analysed is limited. We can however note that there is a bump in the distribution around 55$^\circ$ as well as a clear overestimate of high inclinations above 80$^\circ$. K19 stated that the mode visibilities are higher for modes in singlets or doublets than for modes in triplets, making it easier to identify oscillation modes for these configurations. Under this assumption, the K19 sample is expected to have a deficit of intermediate inclinations. The isotropic distribution of inclinations results in a low probability of observing a star with $i$ close to 0$^\circ$. An overestimate of stars with $i$ close to 90$^\circ$ is therefore expected if the observational bias induced by the relative heights of oscillation modes is not corrected. Figure~\ref{fig-Kuszlewicz-reconstructed} may suggest that K19 could not overcome this bias, which could potentially result in a non-isotropic sample of stars for which they derive the inclination angle.

\begin{figure}
\centering
\includegraphics[width=10cm]{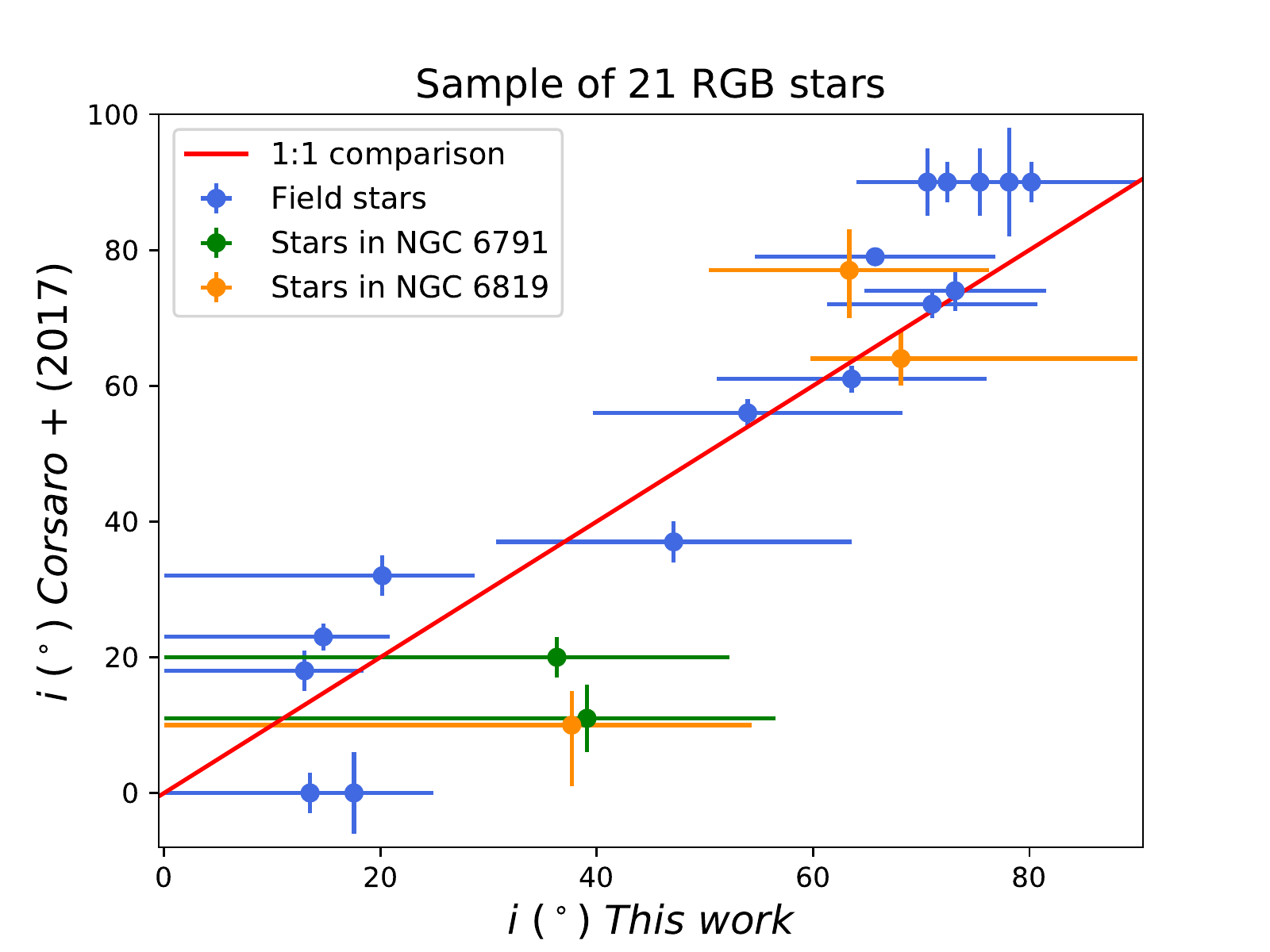}
\caption{Star-by-star comparison between \cite{Corsaro} measurements and ours for RGB stars. Field stars are in blue, stars from NGC 6791 are in green, and stars from NGC 6819 are in orange. The red line represents the 1:1 comparison.}
\label{fig-Corsaro-star-by-star}
\end{figure}

\begin{figure}
\centering
\includegraphics[width=10cm]{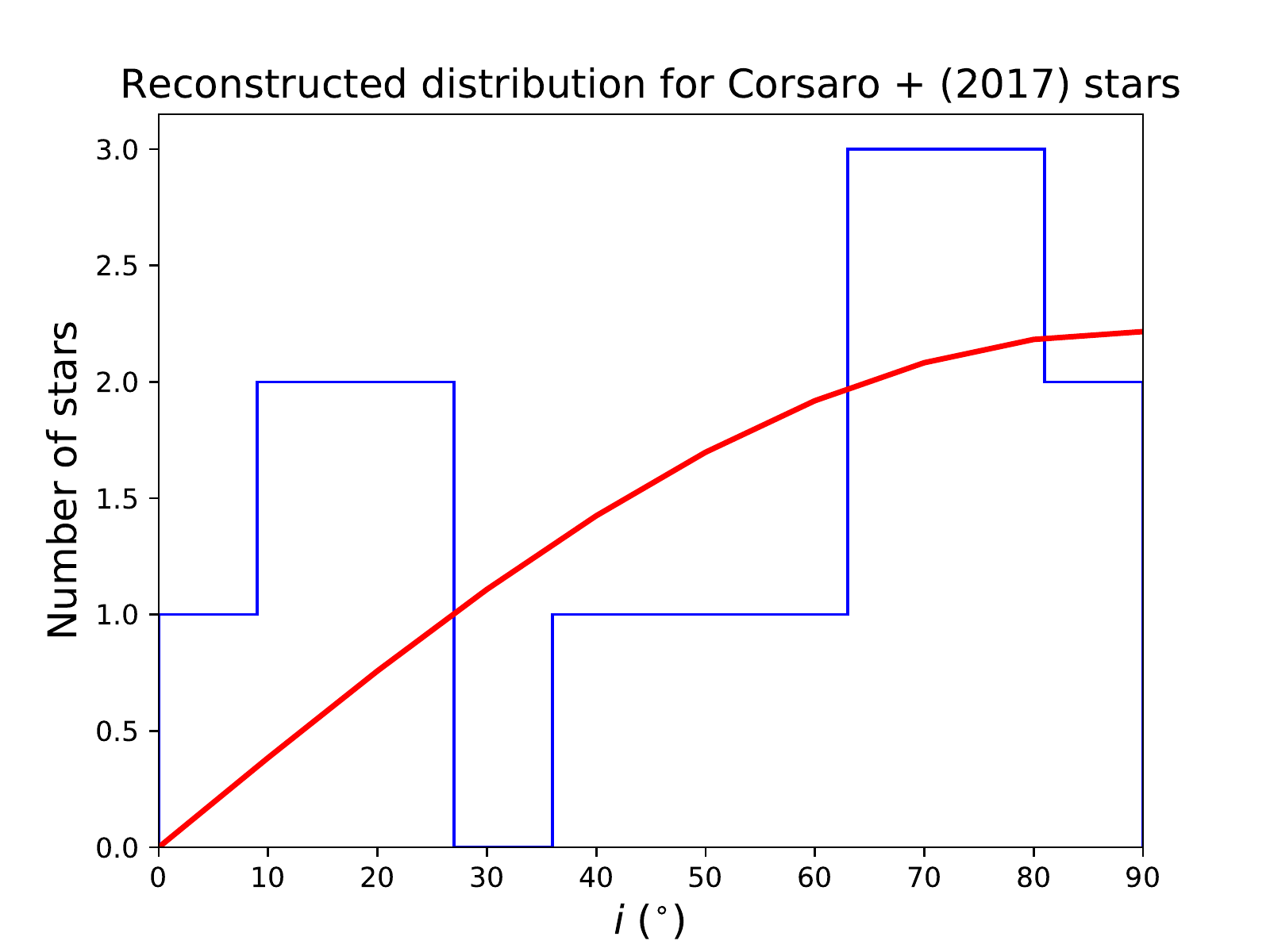}
\caption{Reconstructed distribution of stellar inclinations such as Fig.~\ref{fig-i-pdf} for the 16 RGB stars we have in common with \cite{Corsaro}. The red line represents an isotropic distribution in $\sin(i)$ derived from a fit to the histogram.}
\label{fig-Corsaro-reconstructed}
\end{figure}

\begin{figure}
\centering
\includegraphics[width=10cm]{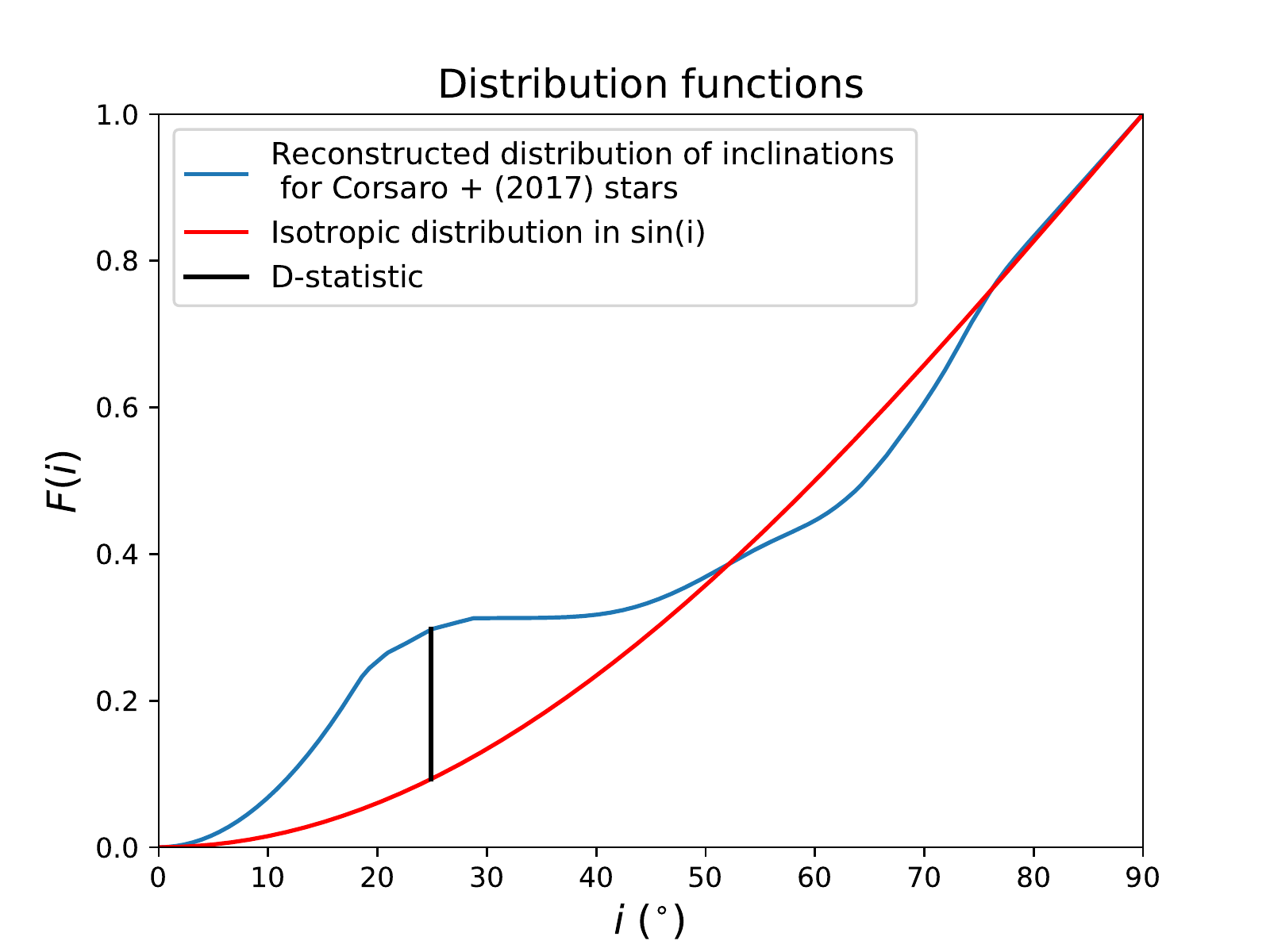}
\caption{Cumulative distribution functions of the sample of 16 RGB field stars from \cite{Corsaro} (blue) computed from the inclination distribution in Fig.~\ref{fig-Corsaro-reconstructed}, and of a purely isotropic distribution in $\sin(i)$ (red). The D-statistic is highlighted in black.}
\label{fig-Corsaro-distribution}
\end{figure}

\subsection{Statistical analysis}\label{stat-Kuszlewicz}

In order to quantify the extent to which the reconstructed distribution of inclinations in Fig.~\ref{fig-Kuszlewicz-reconstructed} is representative of a sample of 72 stars selected among an isotropic inclination distribution, we performed a K-S test in the same way as in Sect.~\ref{K-S-us}.

We computed the D-statistic for the CDF of our sample of 72 stars with respect to the CDF of a purely isotropic distribution and found $D = 16.33$ \% (Fig.~\ref{fig-Kuszlewicz-distribution}).
For the rejection level $\alpha = 5$ \%, we have $c(\alpha) = 1.36$ from the K-S table and we obtain $D\ind{H\ind{0}} = 16.03$ \% for our sample of size $n = 72$ (Eq.~(\ref{eqt-D-bis})). This value is below $D = 16.33$ \% computed for our sample. This indicates that the $H\ind{0}$ hypothesis is rejected at the 5 \% level, and we can state that the inclination distribution we have is not isotropic.

These results from the K-S test are consistent with the analysis of K19, who found $D \simeq 17$ \% for the 89 stars they analysed (Fig. 8 of K19). The corresponding D-statistic necessary to reject the $H\ind{0}$ hypothesis at the 2 \% level is $D\ind{H\ind{0}} = 16.11$ \% for a sample of size $n = 89$ (Eq.~(\ref{eqt-D-bis})). The $H\ind{0}$ hypothesis is therefore also rejected at the 2 \% level for the total sample analysed by K19.

This statistical analysis emphasises the fact that K19 could only derive inclinations for a biased sample, which is intrinsically non-isotropic. This is not surprising because they were able to fit 89 stars out of their sample of 123 stars. This bias comes from the relative mode visibilities that lead to an easier identification of mixed modes when the inclination is close to 55$^\circ$ or close to 90$^\circ$. This evidence does not diminish the interest of their work.

\section{Comparison with \cite{Corsaro} results}\label{Corsaro}

C17 derived inclination measurements for 84 RGB and clump stars. Of these, 36 are field stars, 25 belong to the open cluster NGC 6791, and 23 to the open cluster NGC 6819. These latter authors find that low inclinations are favoured in the two open clusters NGC 6791 and NGC 6819, and therefore that inclinations do not follow the expected isotropic distribution. They concluded that stellar spins seem to be preferentially aligned towards the line of sight for those two open clusters. This is a relatively unexpected result that was contradicted by \cite{Mosser_2018}, who reanalysed C17 stars belonging to NGC 6819 and found that the distribution of inclinations follows an isotropic trend.

The method of \cite{Gehan_2018} is not suitable for analysing clump stars. Indeed, oscillation modes present larger mode line widths combined with lower rotational splitting values compared to RGB stars, making rotational components very hard to disentangle because they are very close to each other in stretched-period échelle diagrams (extension of the upper part of Fig. 3 of \cite{Gehan_2018}). We therefore focus here on the comparison between our results and those of C17 for RGB stars only. C17 have 36 RGB stars in their sample, including 19 field stars, 13 stars belonging to NGC 6791, and 4 stars belonging to NGC 6819. We were able to derive inclination measurements for 21 RGB stars analysed by C17, including 16 field stars, 3 stars from NGC 6819, and 2 stars from NGC 6791 (Table~\ref{table:Corsaro} and Figs in Sect.~\ref{appendix-spectra}). We were not able to form conclusions for all stars analysed by C17 because some of them have a low HBR for g-m modes, in particular for the open cluster NGC 6791.

\subsection{Raw stellar inclinations}\label{raw-Corsaro}

The measurements we derived for RGB stars are in global agreement with those obtained by C17, not only for field stars but also for cluster stars (Fig.~\ref{fig-Corsaro-star-by-star}). As we already noted for K19 results, uncertainties derived by C17 are much smaller than ours, while corresponding to a 1-$\sigma$ confidence interval. These differences are likely due to the different approaches used: C17 used a Bayesian approach including a prior on the inclination, while we used a frequentist approach with no prior to derive our measurements. Additionally, C17 were able to measure inclinations as low as 0$^\circ$ for two RGB stars that we analysed and as high as 90$^\circ$ for five stars (Fig.~\ref{fig-Corsaro-star-by-star}), which is not possible as explained in in Sect.~\ref{high-i}.

\begin{figure}
\centering
\includegraphics[width=10cm]{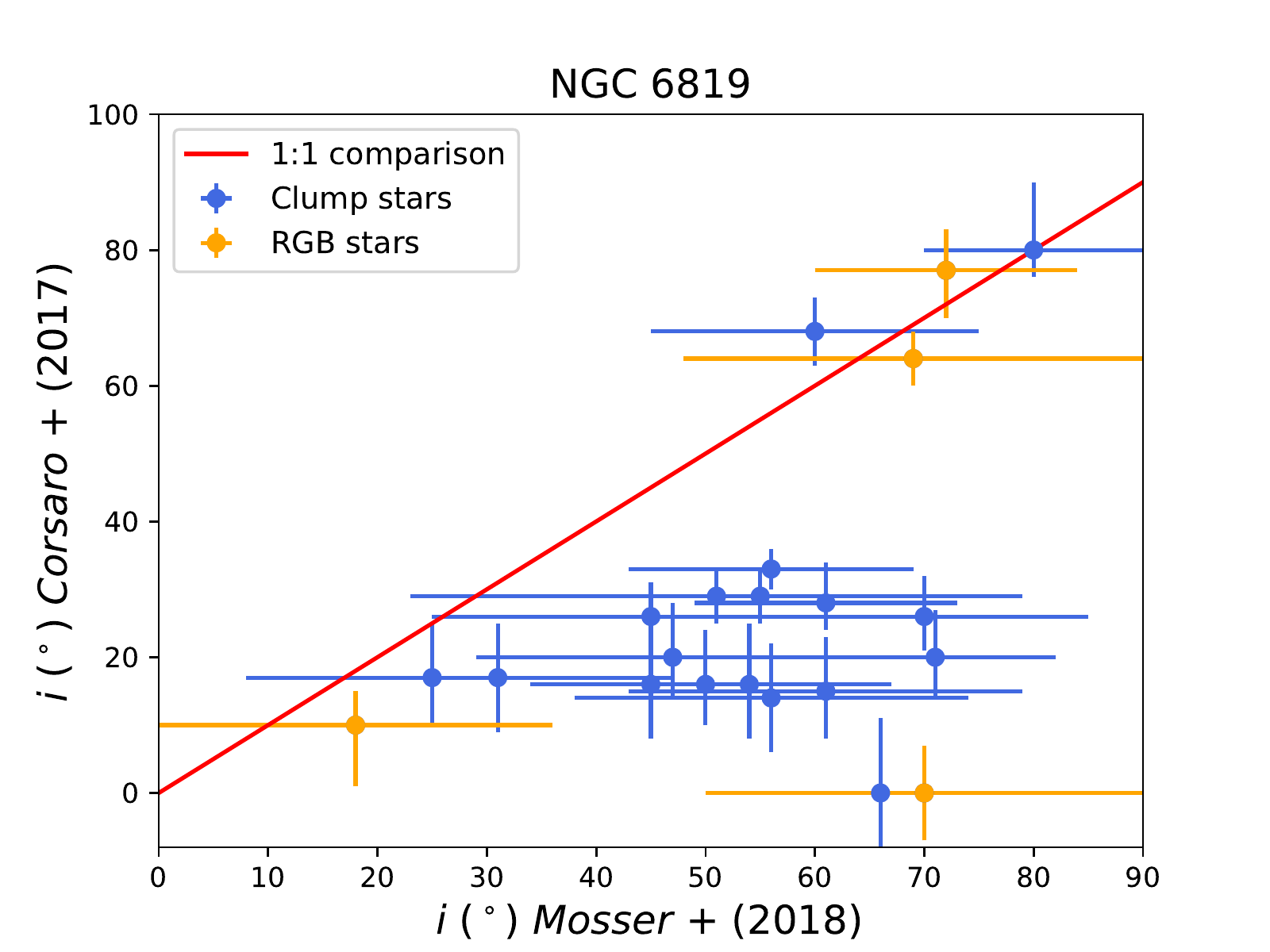}
\caption{Star-by-star comparison between \cite{Corsaro} and \cite{Mosser_2018} measurements for stars belonging to NGC 6819. Clump stars are in blue, and RGB stars are in orange. The red line represents the 1:1 comparison.}
\label{fig-Corsaro-Mosser}
\end{figure}

\subsection{Statistical analysis}\label{stat-Corsaro}

C17 claimed that the distribution of inclinations they derive for their sample of 36 RGB and clump field stars is in agreement with an isotropic distribution. We obtained the reconstructed histogram of inclinations using PDFs for each of the 16 field stars from the C17 sample (Fig.~\ref{fig-Corsaro-reconstructed}). In order to assess the extent to which the reconstructed distribution of inclinations is representative of a sample of 16 stars selected among an isotropic inclination distribution, we performed a K-S test in the same way as in Sect.~\ref{K-S-us} and Sect.~\ref{stat-Kuszlewicz}.

We computed the D-statistic for the CDF of our sample of 16 field stars with respect to the CDF of a purely isotropic distribution and found $D = 20.42$ \% (Fig.~\ref{fig-Corsaro-distribution}).
For our sample of size $n = 16$, the K-S table directly gives \citep{Smirnov} $D\ind{H\ind{0}} = 29.47$ \% for the low rejection level $\alpha = 10$ \%, which is above $D = 20.42$ \% computed for our sample.
This indicates that the $H\ind{0}$ hypothesis is not rejected to the 10 \% level. The inclination distribution for C17 stars therefore does not exhibit significant deviation from isotropy.
Therefore, there is no obvious observational bias in the sample selection, as stated by C17.

\begin{table*}
\caption{Properties of RGB stars with a planet candidate(s).}
\label{table:candidates}
\centering
\begin{tabular}{c c c c}
\hline\hline
KIC & $i$ ($^\circ$) & $\sigma \ind{i, +}$ ($^\circ$) & $\sigma \ind{i, -}$ ($^\circ$)\\
\hline
4953262 & 39.9 & 25.0 & 25.0\\
5110453 & 42.0 & 16.8 & 16.8\\
5115688 & 30.4 & 13.2 & 30.4\\
6425377 & 44.0 & 13.0 & 13.0\\
7198587 & 68.8 & 21.2 & 8.0\\
7458743 & 76.7 & 13.3 & 6.9\\
8016650 & 33.0 & 14.4 & 33.0\\
9145861 & 74.7 & 6.7 & 6.7\\
9475697 & 20.2 & 8.5 & 20.2\\
10790401 & 67.2 & 22.8 & 8.0\\
11097752 & 63.6 & 13.3 & 13.3\\
11911929 & 47.33 & 10.7 & 10.7\\
\hline
\end{tabular}
\end{table*}

\begin{table*}
\caption{Properties of RGB stars with a confirmed planet(s).}
\label{table:planets}
\centering
\begin{tabular}{c c c c c c c}
\hline\hline
Name & KIC & Number of planets & $\N$ & $i$ ($^\circ$) & $\sigma \ind{i,+}$ ($^\circ$) & $\sigma \ind{i,-}$ ($^\circ$)\\
\hline
Kepler-56 & 6448890 & 3 & 3.7 & 51.7 & 12.5 & 12.5\\
Kepler-91 & 8219268 & 1 & 10.4 & 73.5 & 16.5 & 6.9\\
Kepler-432 & 10864656 & 2 & 3.1 & 84.1 & 5.9 & 2.6\\
\hline
\end{tabular}
\end{table*}

\subsection{Inclinations for open clusters}\label{cluster-Corsaro}

For the RGB stars analysed here, we obtain inclination measurements in agreement with those derived by C17 (Fig.~\ref{fig-Corsaro-star-by-star}). However,  \cite{Mosser_2018}  obtained inclination measurements for red giants of the open cluster NGC 6819 that contradict the results of C17  (Fig.~\ref{fig-Corsaro-Mosser}). Part of the explanation might come from the fact that the great majority of these NGC 6819 stars are clump stars, for which C17 overestimate low inclinations (Fig.~\ref{fig-Corsaro-Mosser}). For three RGB stars out of four, results obtained by C17 and \cite{Mosser_2018} are in agreement. Figure~\ref{fig-Corsaro-Mosser} therefore suggests that mixed modes are not always correctly identified by C17 in the case of clump stars. This is highlighted by Supplementary Figure 1 of C17 where only $m=0$ modes are identified, resulting in an inclination of $20^{+8}\ind{-6}$$^\circ$, while Fig. 16 of \cite{Mosser_2018} reveals that the three components with $m=\{-1, 0, 1\}$ are visible and the corresponding inclination is $47 \pm 18 ^\circ$. This mode misidentification is not surprising as clump stars present larger line widths together with smaller rotational splittings compared to RGB stars. In these conditions, it is  more probable to identify only one rotational component while two or three are actually visible, resulting in an underestimated inclination. The conclusion of C17 regarding aligned stellar spins in clusters must therefore be considered with caution. As our method is not suitable for clump stars, we cannot directly check this hypothesis by deriving $i$ measurements for clump stars in the open clusters NGC 6819 and 6791. This merits a dedicated study with another method to analyse clump stars.

\section{Red giants with a planet candidate(s) or a confirmed planet(s)}\label{planets}

As mentioned in Sect.~\ref{introduction}, inclination measurements for planet-hosting stars provide information on the obliquity of planetary systems (Eq.~\ref{eqt-obliquity}). For stars with transiting planet(s), we expect that,  on average, the lower the inclination, the larger the obliquity, and therefore the higher the degree of spin-orbit misalignment between the host star and the planet(s). In this context, we used the NASA Exoplanet Archive cumulative table listing Kepler Objects of Interest (KOI)\footnote{https://exoplanetarchive.ipac.caltech.edu/cgi-bin/TblView/nph-tblView?app=ExoTbls\&config=cumulative} to lead a systematic search for RGB stars in our sample with a confirmed planet(s) and for stars with a planet candidate(s) that have not yet been confirmed or dismissed.

\subsection{Inclination and evolutionary stage}\label{i-N}

\begin{figure}
\centering
\includegraphics[width=10cm]{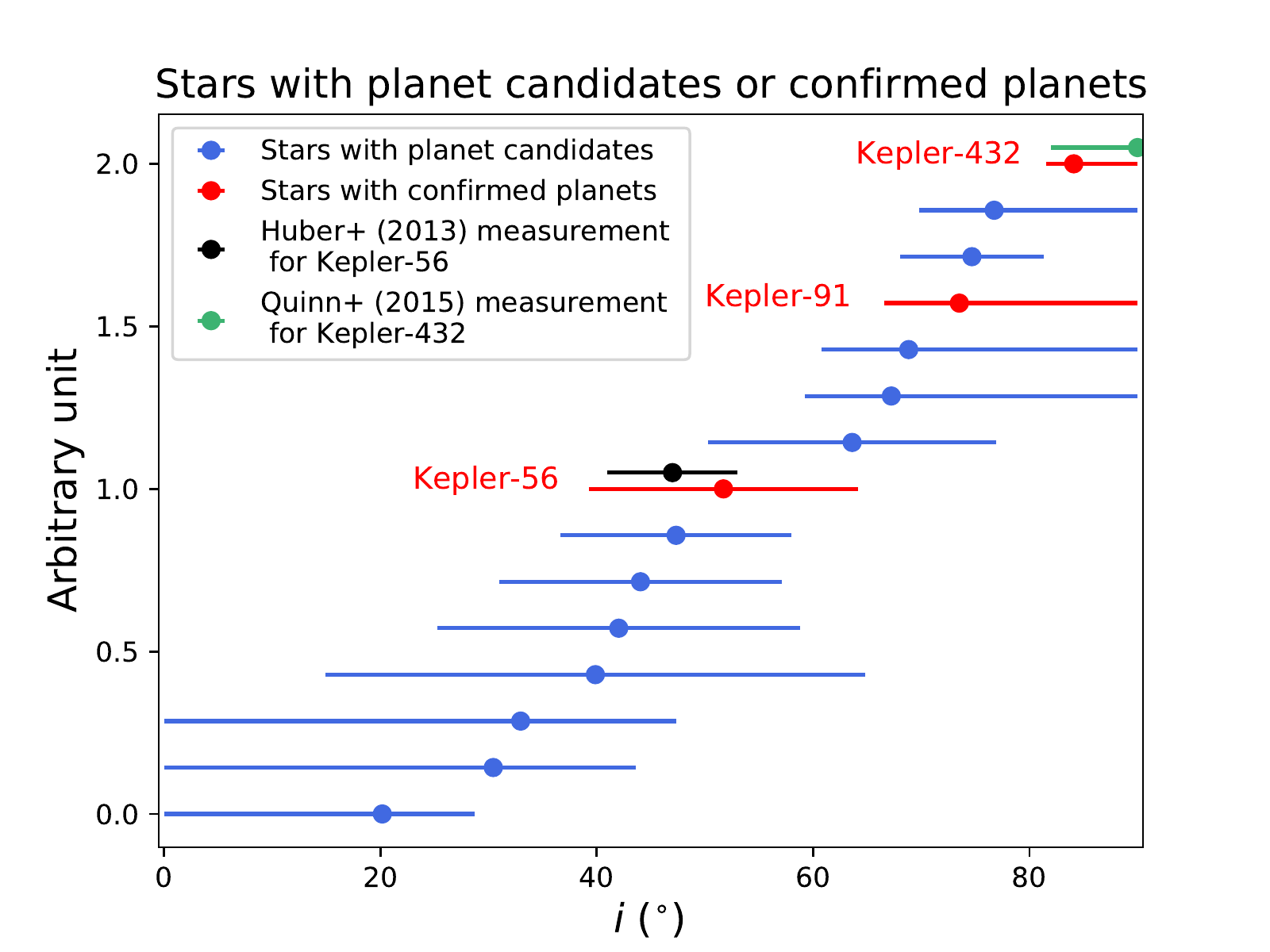}
\caption{Measured inclinations for stars with a planet candidate(s), in blue, and with a confirmed planet(s), in red. The y-axis is arbitrary. The \cite{Huber_2013} measurement for Kepler-56 with misaligned planets is shown in black, with the \cite{Quinn_2015} measurement for Kepler-432 in green.}
\label{fig-exoplanets}
\end{figure}

Among all the stars analysed in this study, we found 12 RGB stars in our sample with a planet candidate(s). These stars have inclinations distributed in the $[0, 90^\circ]$ range, with five stars presenting intermediate inclinations and three stars presenting low inclinations, being possibly pole-on (Fig.~\ref{fig-exoplanets} and Table~\ref{table:candidates}). We suggest that these 12 RGB stars should be analysed in priority to assess whether or not they have a confirmed transiting planet(s). If these stars turn out to host a planet(s), it will be possible to put constraints on the obliquity of the planetary orbit because we have measured their inclination angle. The eight aforementioned stars with low and intermediate inclinations should in particular present large obliquities should a transiting planet(s) be confirmed. It is interesting to note that these stars present diverse evolutionary stages on the RGB, with mixed-mode density values between $\N = 3.8$ and $10.7$ (Fig.~\ref{fig-Dnu-N}). Any planet detection around these stars would provide constraints on planetary evolution and survival around evolved stars, until relatively far up on the RGB where stars already started to expand significantly. We also find one RGB star in
our sample belonging to an eclipsing binary system, KIC 8564976, which was also known as KOI-3890 before its binary nature was confirmed by \cite{Kuszlewicz_bis}. We find an inclination of $i = 74.6^{+15.4}\ind{-7.7}$$^\circ$ for this star, consistent with the value of $i=87.3^{+2.7}\ind{-1.1}$$^\circ$ found by \cite{Kuszlewicz_bis}.

Additionally, we find no stars with confirmed planet(s) among stars analysed in this study. We therefore considered the seven \textit{Kepler} evolved stars hosting exoplanet(s) presented by \cite{Lillo-Box_2016} and were able to measure the inclination of three of these stars (Fig.~\ref{fig-exoplanets} and Table~\ref{table:planets}). Two stars have large inclination values consistent with a star seen almost equator-on, and therefore the planetary orbit axis is likely to be approximately aligned with the star's rotation axis. Our measurement of $i = 84.1^{+5.9}\ind{-2.6}$$^\circ$ for Kepler-432 is compatible with \cite{Quinn_2015}, who measured $i = 90^{+0}\ind{-8}$$^\circ$ (Fig.~\ref{fig-exoplanets}). We measured  an inclination of $i = 51.7 \pm 12.5^\circ$ for Kepler-56, which is consistent with \cite{Huber_2013} who measured $i = 47 \pm 7 ^\circ$ (Fig.~\ref{fig-exoplanets}), from which they deduced a lower limit for the obliquity of $\psi > 37^\circ$. Unsurprisingly, two of these three host stars are on the very low RGB, with mixed-mode density values of $\N = 3.1$ and $3.7$ respectively (Fig.~\ref{fig-Dnu-N} and Table~\ref{table:planets}). However, the third star (Kepler-91) is more evolved, with $\N = 10.4$ (Fig.~\ref{fig-Dnu-N} and Table~\ref{table:planets}). Seismic scaling relations give $R \sim 7.33 \, \Rsol$ for this star, which is compatible with $R = 6.30 \pm 0.16 \, \Rsol$ indicated in \cite{Lillo-Box_2016}. This confirms that planets can survive far up on the RGB and not only on the early RGB phase. 

\subsection{Planetary orbit angle for a transiting planet}

One might be tempted to consider that $i\ind{p} \simeq 90^\circ$ for transiting planets, and therefore that Eq.~(\ref{eqt-obliquity}), which defines the obliquity, can be simplified to \citep{Kamiaka}
\begin{equation}\label{eqt-obliquity-approx}
\cos \psi \simeq \sin i \, \cos \lambda.
\end{equation}
However, a transit can occur if the inclination of the planetary orbit follows  
\begin{equation}
\left( 90^\circ - \theta \right) \leq i\ind{p} \leq \left( 90^\circ + \theta \right),
\end{equation}
with, assuming that the planet radius is much smaller than the stellar radius,
\begin{equation}\label{eqt-theta}
\theta = \arcsin \left( \frac{R\ind{\star}}{a} \right),
\end{equation}
where $R\ind{\star}$ is the stellar radius and $a$ is the semi-major axis of the planet \citep{Beatty}.

When considering transiting planets around Sun-like stars, Eq.~(\ref{eqt-obliquity-approx}) is mostly valid, with $\theta$ taking maximum values around $5.7^\circ$ for hot Jupiters that have typically $R\ind{\star}/a = 1/10$ \citep{Beatty}. We nevertheless note that, although we can still safely consider that $\sin \, i\ind{p} \simeq 1$ in the left-hand term of Eq.~(\ref{eqt-obliquity}) for $\theta$ values as high as $5.7^\circ$, we have $\cos \, i\ind{p} \simeq 0.1$ for $\theta = 5.7^\circ$. We therefore advocate caution when considering that $\cos \, i\ind{p} \simeq 0$ in the right-hand term of Eq.~(\ref{eqt-obliquity}) for close-in planets transiting Sun-like stars, such as hot Jupiters.

\begin{figure}
\centering
\includegraphics[width=10cm]{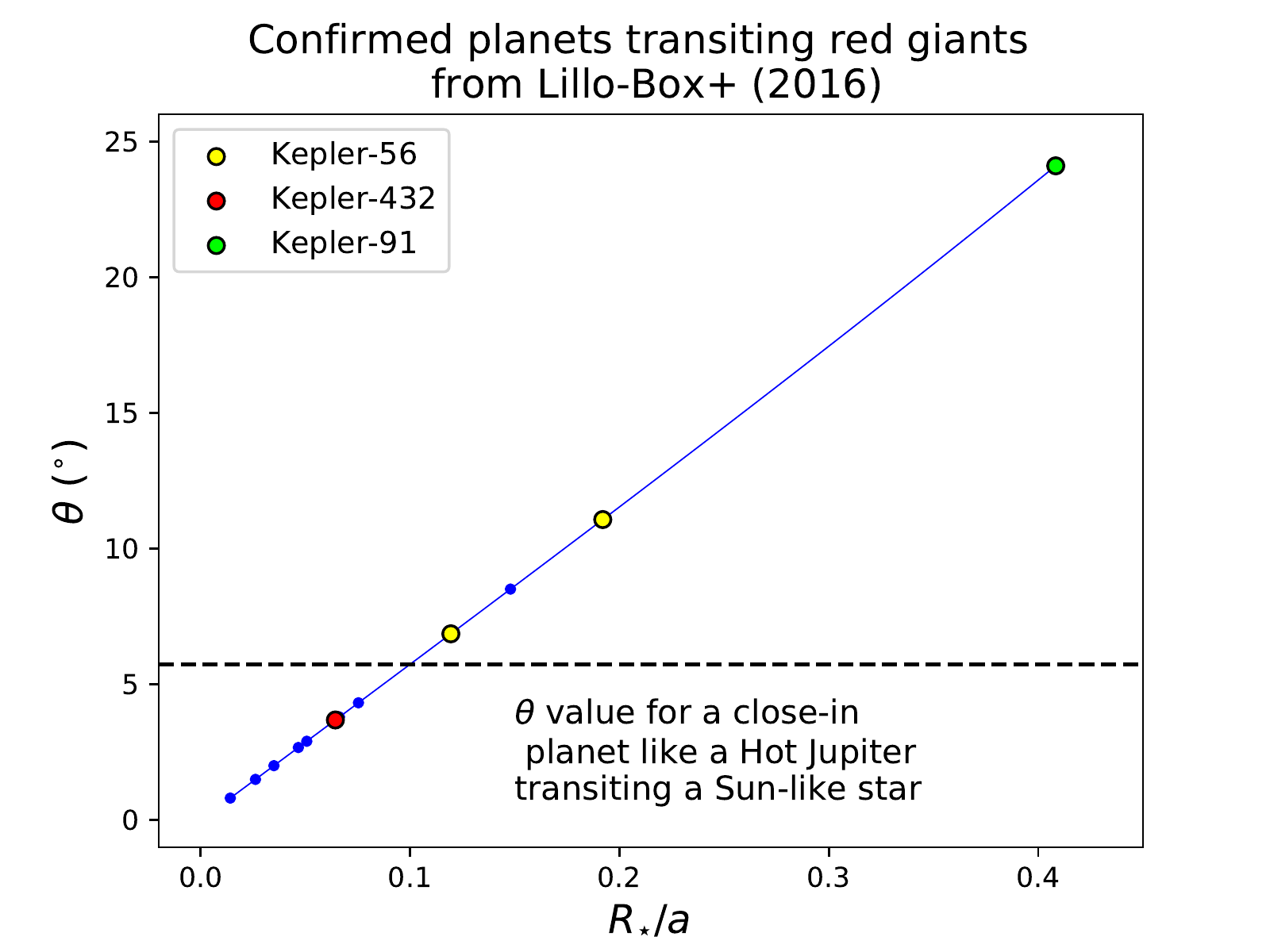}
\caption{Deviation angle $\theta$ (Eq.~\ref{eqt-theta}) as a function of the ratio between the stellar radius and the planetary semi-major axis for \textit{Kepler} RGB stars with confirmed transiting planets listed in \cite{Lillo-Box_2016}. Planets around RGB stars analysed in this study are coloured according to their host star. The black dashed line indicates the maximum $\theta$ value for planets transiting Sun-like stars computed for close-in planets like hot Jupiters, $\theta \simeq 5.7 ^\circ$.}
\label{fig-inclination_planetary_orbit_transit}
\end{figure}

Nevertheless, Eq.~(\ref{eqt-obliquity-approx}) can be inappropriate in many cases for transiting planets around evolved stars with larger radii. We considered planets around red giants listed by \cite{Lillo-Box_2016} and used the Extrasolar Planet Encyclopedia\footnote{http://exoplanet.eu/catalog/\detokenize{all_fields/}} to keep those that were detected by the transit method. All \textit{Kepler} planets in \cite{Lillo-Box_2016} are transiting planets, except Kepler-56d and Kepler-432c. Of the 12 transiting planets from \cite{Lillo-Box_2016}, 4 have $\theta$ values above the maximum value of $\theta = 5.7^\circ$ for Sun-like stars, including 3 planets that are transiting RGB stars analysed in this study (Fig.~\ref{fig-inclination_planetary_orbit_transit}). The planet around the most evolved RGB star (Kepler-91), which also has the largest radius, has the most important $\theta$ value with $\theta = 24.1^\circ$. As a consequence, while one can safely apply Eq.~(\ref{eqt-obliquity-approx}) to all planets around Sun-like stars up to close-in planets like hot Jupiters, this is no longer systematically true for more evolved stars with larger radii, and one has to consider Eq.~(\ref{eqt-obliquity}) as a whole in many cases.

\section{Conclusion\label{conclusion}}

This work highlights the biases that affect stellar inclination measurements and provides a way to infer the underlying statistical distribution of inclinations for a given population of stars.

We developed a general and automated approach to derive seismic measurements of the stellar inclination that can be applied to any solar-type pulsator for which oscillation modes are identified. We analysed RGB stars observed by \textit{Kepler}, including mainly field stars. Raw inclinations that are measured are not isotropically distributed, in contrast to what is expected for random inclinations. In particular, values of $i \lesssim 10 ^\circ$ and $i \gtrsim 85 ^\circ$ cannot be directly measured, and this observational limitation results in a biased distribution. The expected isotropic distribution is actually recovered when taking into account individual uncertainties on the inclination measurement for each star using PDFs, validating the method setup in this study. Our results emphasise the fact that unveiling the statistical significance of the inclination distribution of a given sample of stars is difficult, and that considering uncertainties in an accurate way is fundamental to retrieving the true underlying distribution.

Additionally, we were able to confirm that the distribution of inclinations derived by \cite{Kuszlewicz} for stars on the lower RGB is not isotropic and results from a bias in sample selection. We also checked the results obtained by \cite{Corsaro} and derived inclination measurements that are in agreement for field stars on the RGB. For the open clusters NGC 6791 and NGC 6819, we were only able to analyse a handful of RGB stars. Results obtained by \cite{Mosser_2018} for red giants in NGC 6819 suggest that mixed-mode misidentification is more common for clump stars analysed by \cite{Corsaro} than for RGB stars, calling into question the stellar spin alignment inferred by \cite{Corsaro} for the two open clusters NGC 6791 and NGC 6819. The analysis of clump stars is beyond the scope of this paper and merits a dedicated study to check this assumption in detail.

Moreover, we measured the inclination angle of 12 RGB stars with a planet candidate or candidates. These should be analysed in priority to assess whether or not they host planets as it would be possible to put constraints on the obliquity of the planetary orbit, with eight systems that should present large obliquities. We also derived the inclination of three RGB stars with a confirmed planet or planets, and our results are in agreement with previous measurements. Kepler-91 and Kepler-432 have large inclination values consistent with a low obliquity, and we confirm that the multiplanetary system Kepler-56 has an intermediate inclination angle, for which \cite{Huber} inferred a high obliquity corresponding to misaligned planets. We find that  Kepler-56 and  Kepler-432 unsurprisingly lie on the lower RGB. However, Kepler-91 is more evolved and has a larger radius, confirming that planets can survive relatively far up on the RGB.

Finally, we show that considering that the inclination angle of the planetary orbit can be safely approximated to 90$^\circ$ in the case of transiting planets is not always valid when the host star is a red giant with a larger radius than that of main sequence stars, which greatly impacts the determination of the obliquity. This assumption should therefore be systematically verified before deriving an estimate of the obliquity for planets transiting red giants.

This work revisits the problem of inferring the distribution of inclinations in large stellar samples. Such studies applied to open clusters can provide valuable insight into the physical conditions driving star formation.
\textit{Kepler} observed only four open clusters with large distance moduli \citep{Bossini_2019}. The undergoing TESS space mission \citep{Aguirre} has a near full-sky coverage, including a lower limit of about 600 open clusters \citep{Bouma}. We can therefore expect to have a significant number of open clusters observed by TESS with smaller distance moduli, potentially presenting higher signal-to-noise ratios compared to \textit{Kepler} clusters. As TESS has now two years of observations and the extended mission is now approved, we can expect a sufficiently high resolution to disentangle individual mixed modes in the oscillation spectra of red giants belonging to open clusters. This would allow us to unveil the physical conditions during star formation for tens and potentially hundreds of open clusters.

\begin{acknowledgements}
C. G. thanks Tiago Campante for helpful discussions as well as D. Stello, J. S. Kuszlewicz and K. J. Bell for constructive comments. This work was supported by FCT - Fundação para a Ciência e a Tecnologia through national funds (PTDC/FIS-AST/30389/2017), by FEDER - Fundo Europeu de Desenvolvimento Regional through COMPETE2020 - Programa Operacional Competitividade e Internacionalização (POCI-01-0145-FEDER-030389), and by FCT/MCTES through national funds (PIDDAC) by these grants UIDB/04434/2020 and UIDP/04434/2020.  M. S. C. is supported by FCT through a contract (CEECIND/02619/2017).
\end{acknowledgements}

\bibliographystyle{./bibtex/aa} 
\bibliography{./bibtex/biblio}

\begin{appendix}
\section{Inclination angles derived for \cite{Kuszlewicz} and \cite{Corsaro} stars}

We provide the inclination angle, the associated uncertainties, and the number of detected rotational components we derived for the 72 \cite{Kuszlewicz} stars in Table~\ref{table:Kuszlewicz} and for the 21 \cite{Corsaro} RGB stars in Table~\ref{table:Corsaro}.

\section{Examples of oscillation spectra and stretched period échelle diagrams}\label{appendix-spectra}

\begin{figure*}
\centering
\includegraphics[width=13cm]{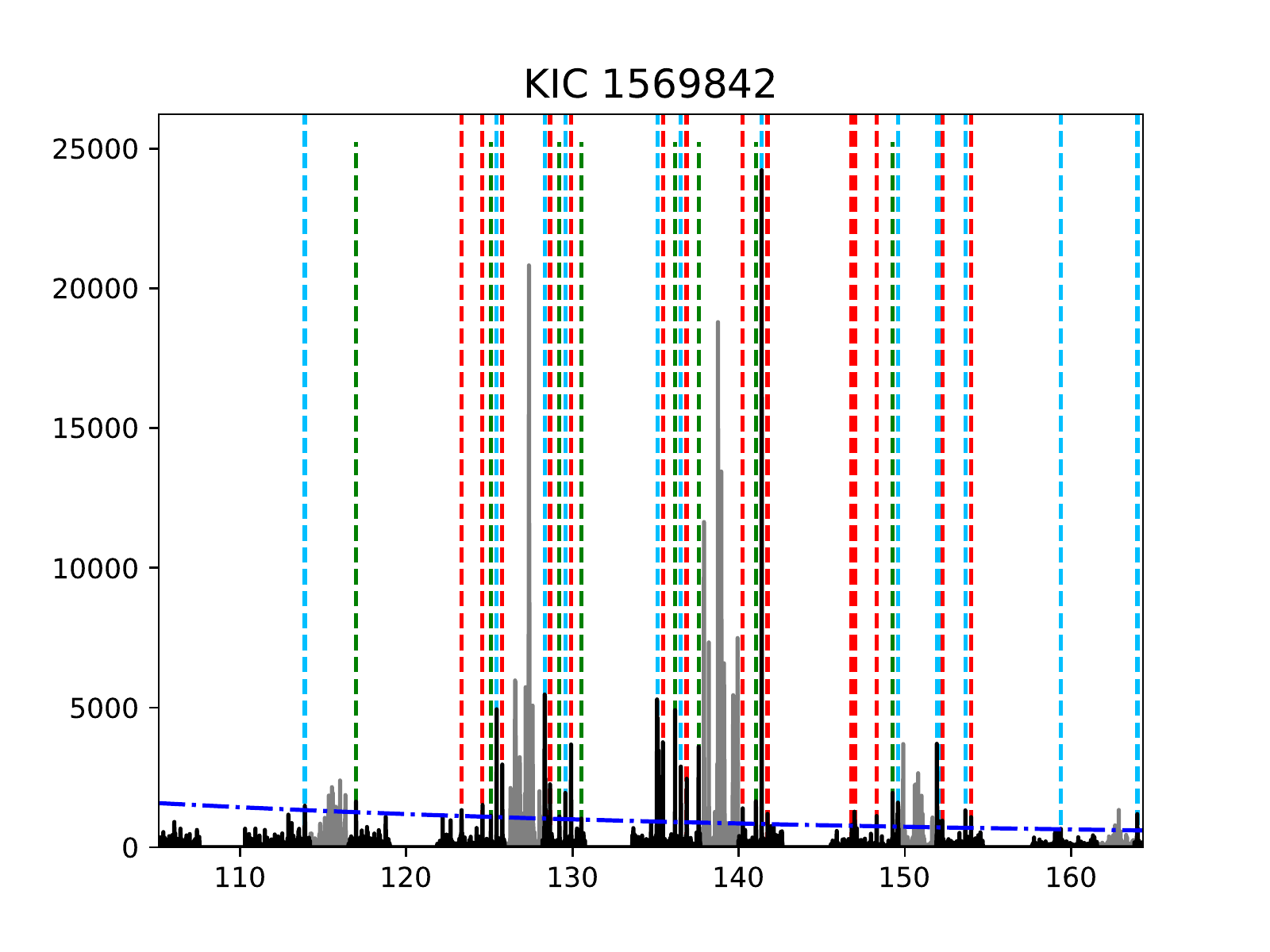}
\includegraphics[width=12cm]{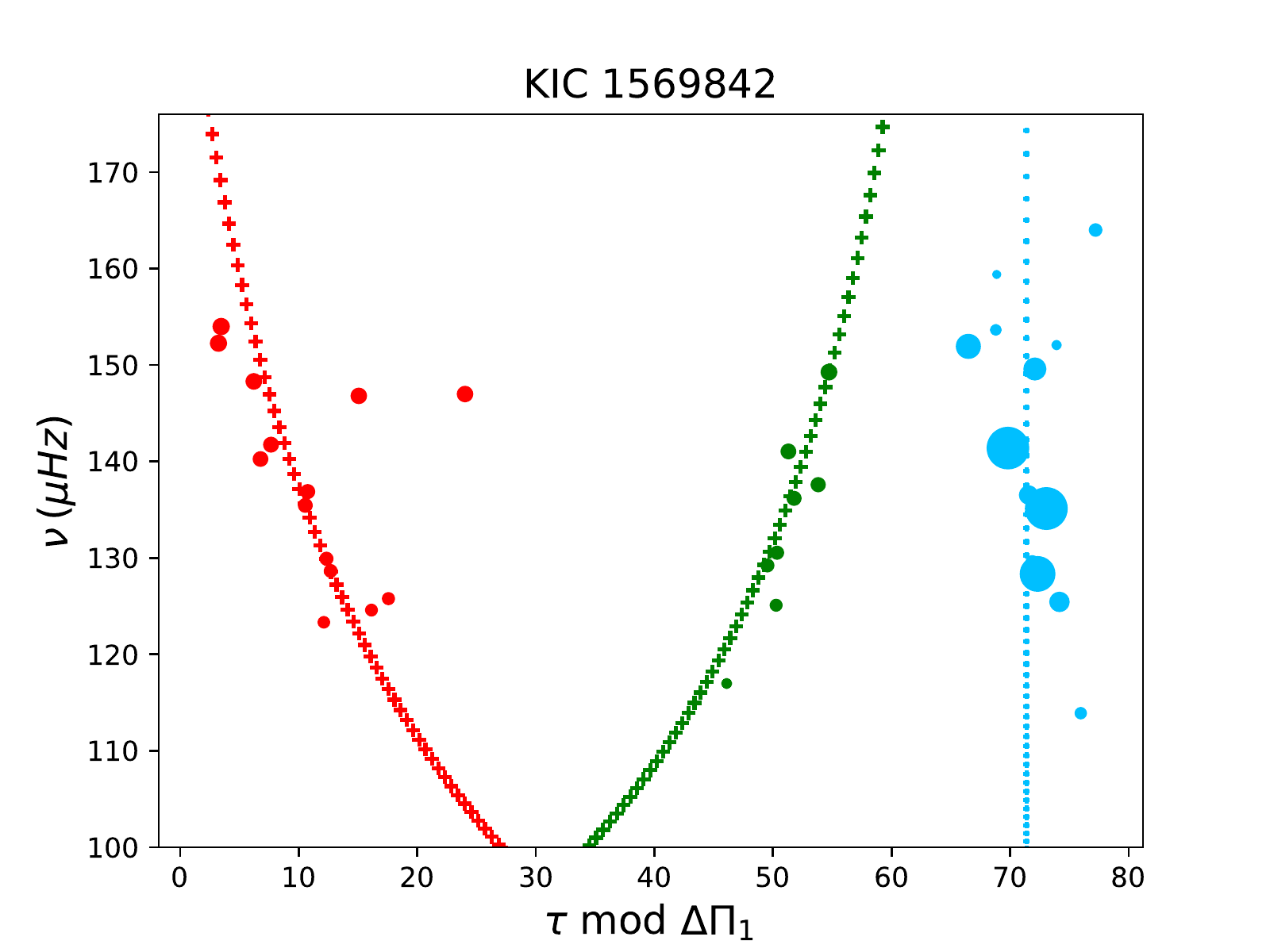}
\caption{KIC 1569842 with three observed rotational components. We found an inclination of $i=32.5 \pm 13.8^\circ$.
 \textit{Upper panel:} oscillation spectrum where frequency intervals around radial and quadrupole modes have been removed. The horizontal blue dashed-line indicates the level above which oscillation modes are considered as significant. Pressure-dominated mixed-modes are in grey, gravity-dominated mixed-modes are in black. Modes with azimuthal orders $m=\{-1,0,1\}$ are indicated by vertical dashed lines in green, light blue and red, respectively. \textit{Lower panel:} échelle diagram in stretched period with same color code. Observed modes are represented by dots and the symbol size varies as the measured power spectral density. The fit of rotational components are represented by crosses with the color coding the azimuthal order.}
\label{fig-me-1}
\end{figure*}

\begin{figure*}
\centering
\includegraphics[width=13cm]{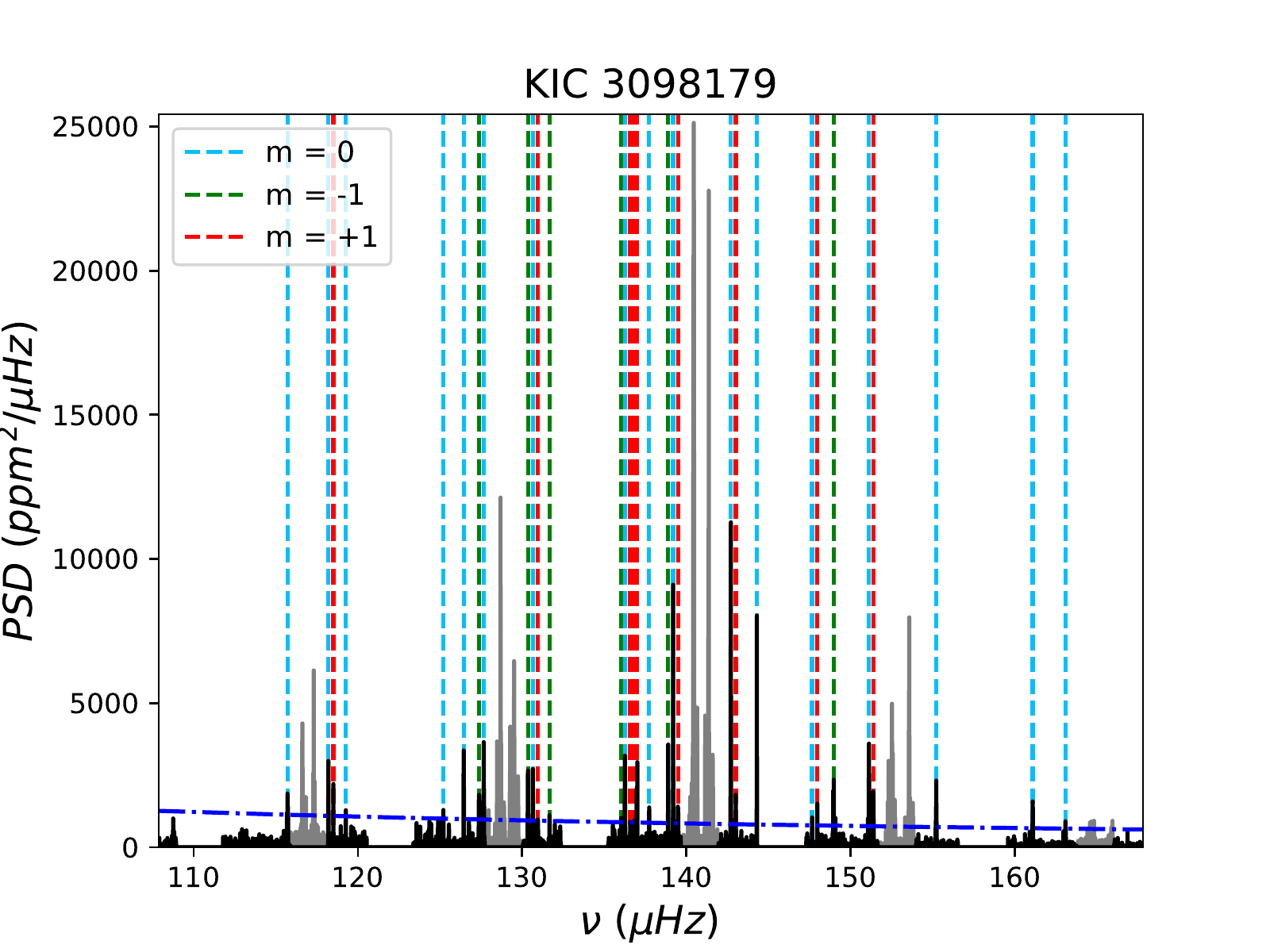}
\includegraphics[width=12cm]{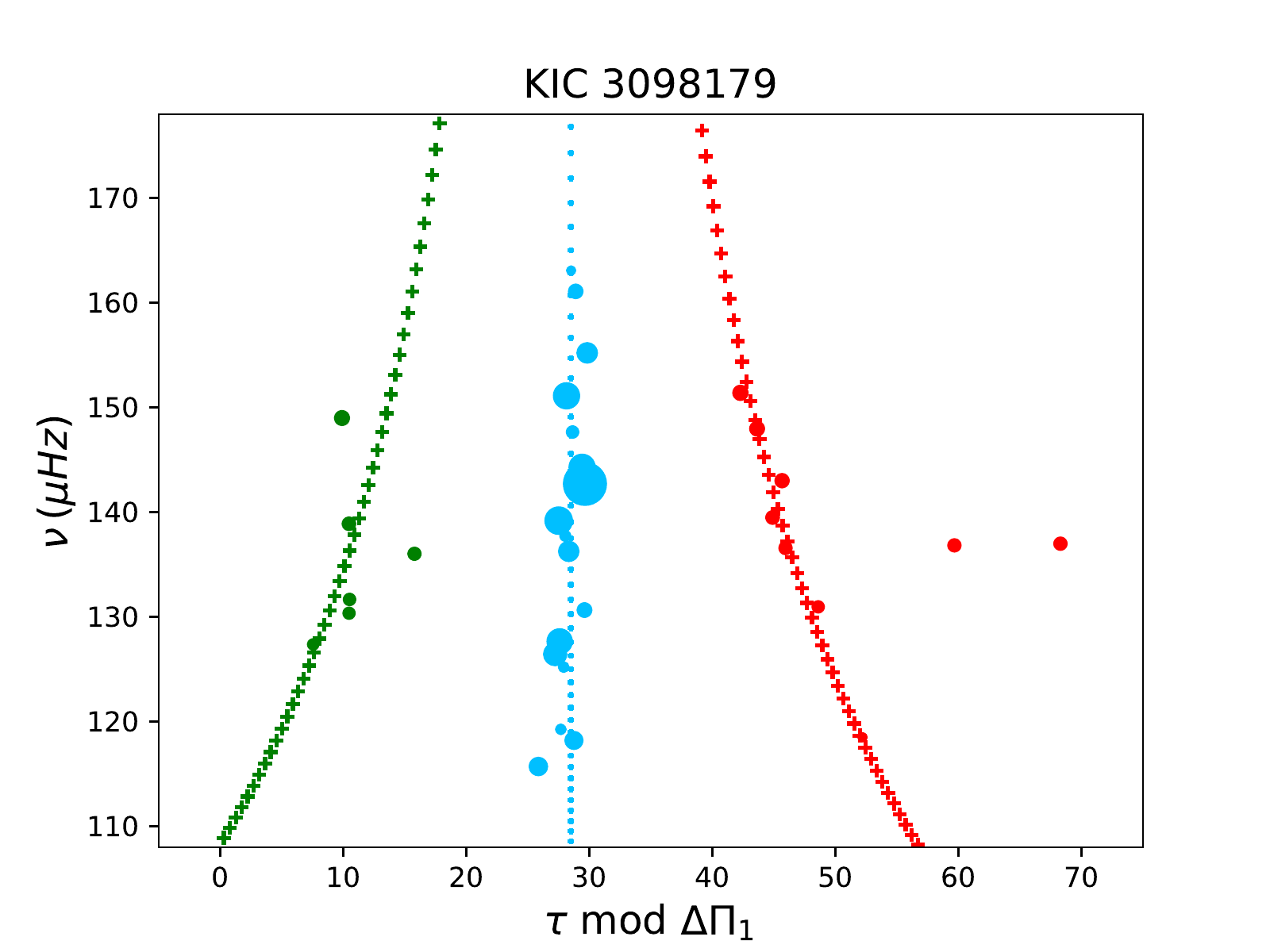}
\caption{KIC 3098179 with three observed rotational components. We found an inclination of $i=39.9 \pm 17.3^\circ$.}
\label{fig-me-2}
\end{figure*}

\begin{figure*}
\centering
\includegraphics[width=13cm]{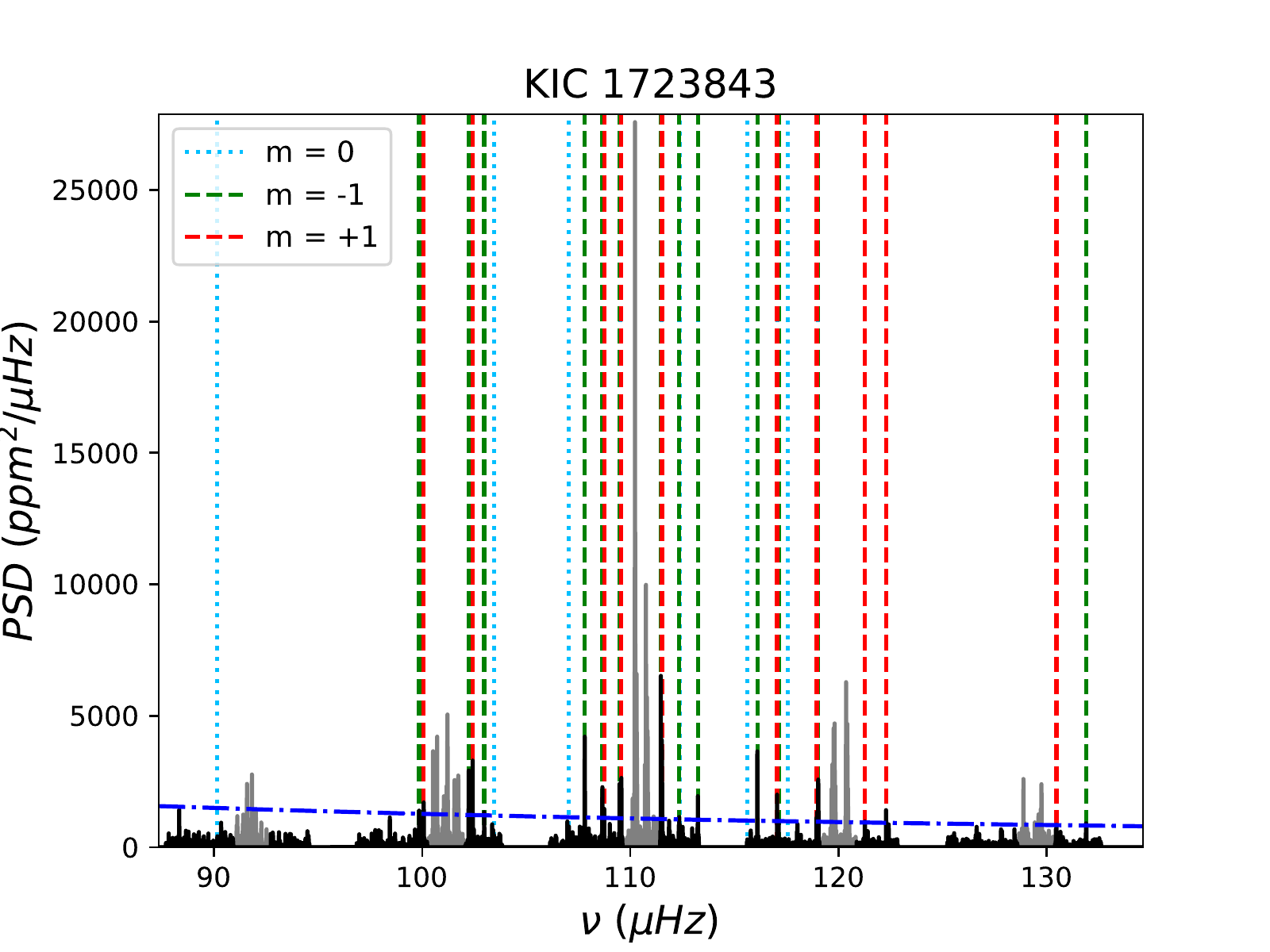}
\includegraphics[width=12cm]{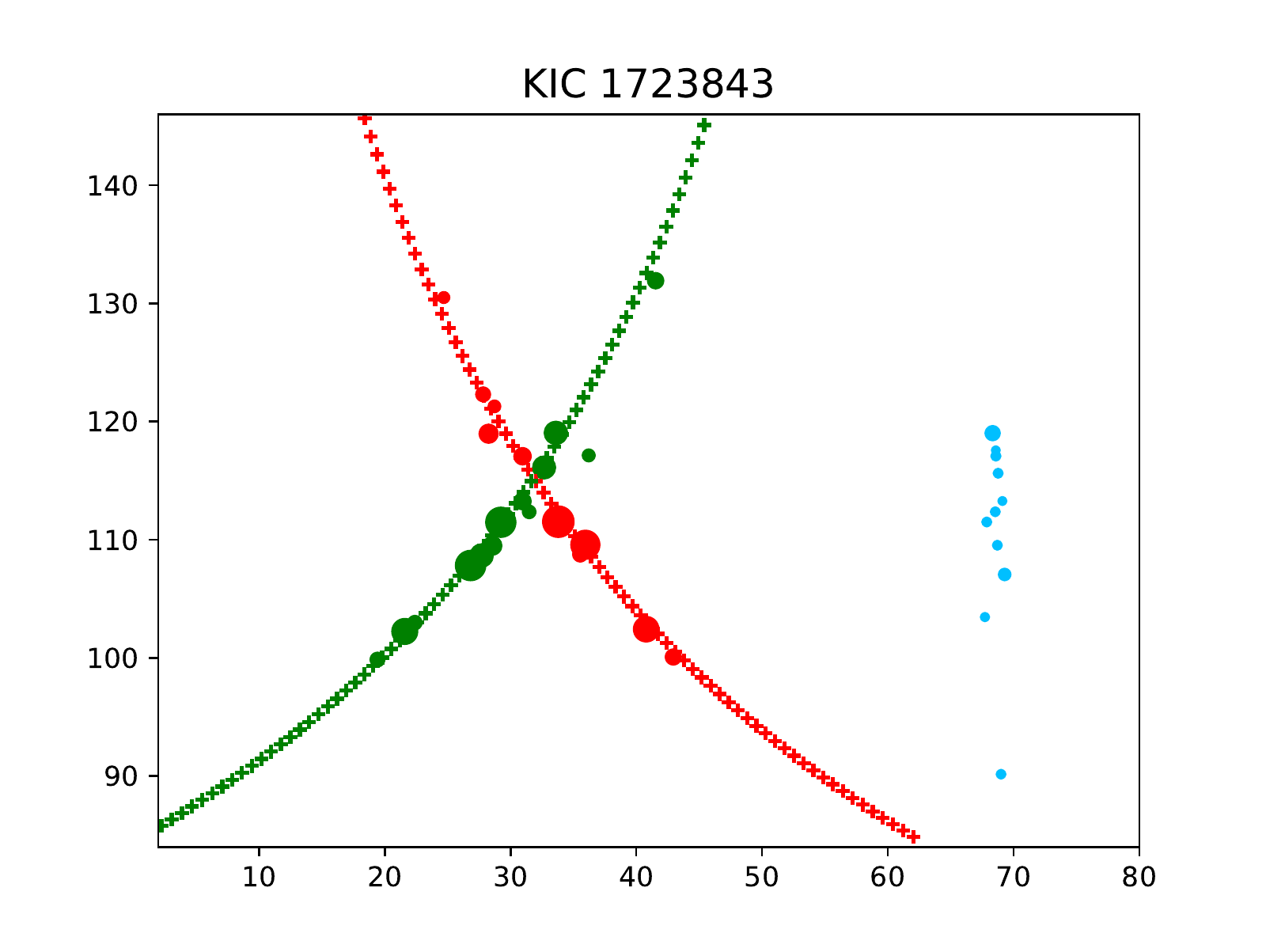}
\caption{KIC 1723843 with two observed rotational components. Modes with $m=0$ have been identified a posteriori considering that the $m=0$ ridge is median with respect to $m = \pm \, 1$ ridges in the échelle diagram. We thus represented them by dotted lines in the spectrum to distinguish them from $m = \pm \, 1$ that are directly visible. We found an inclination of $i=74.3 \pm^{15.7}\ind{7.2}$$^\circ$.}
\label{fig-me-3}
\end{figure*}

\begin{figure*}
\centering
\includegraphics[width=13cm]{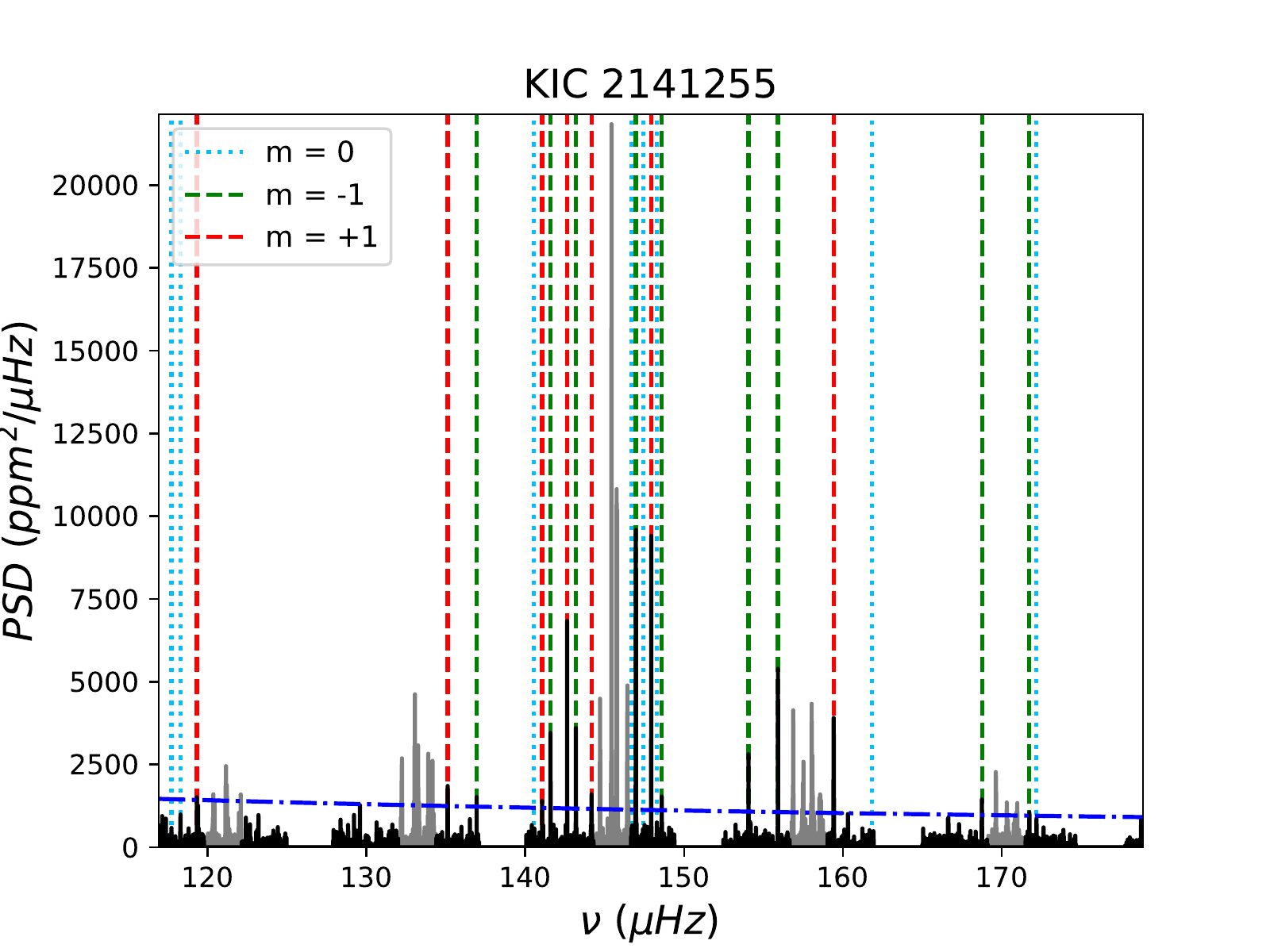}
\includegraphics[width=12cm]{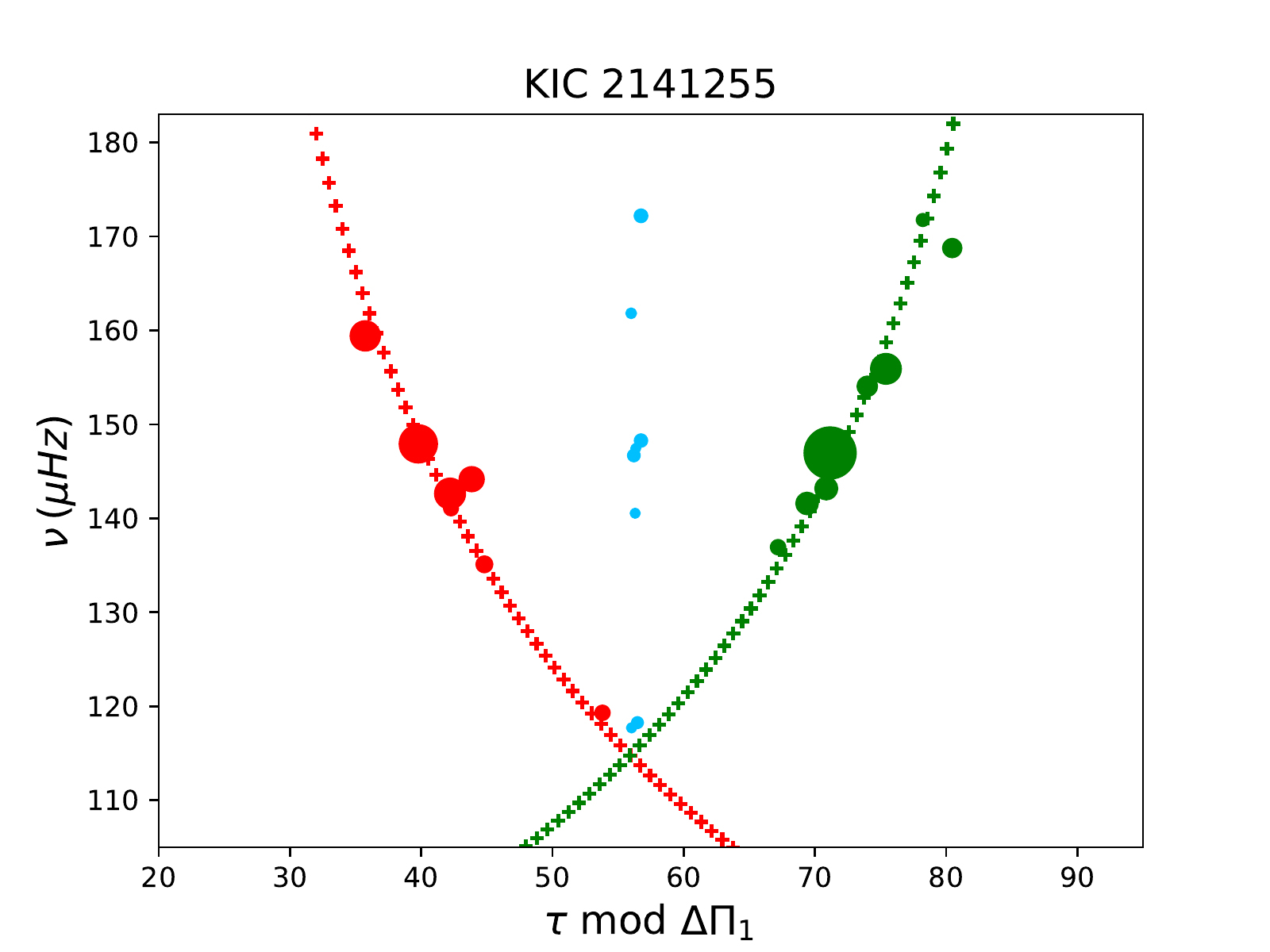}
\caption{KIC 2141255 with two observed rotational components. We found an inclination of $i=76.4 \pm^{13.6}\ind{8.2}$$^\circ$.}
\label{fig-me-4}
\end{figure*}

\begin{figure*}
\centering
\includegraphics[width=13cm]{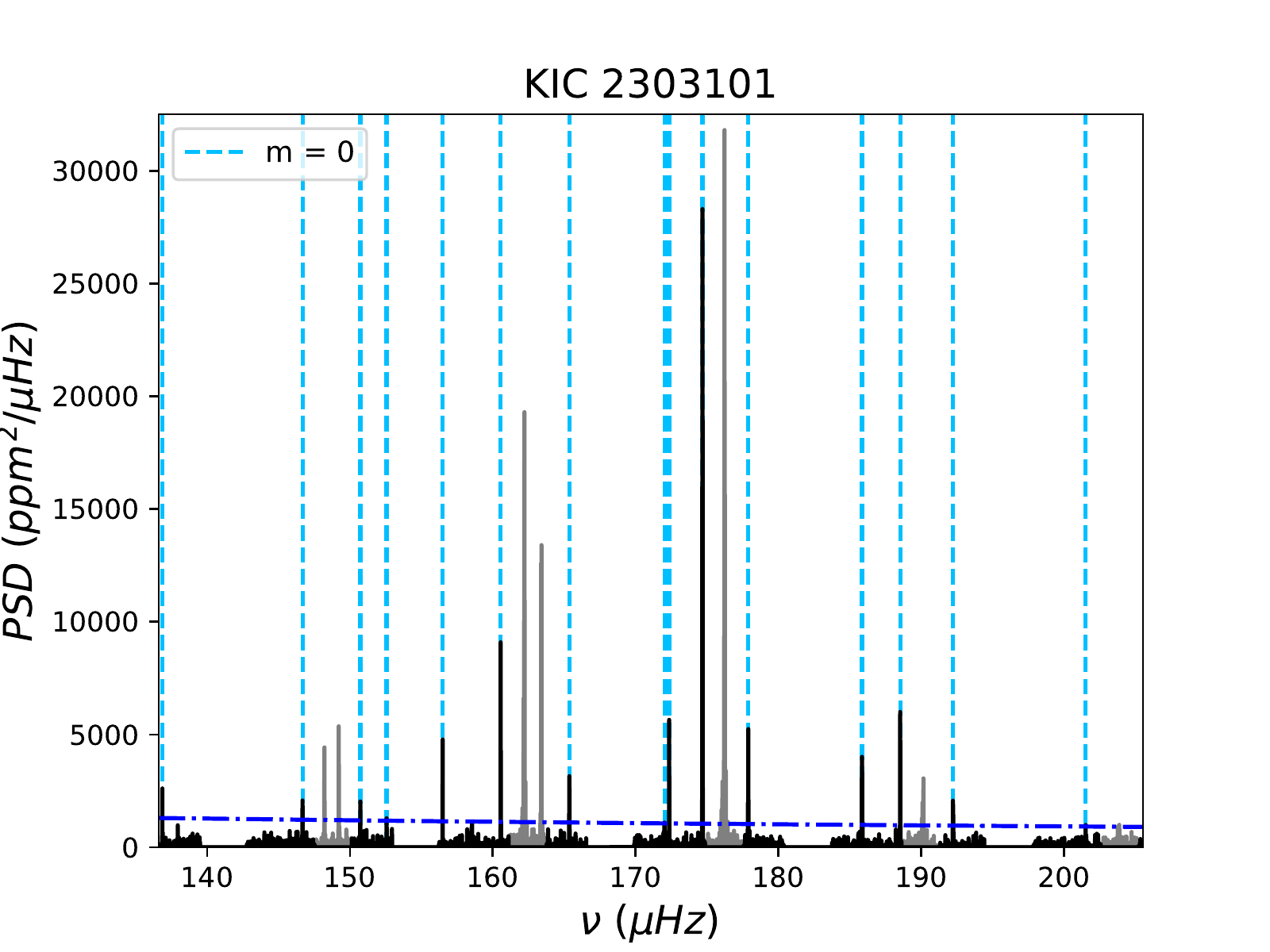}
\includegraphics[width=12cm]{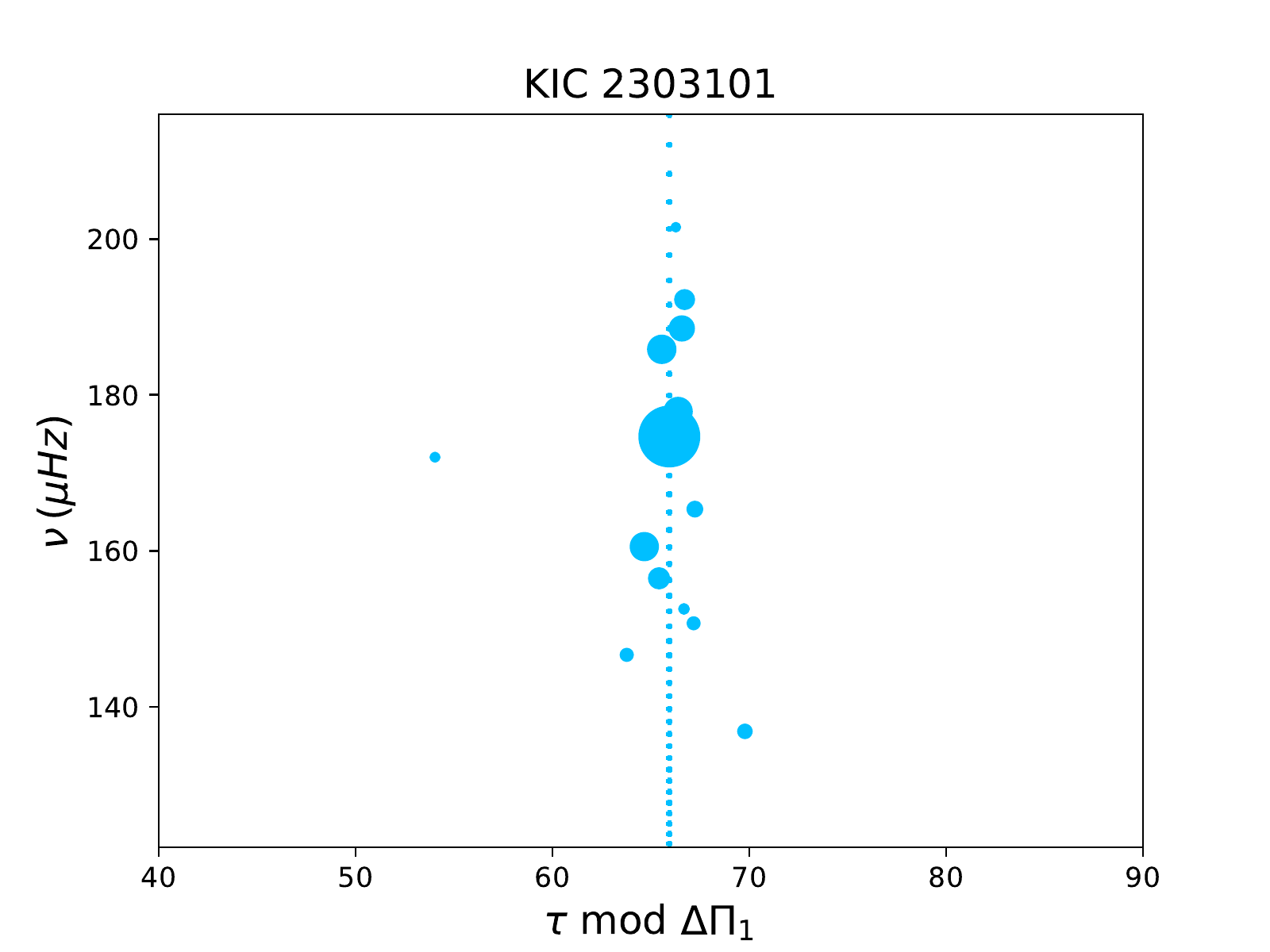}
\caption{KIC 2303101 with one observed rotational component. We found an inclination of $i=19.8 \pm^{8.4}\ind{19.8}$$^\circ$.}
\label{fig-me-5}
\end{figure*}

\begin{figure*}
\centering
\includegraphics[width=13cm]{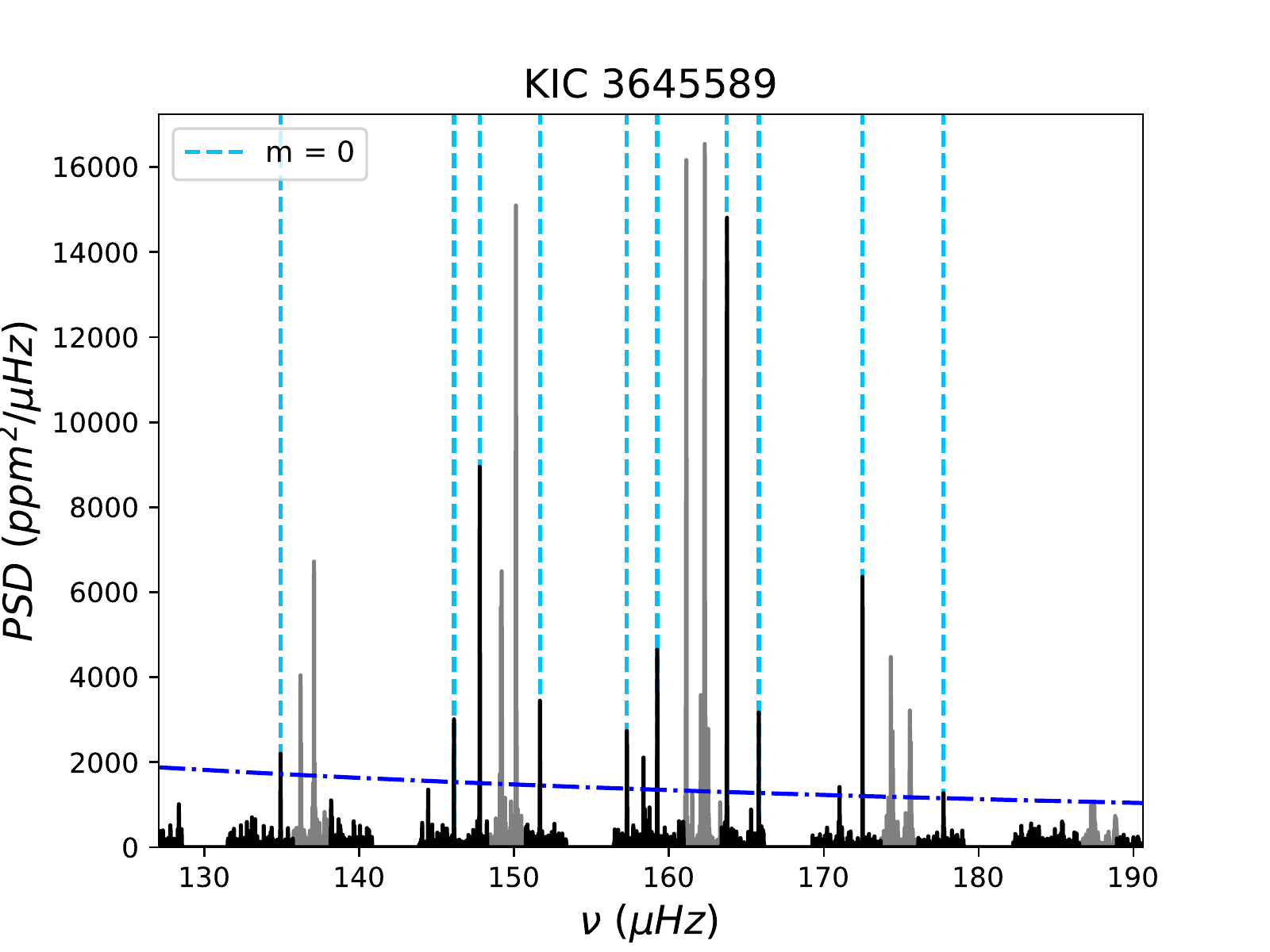}
\includegraphics[width=12cm]{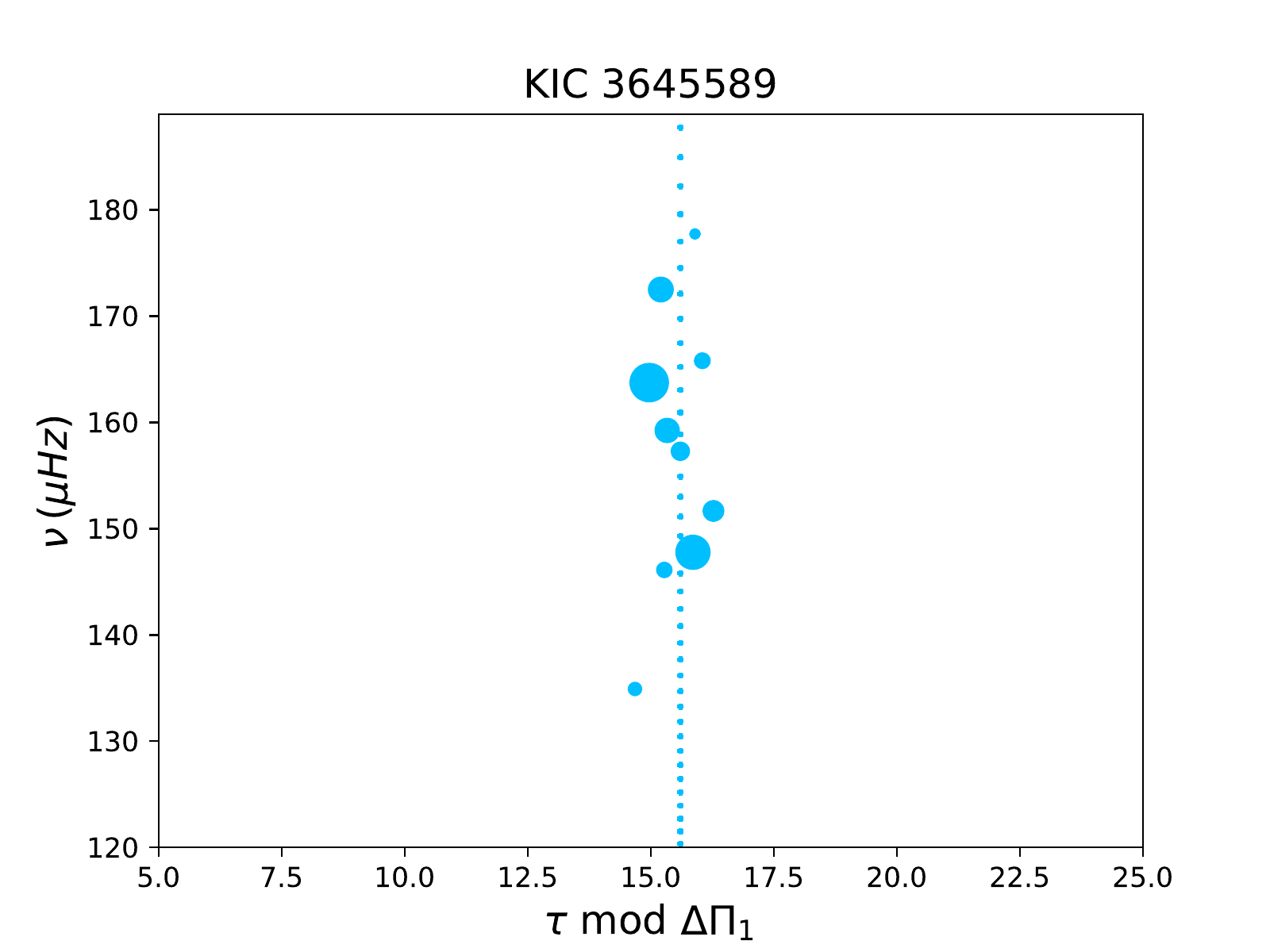}
\caption{KIC 3645589 with one observed rotational component. We found an inclination of $i=18.9 \pm^{8.0}\ind{18.9}$$^\circ$.}
\label{fig-me-6}
\end{figure*}

\begin{figure*}
\centering
\includegraphics[width=13cm]{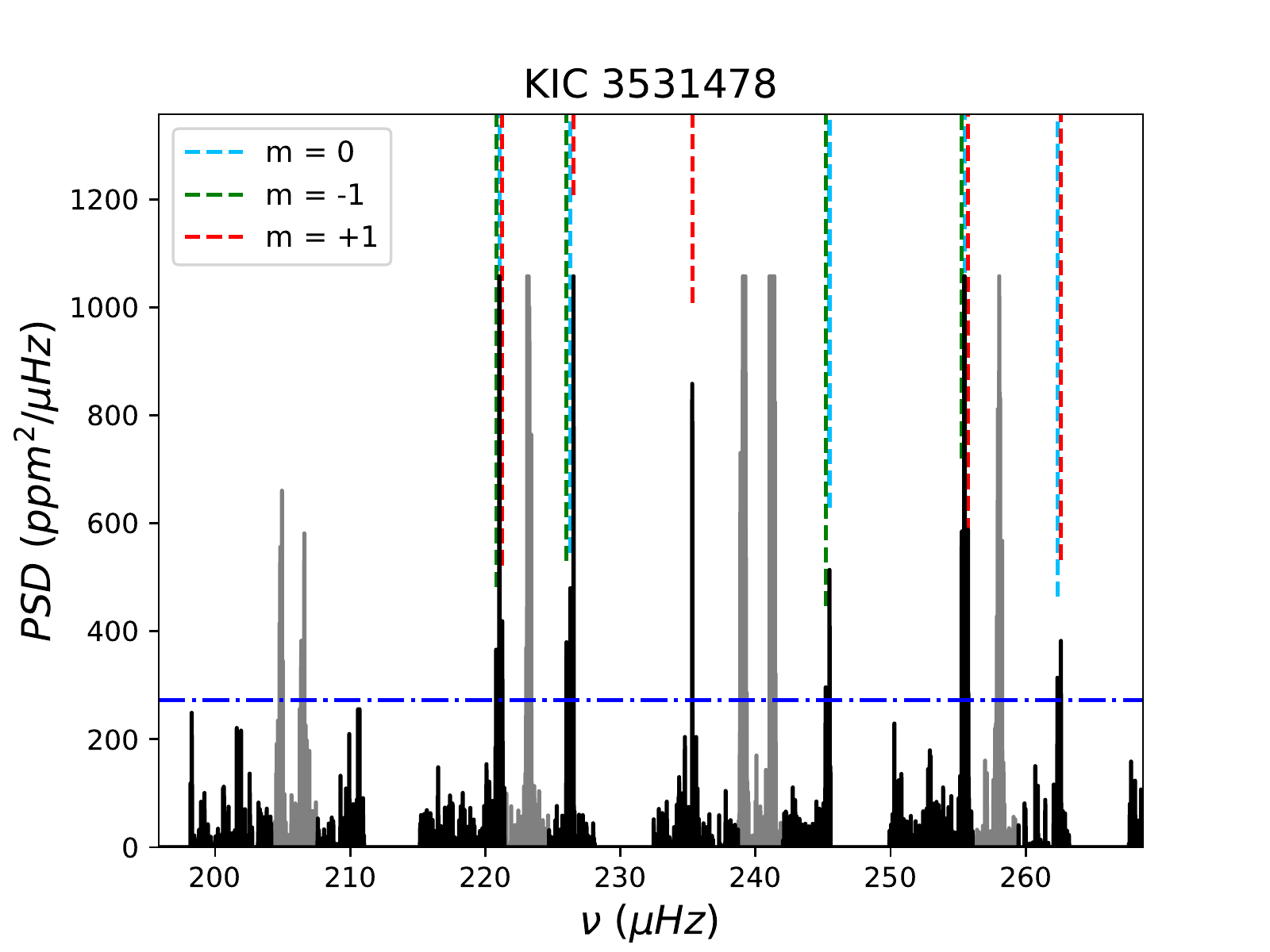}
\includegraphics[width=12cm]{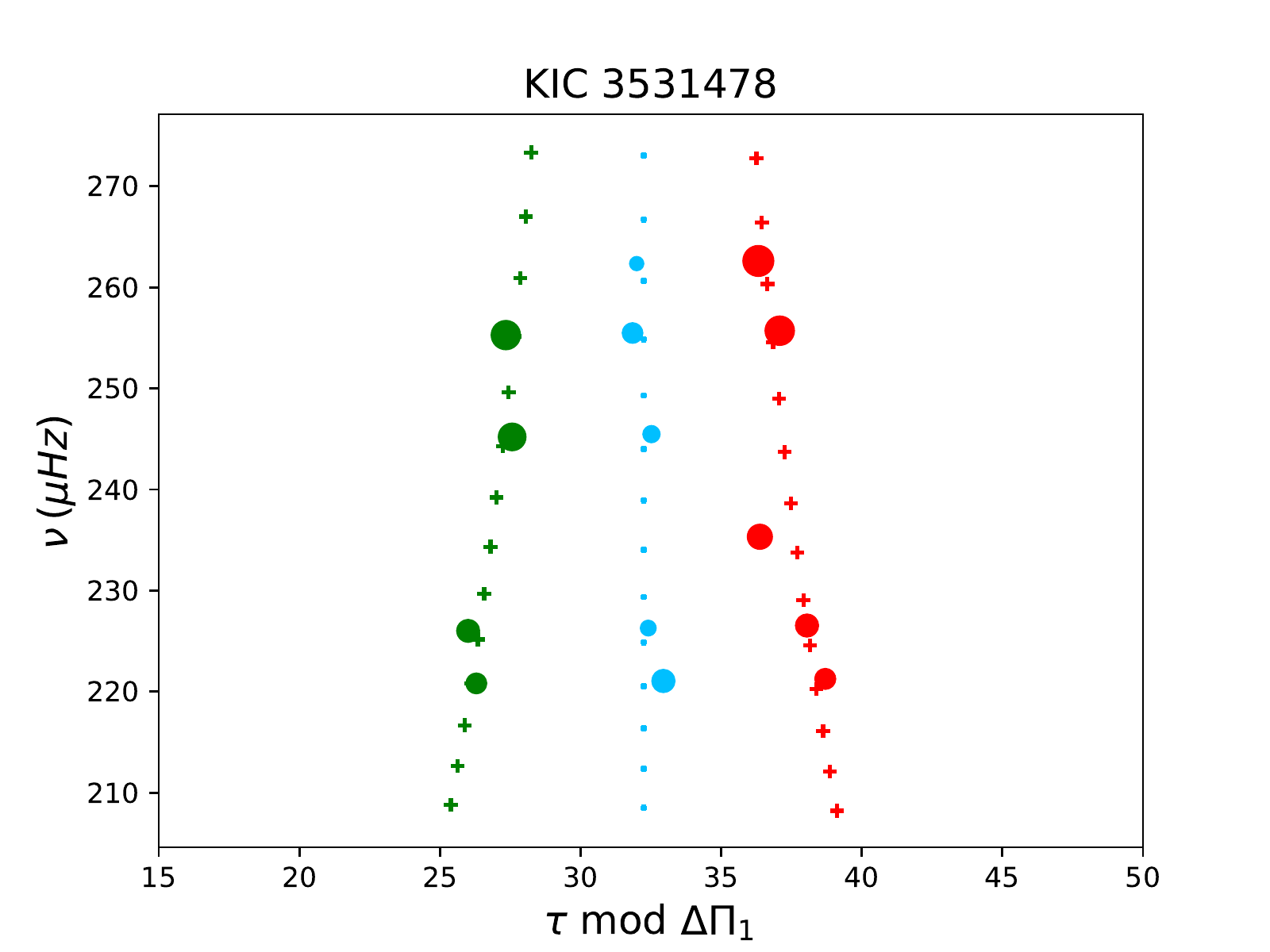}
\caption{Same as Fig.~\ref{fig-me-1} for KIC 3531478 with three observed rotational components. We found an inclination of $i=41.2 \pm 9.3^\circ$ consistent with \cite{Kuszlewicz} measurement of $i=46.8 \pm^{3.2}\ind{3.1}$$^\circ$.}
\label{fig-Kuszlewicz-agreement-1}
\end{figure*}

\begin{figure*}
\centering
\includegraphics[width=12.8cm]{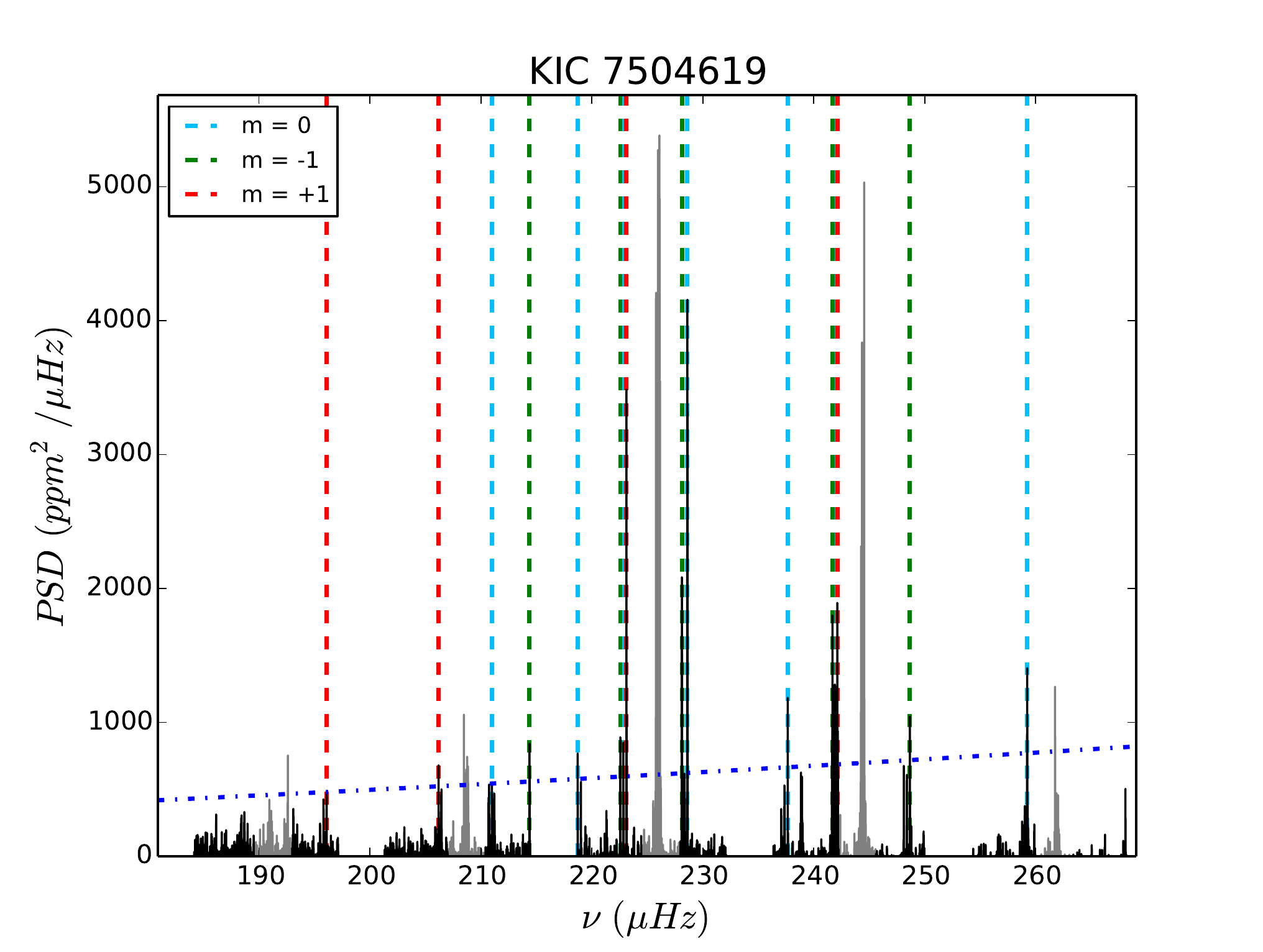}
\includegraphics[width=12cm]{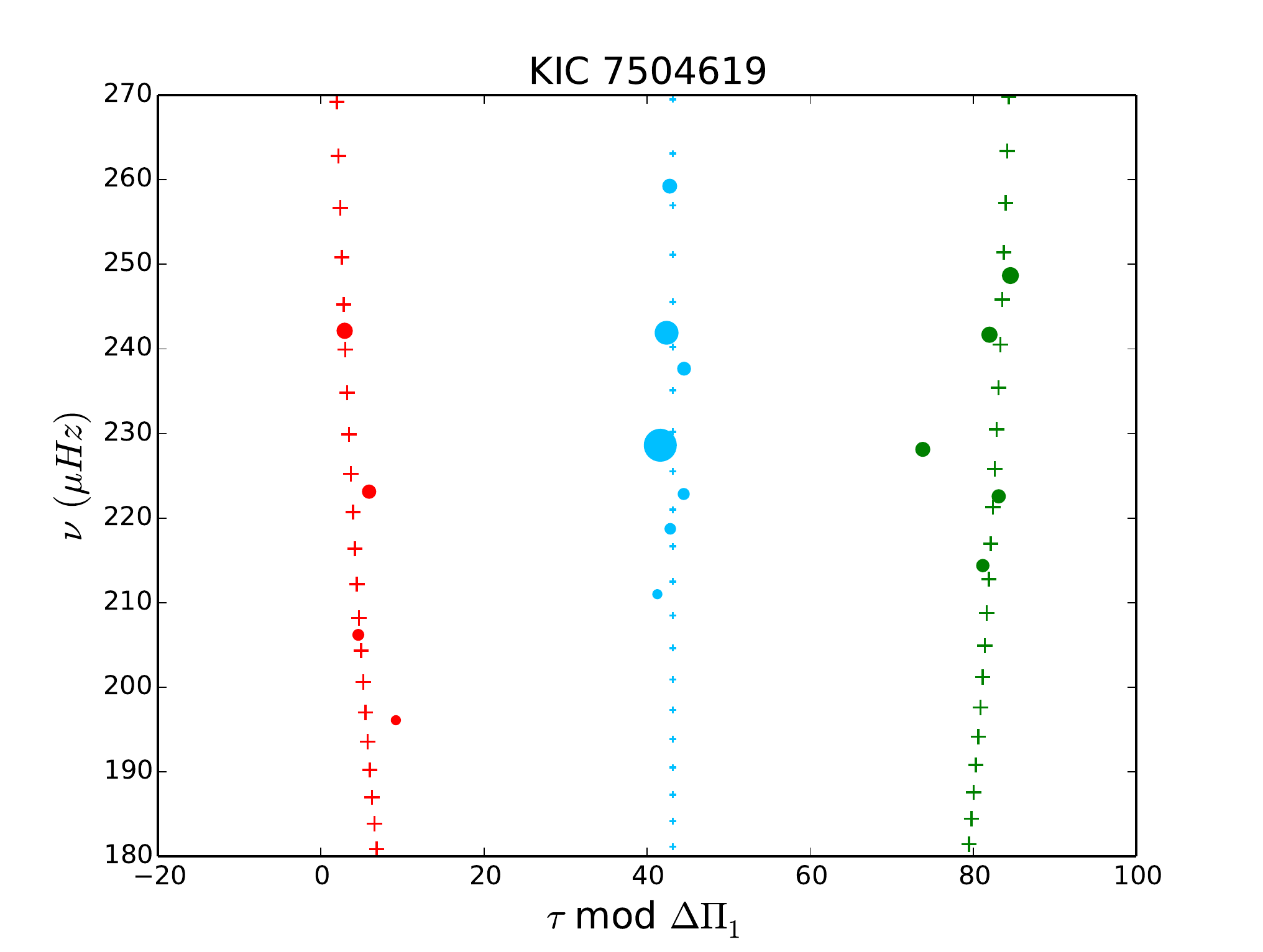}
\caption{Same as Fig.~\ref{fig-me-1} for KIC 7504619 with three observed rotational components. We found an inclination of $i=56.8 \pm{16.6}^\circ$ in agreement with \cite{Kuszlewicz} measurement of $i=59.3 \pm^{2.4}\ind{2.5}$$^\circ$.}
\label{fig-Kuszlewicz-agreement-5}
\end{figure*}

\begin{figure*}
\centering
\includegraphics[width=13cm]{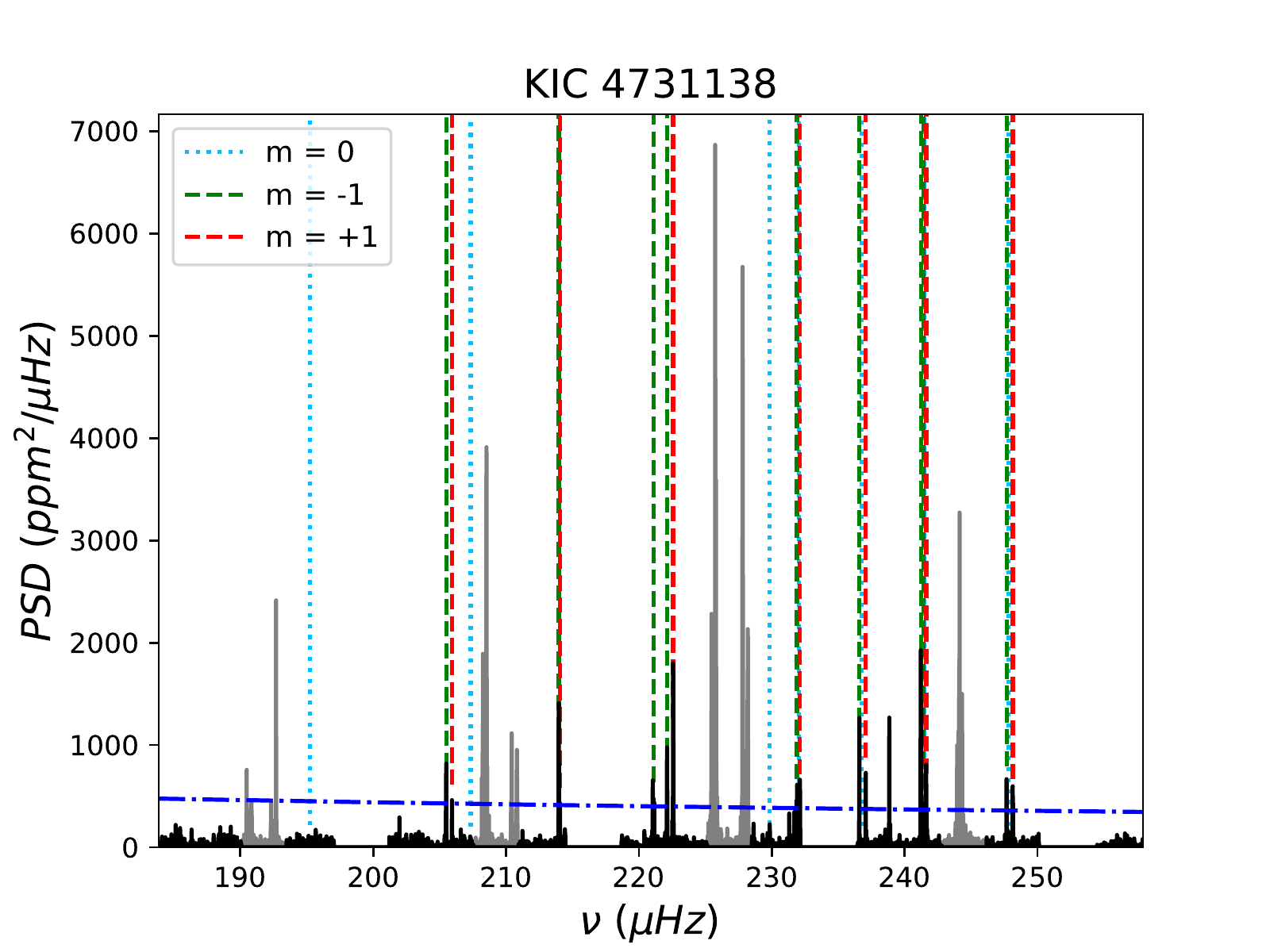}
\includegraphics[width=12cm]{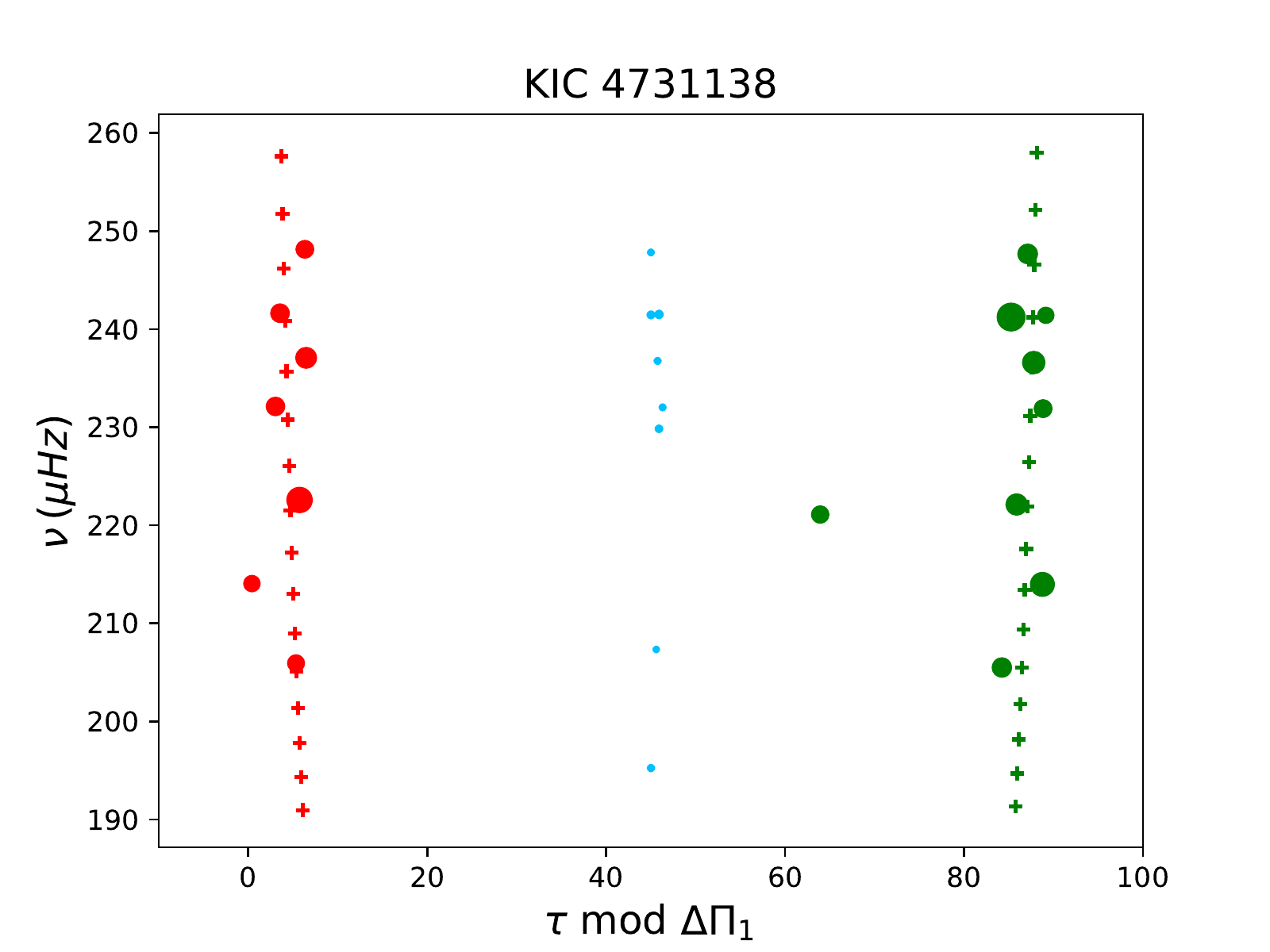}
\caption{Same as Fig.~\ref{fig-me-1} for KIC 4731138 with two observed rotational components. We found an inclination of $i=79.8 \pm^{10.2}\ind{3.1}$$^\circ$ consistent with \cite{Kuszlewicz} measurement of $i=84.5 \pm^{3.3}\ind{3.0}$$^\circ$.}
\label{fig-Kuszlewicz-agreement-2}
\end{figure*}

\begin{figure*}
\centering
\includegraphics[width=12.8cm]{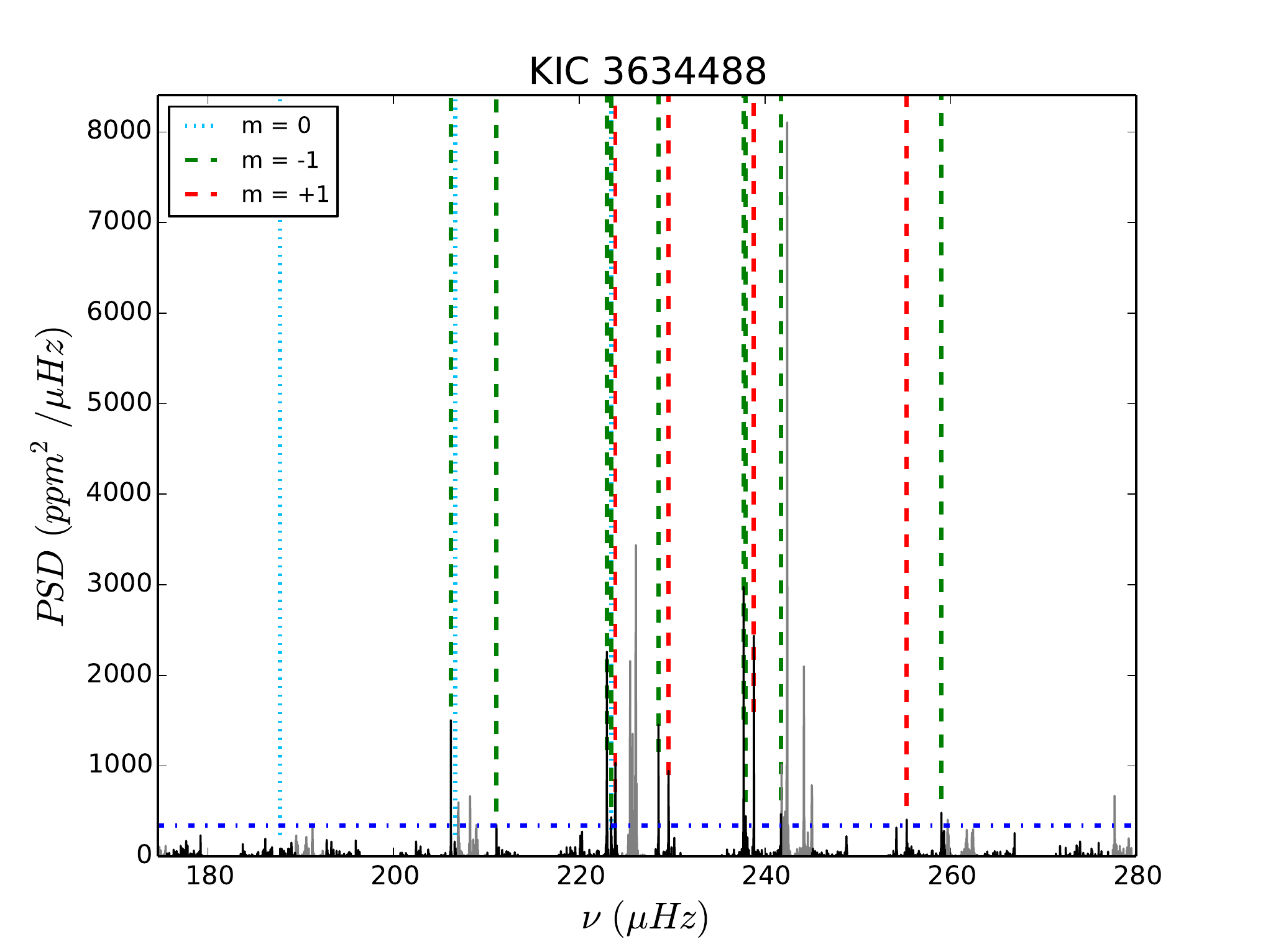}
\includegraphics[width=12cm]{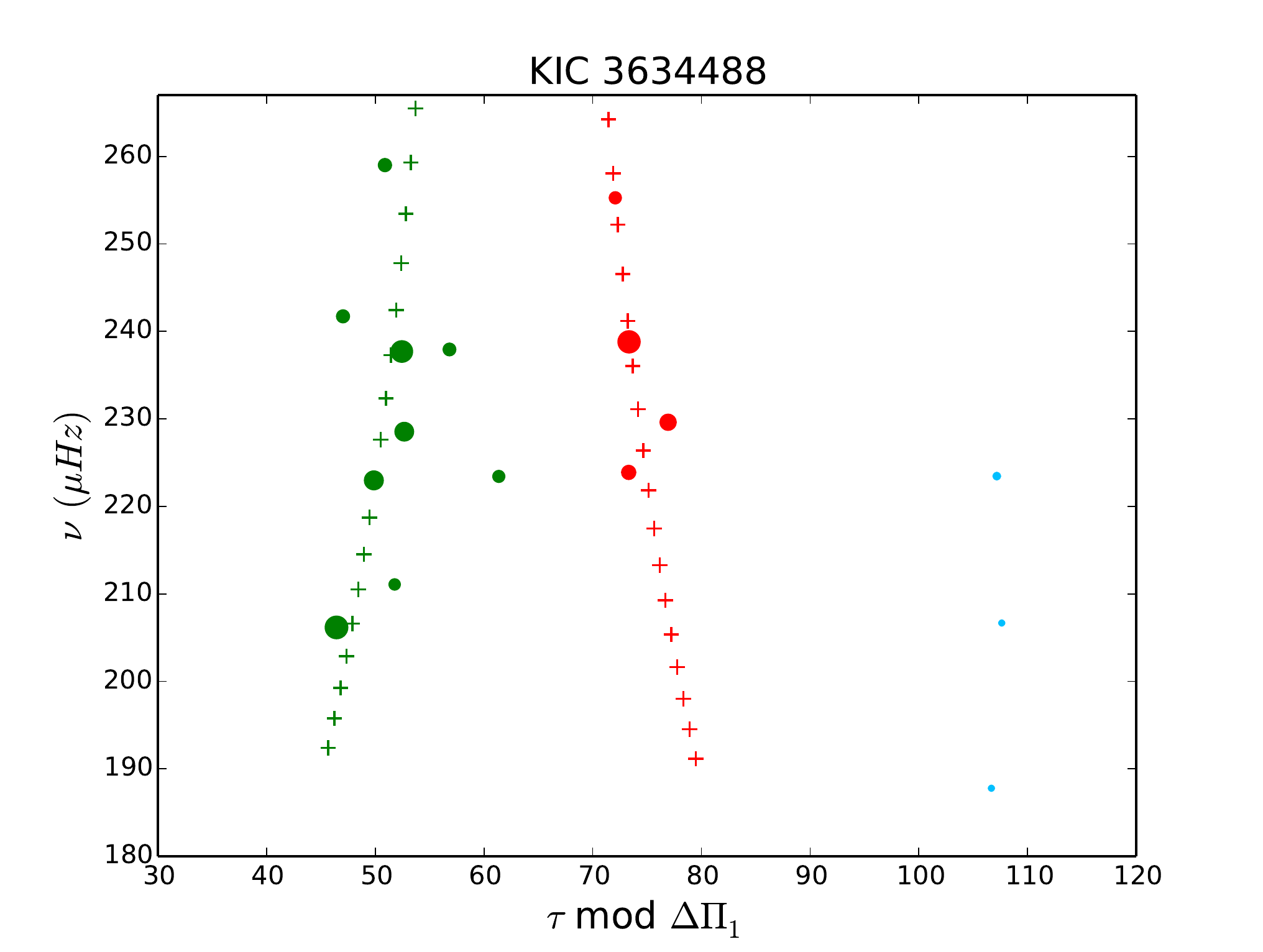}
\caption{Same as Fig.~\ref{fig-me-1} for KIC 3634488 with two observed rotational components. We found an inclination of $i=86.2 \pm^{3.8}\ind{9.7}$$^\circ$ in marginal agreement with \cite{Kuszlewicz} measurement of $i=70.7 \pm^{2.3}\ind{2.4}$$^\circ$.}
\label{fig-Kuszlewicz-agreement-4}
\end{figure*}

\begin{figure*}
\centering
\includegraphics[width=12.8cm]{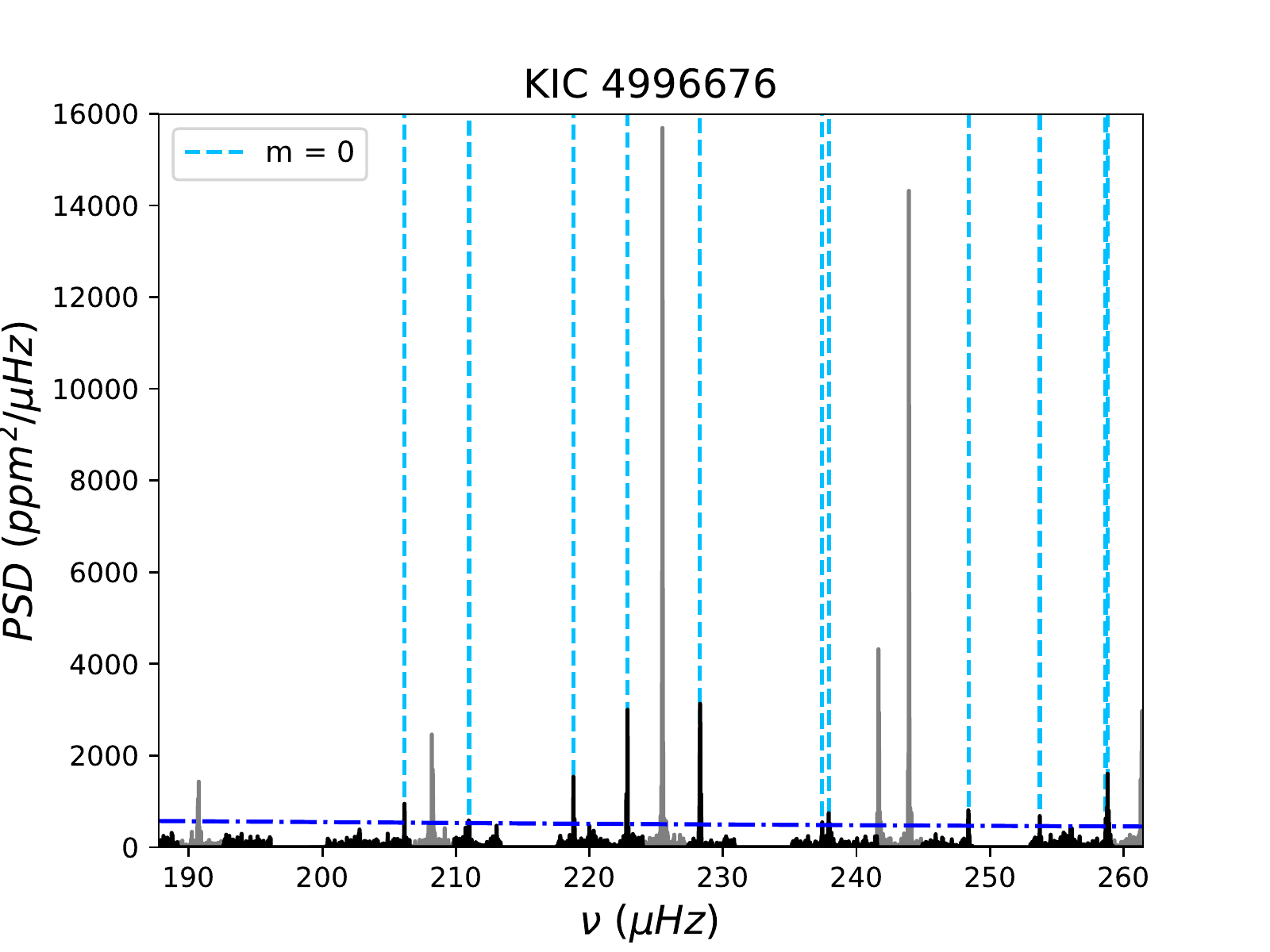}
\includegraphics[width=12cm]{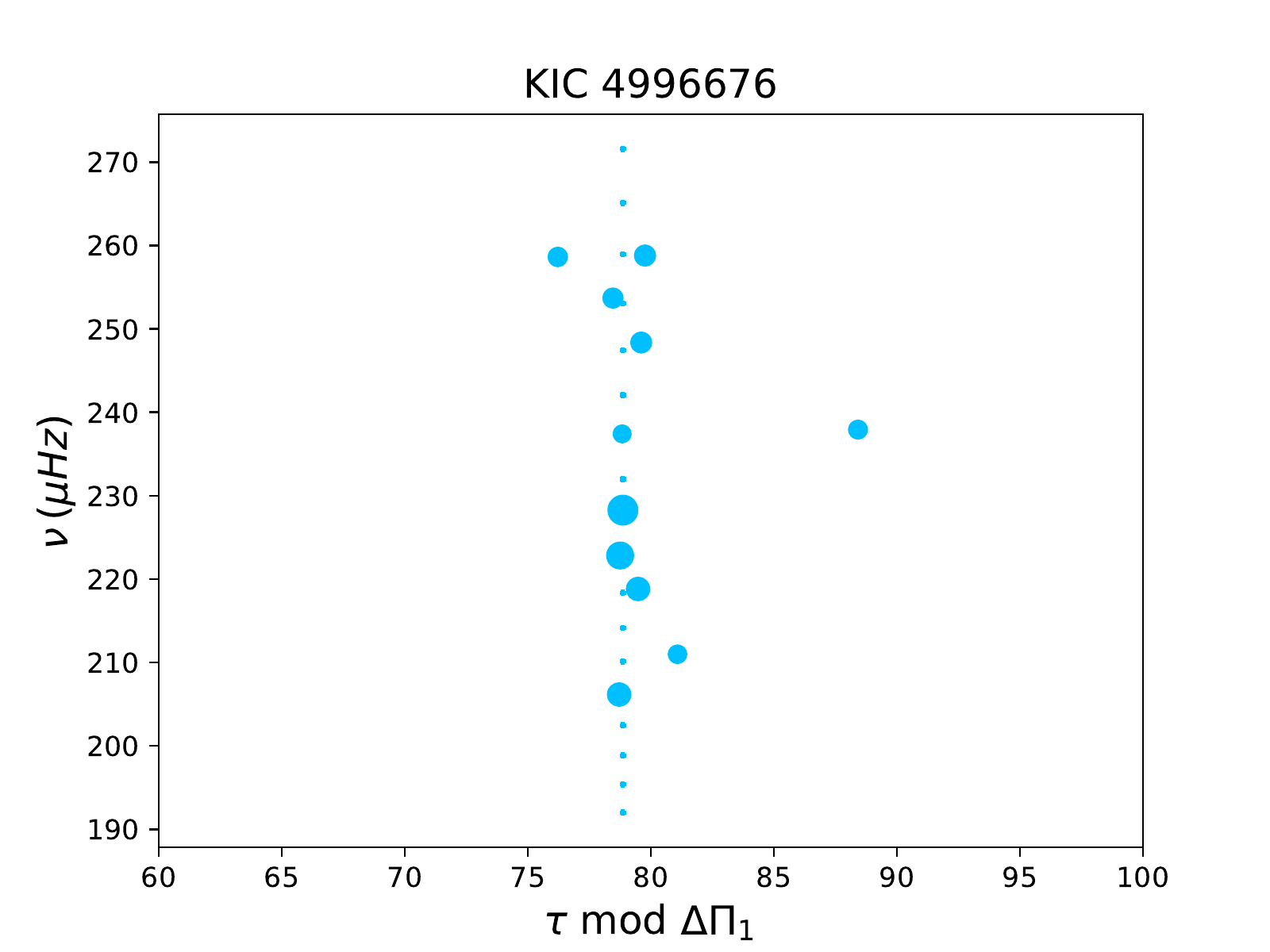}
\caption{Same as Fig.~\ref{fig-me-1} for KIC 4996676 with one observed rotational component. We found an inclination of $i=30.1 \pm^{13.0}\ind{30.1}$$^\circ$ consistent with \cite{Kuszlewicz} measurement of $i=5.7 \pm^{2.6}\ind{4.0}$$^\circ$.}
\label{fig-Kuszlewicz-agreement-3}
\end{figure*}

\begin{figure*}
\centering
\includegraphics[width=12.8cm]{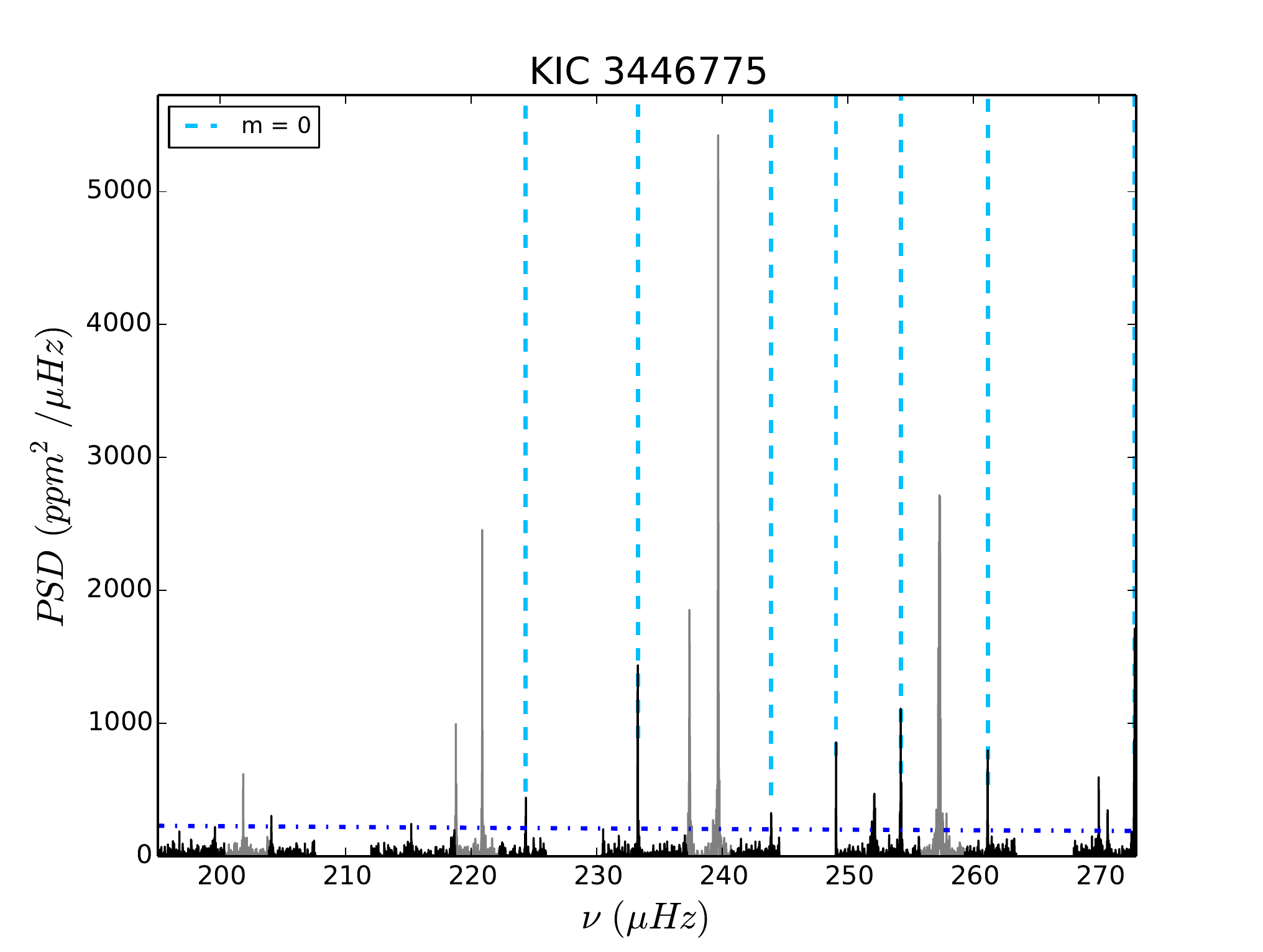}
\includegraphics[width=12cm]{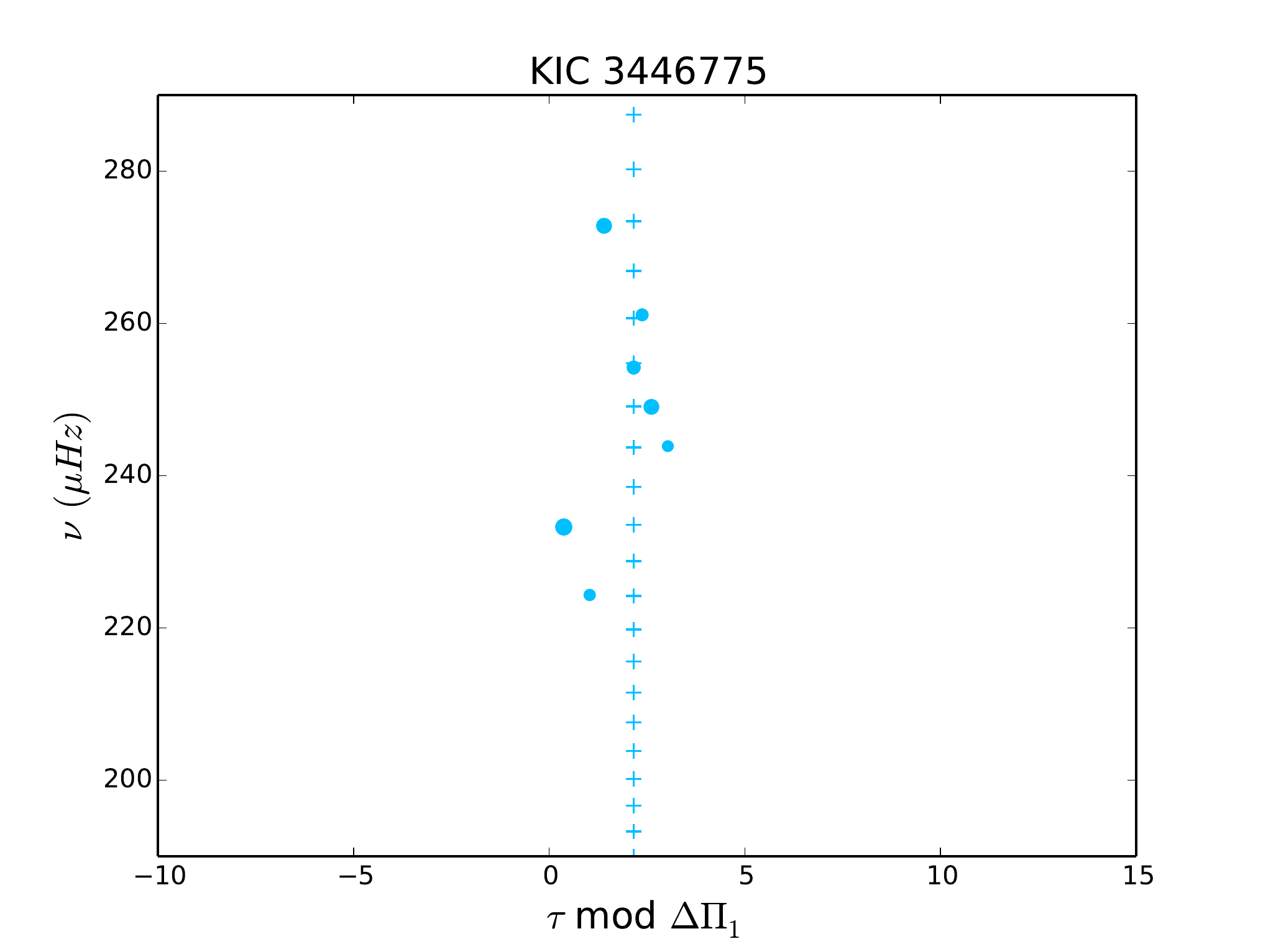}
\caption{Same as Fig.~\ref{fig-me-1} for KIC 3446775 with one observed rotational component. We found an inclination of $i=31.8 \pm^{13.8}\ind{31.8}$$^\circ$ in agreement with \cite{Kuszlewicz} measurement of $i=10.3 \pm{4.3}^\circ$.}
\label{fig-Kuszlewicz-agreement-6}
\end{figure*}

\begin{figure*}
\centering
\includegraphics[width=12.8cm]{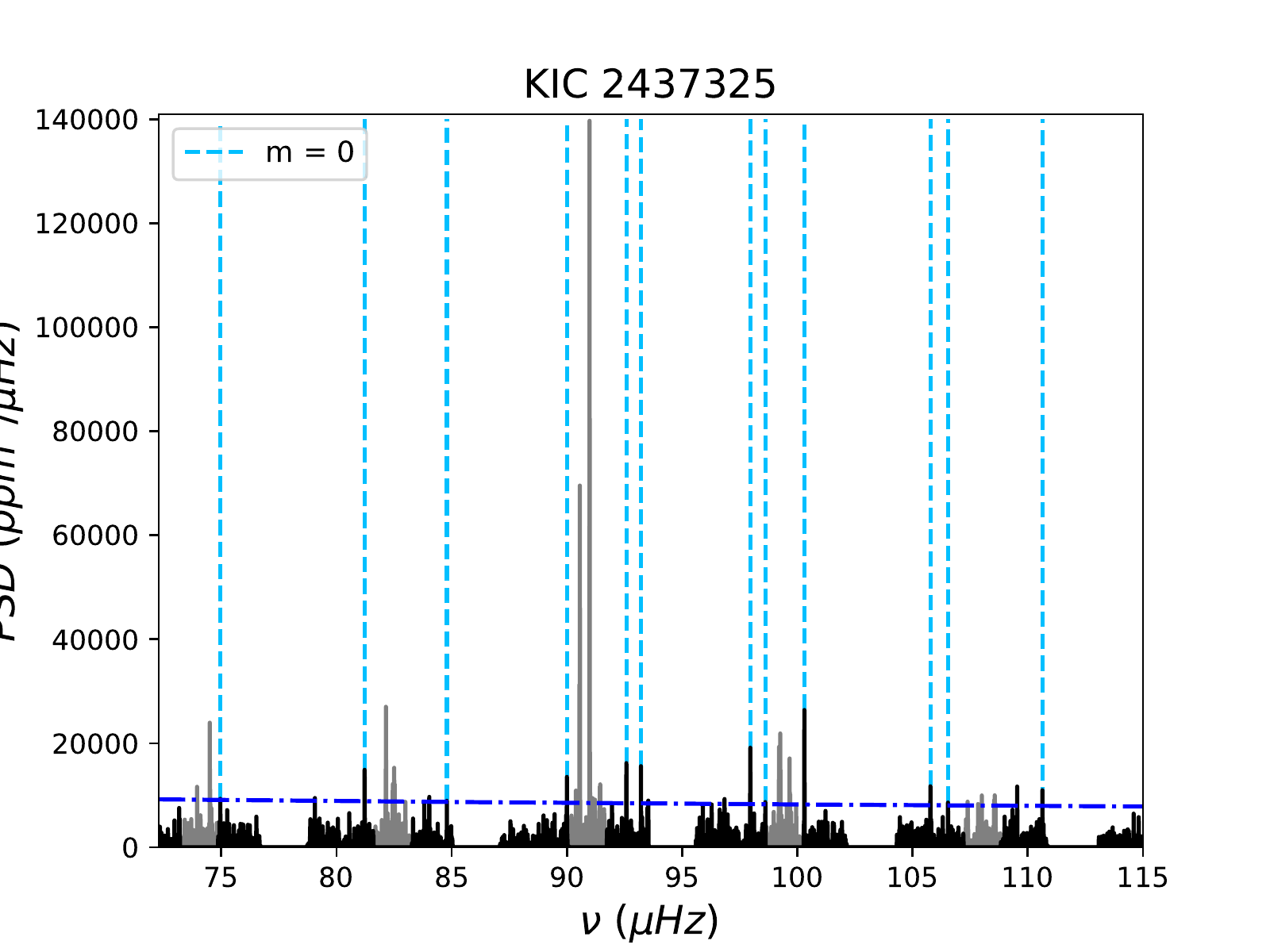}
\includegraphics[width=12cm]{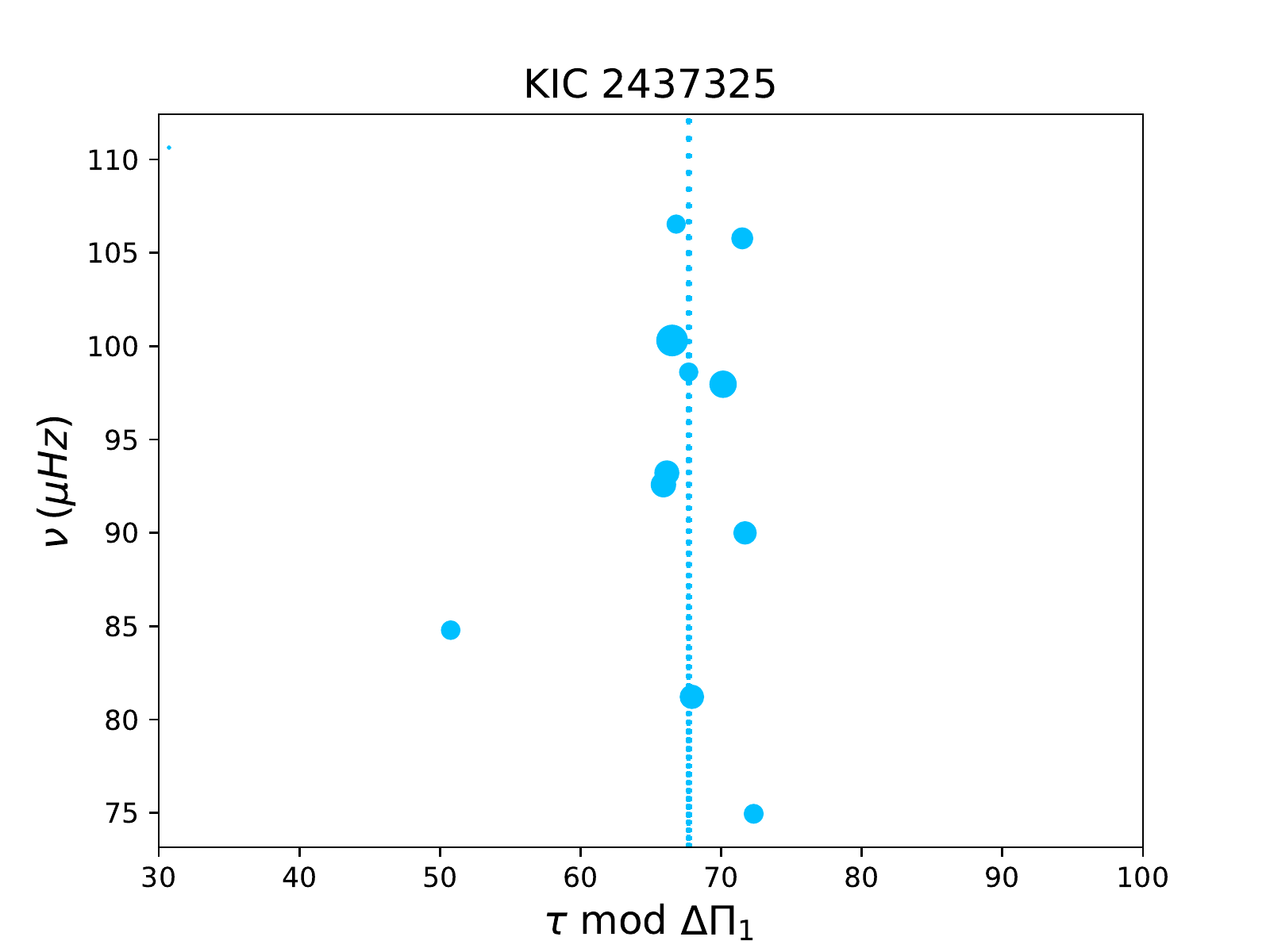}
\caption{Same as Fig.~\ref{fig-me-1} for the NGC 6791 star KIC 2437325 with one observed rotational component. We found an inclination of $i=39.1 \pm^{17.4}\ind{39.1}$$^\circ$ consistent with \cite{Corsaro} measurement of $i=11 \pm 5^\circ$.}
\label{fig-Corsaro-1}
\end{figure*}

\begin{figure*}
\centering
\includegraphics[width=12.8cm]{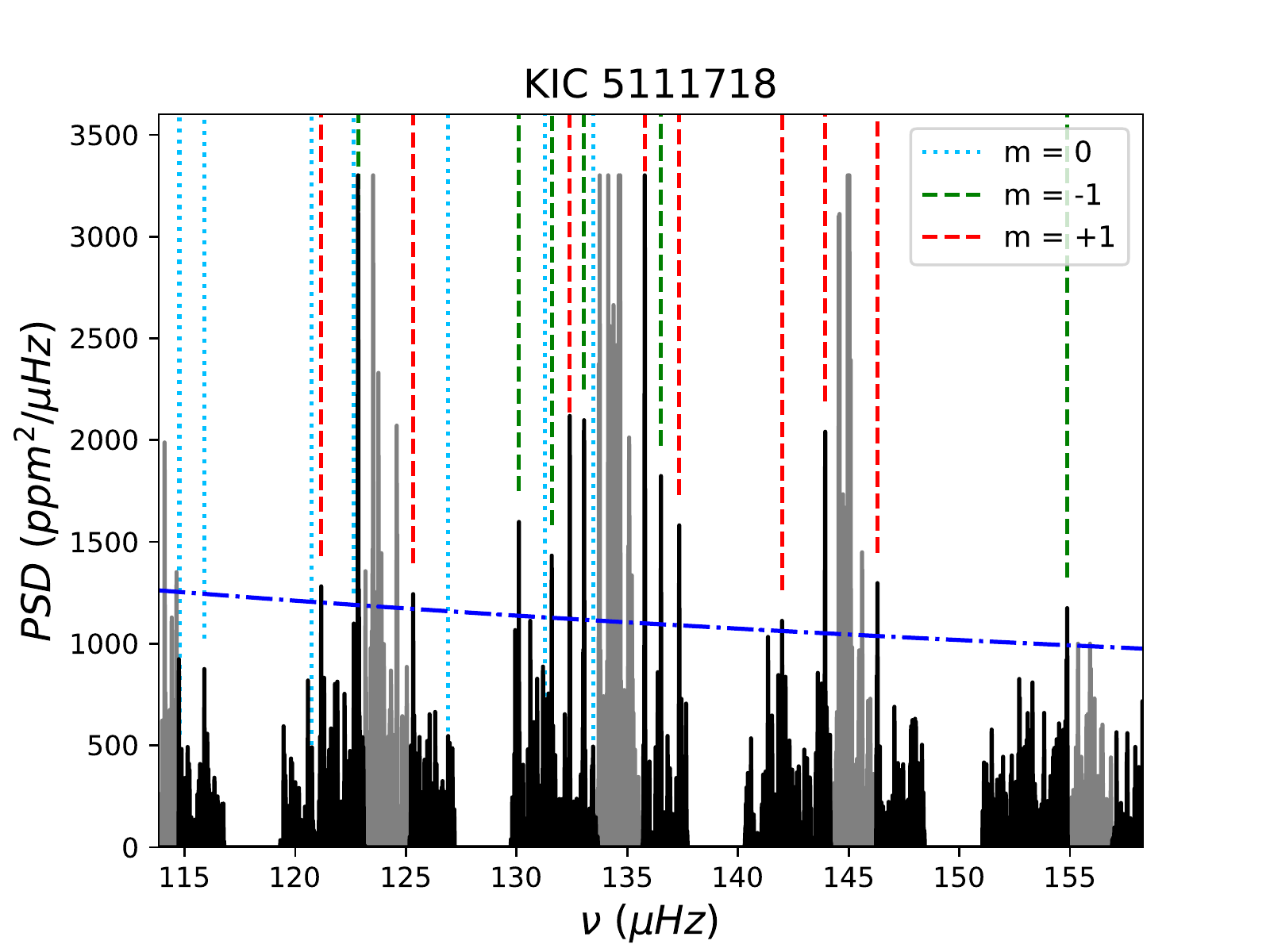}
\includegraphics[width=12cm]{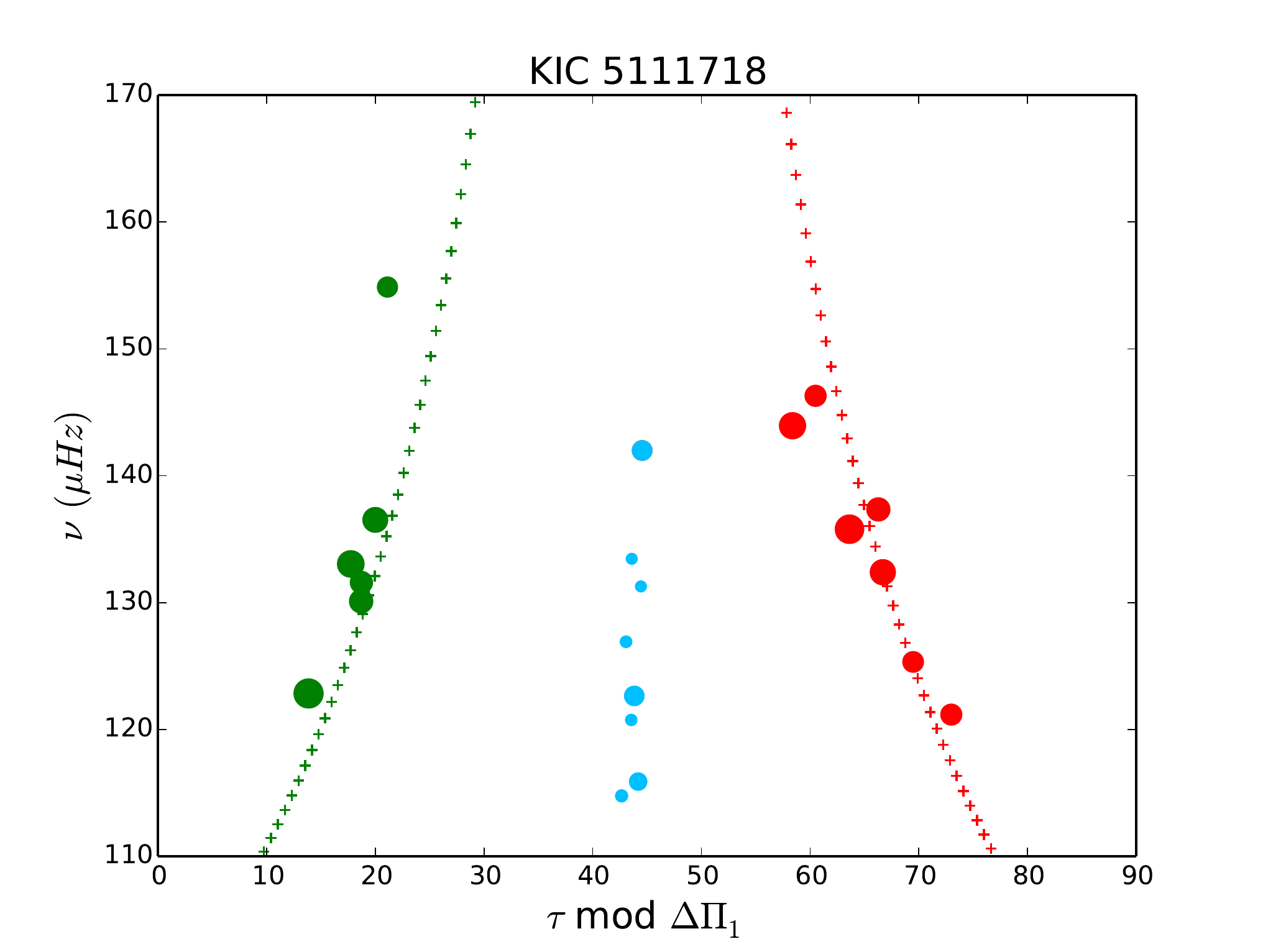}
\caption{Same as Fig.~\ref{fig-me-1} for the NGC 6819 star KIC 5111718 with two observed rotational components. We found an inclination of $i=68.1 \pm^{21.9}\ind{8.4}$$^\circ$ consistent with \cite{Corsaro} measurement of $i=64 \pm 4^\circ$.}
\label{fig-Corsaro-2}
\end{figure*}

\begin{figure*}
\centering
\includegraphics[width=12.8cm]{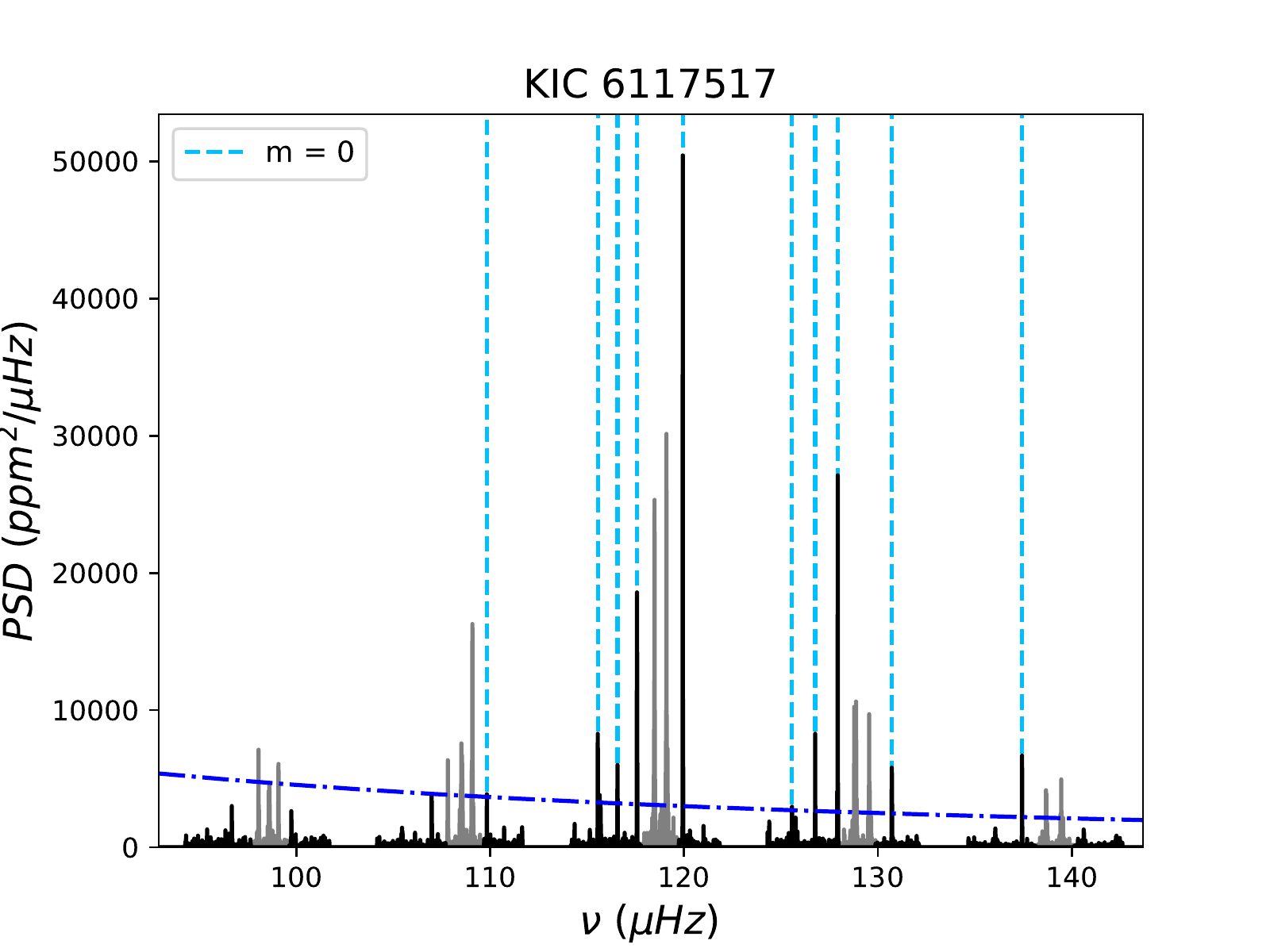}
\includegraphics[width=12cm]{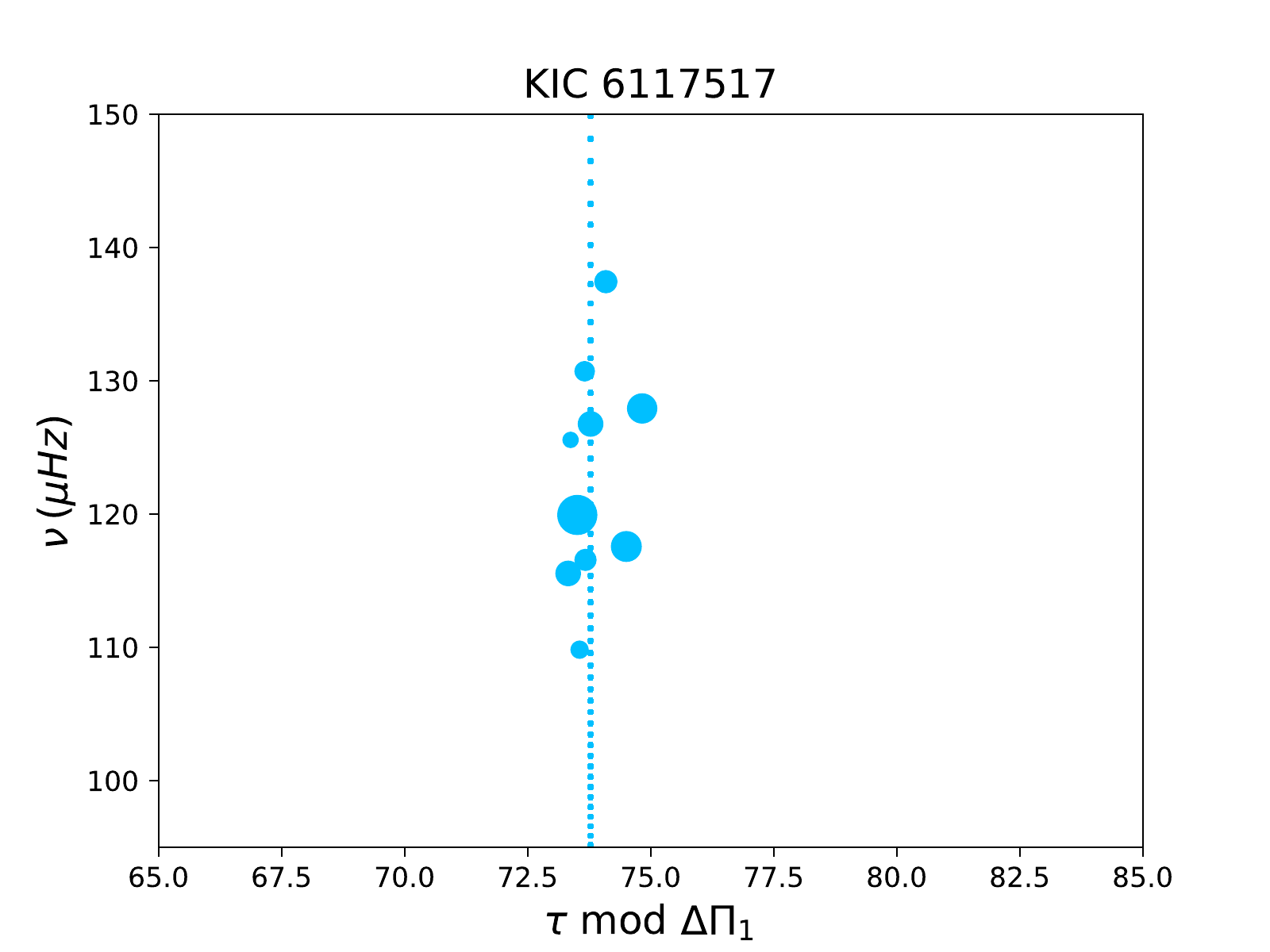}
\caption{Same as Fig.~\ref{fig-me-1} for the field star KIC 6117517 with one observed rotational component. We found an inclination of $i=13.0 \pm^{5.4}\ind{13.0}$$^\circ$ consistent with \cite{Corsaro} measurement of $i=18 \pm 3^\circ$.}
\label{fig-Corsaro-3}
\end{figure*}

\begin{figure*}
\centering
\includegraphics[width=12.8cm]{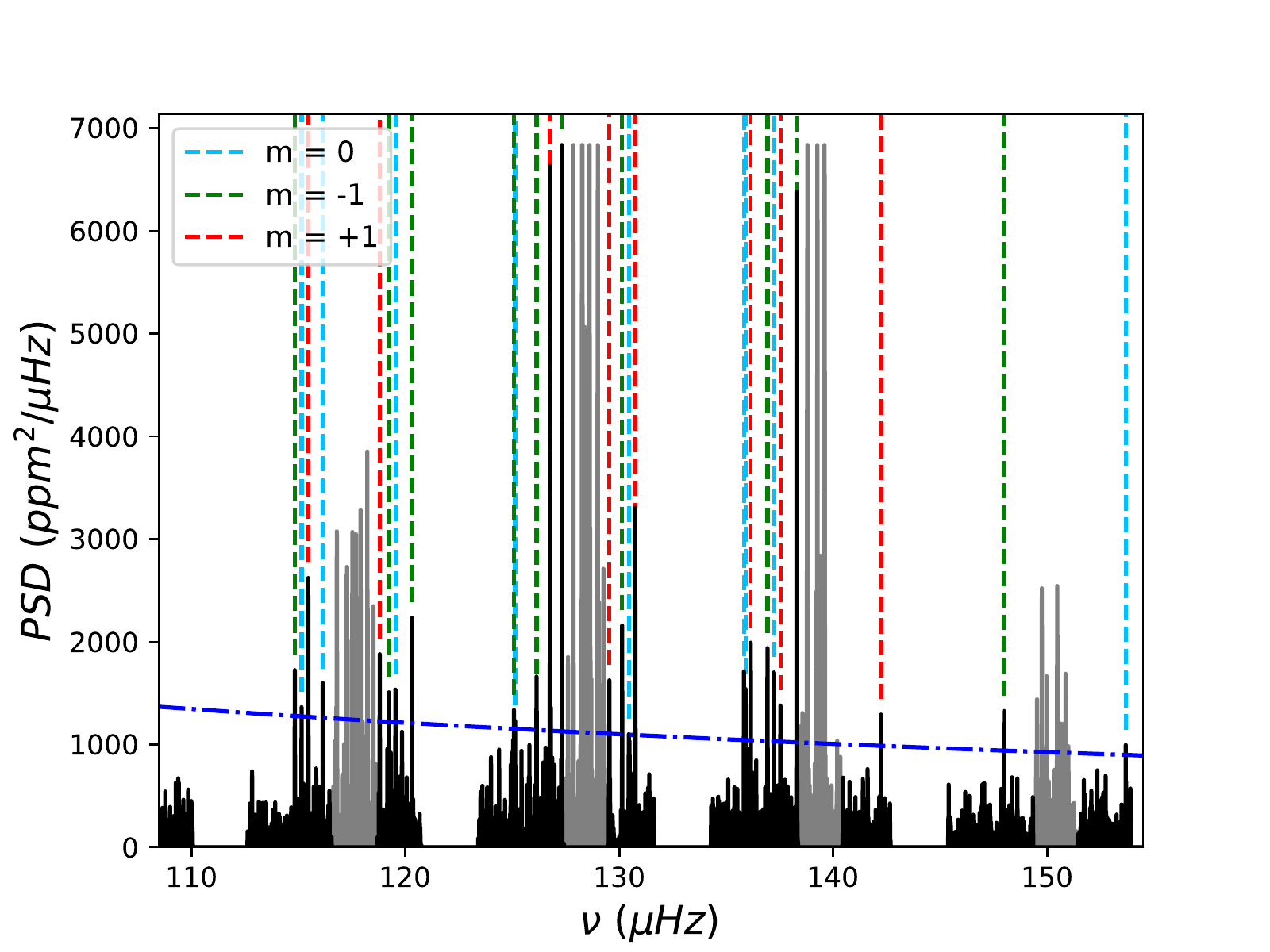}
\includegraphics[width=12cm]{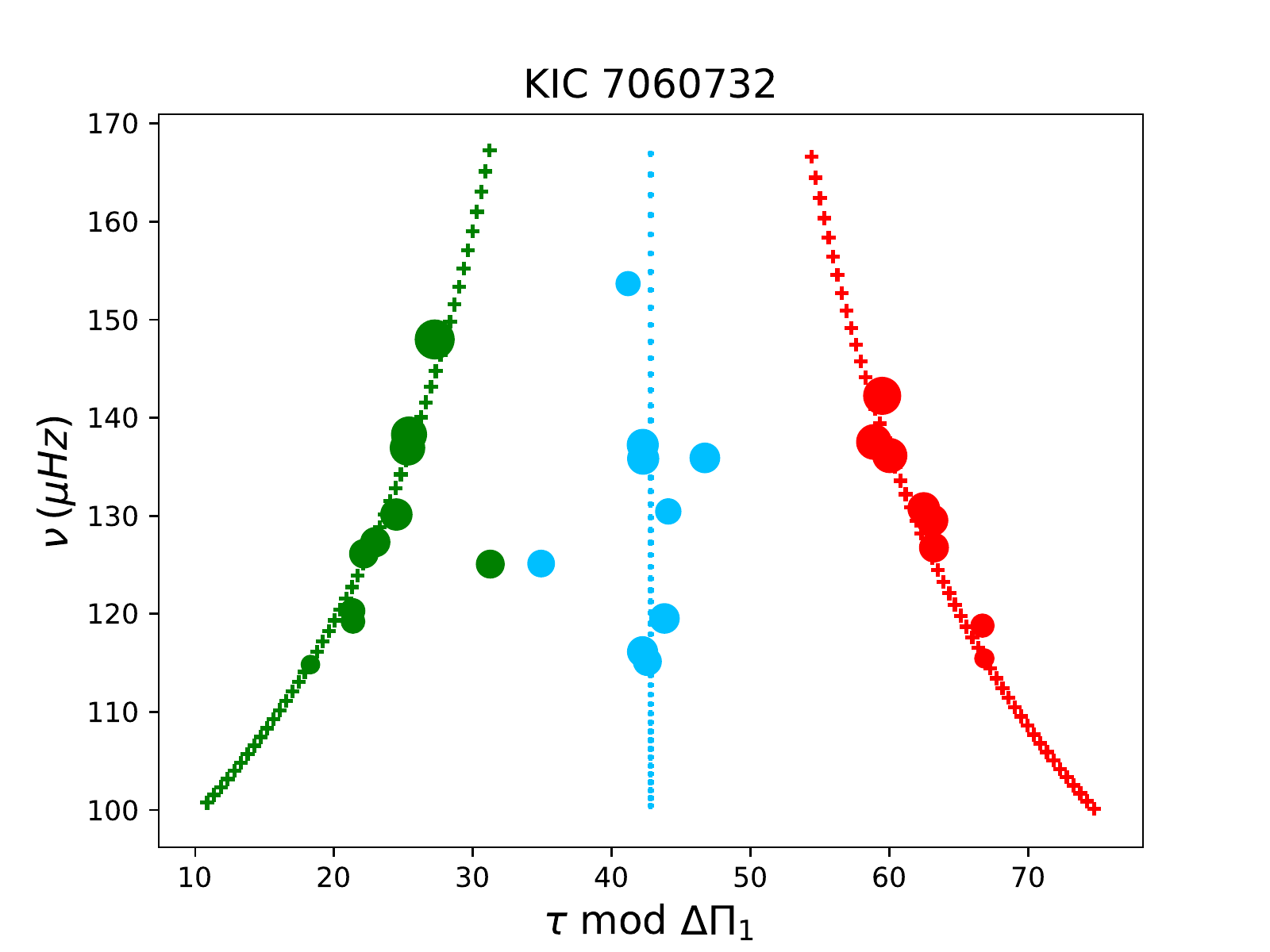}
\caption{Same as Fig.~\ref{fig-me-1} for the field star KIC 7060732 with three observed rotational components. We found an inclination of $i=65.7 \pm 11.1^\circ$ marginally consistent with \cite{Corsaro} measurement of $i=79 \pm 2^\circ$.}
\label{fig-Corsaro-4}
\end{figure*}

\begin{figure*}
\centering
\includegraphics[width=12.8cm]{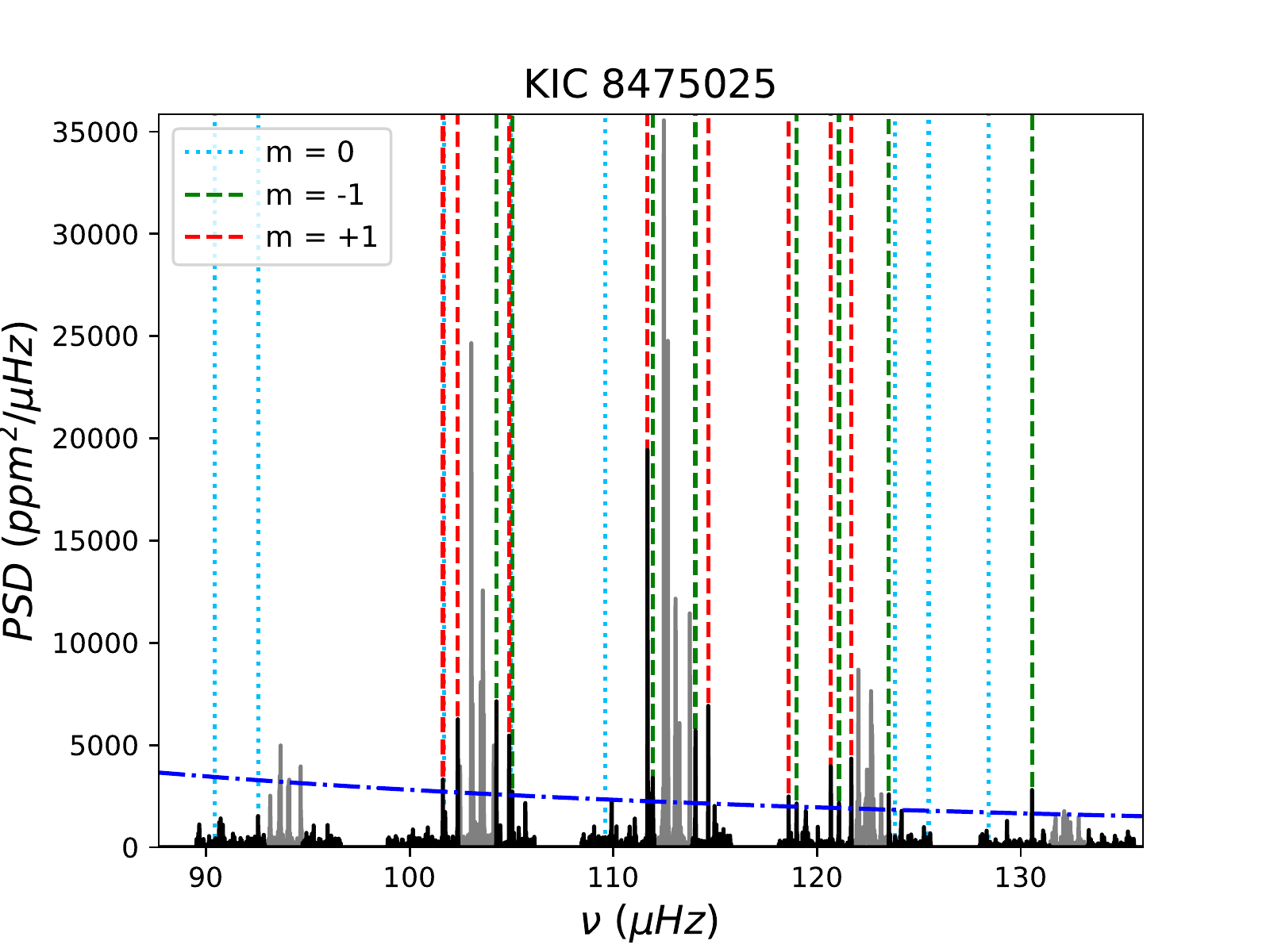}
\includegraphics[width=12cm]{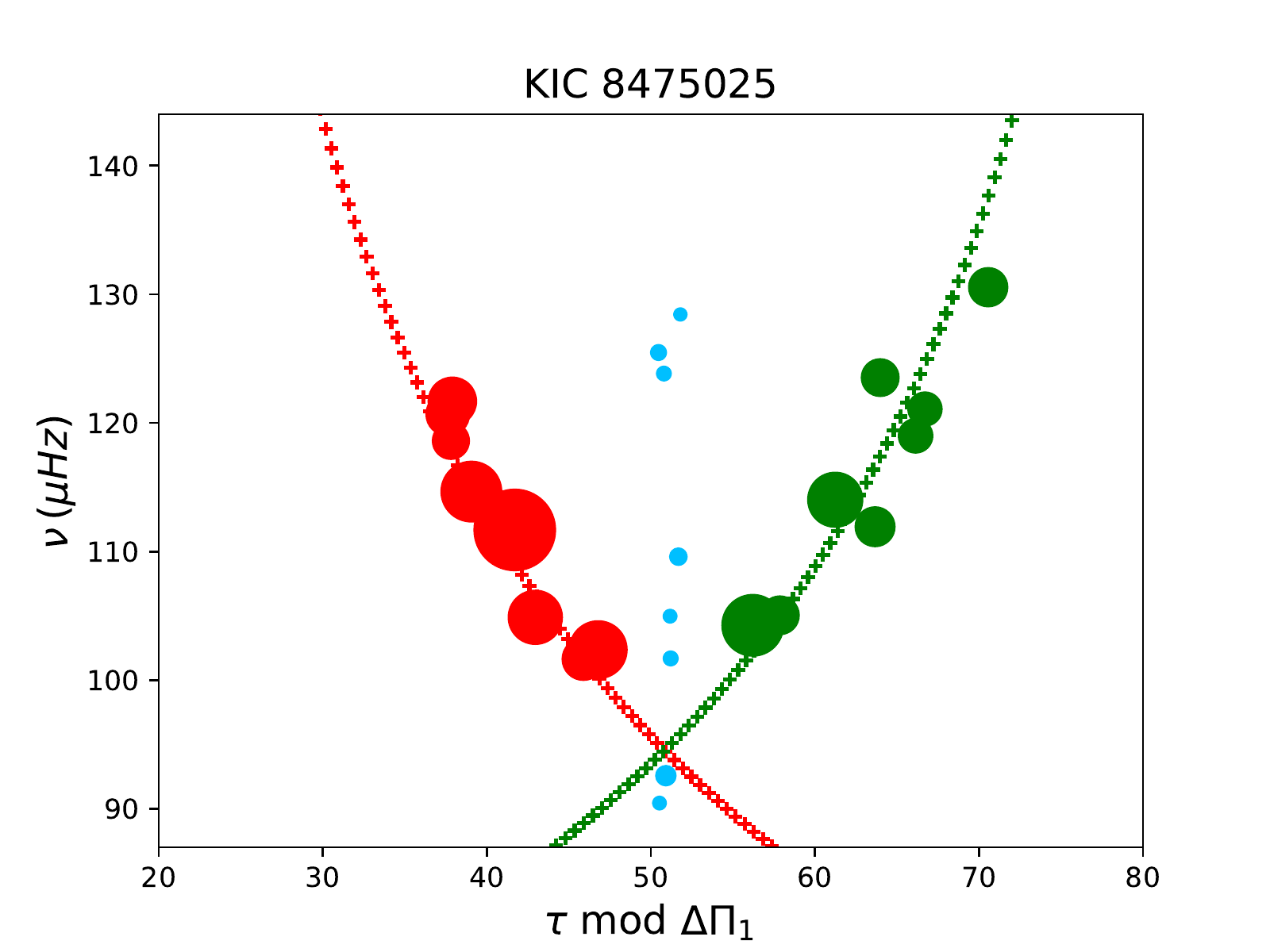}
\caption{Same as Fig.~\ref{fig-me-1} for the field star KIC 8475025 with two observed rotational components. We found an inclination of $i=80.2 \pm^{9.8}\ind{5.1}$$^\circ$ consistent with \cite{Corsaro} measurement of $i=90 \pm 3^\circ$.}
\label{fig-Corsaro-5}
\end{figure*}

The oscillation spectra with the azimuthal order of dipole gravity-dominated mixed modes identified, as well as the corresponding échelle diagrams are presented for:
\begin{itemize}
\item KIC 1569842 with the three $m=\{-1,0,1\}$ rotational components observed (Fig.~\ref{fig-me-1});
\item KIC 3098179 with the three $m=\{-1,0,1\}$ rotational components observed(Fig.~\ref{fig-me-2});
\item KIC 1723843 with only the two $m=\pm \, 1$ rotational components observed (Fig.~\ref{fig-me-3});
\item KIC 2141255 with only the two $m=\pm \, 1$ rotational components observed (Fig.~\ref{fig-me-4});
\item KIC 2303101 with only the $m=0$ rotational component observed (Fig.~\ref{fig-me-5});
\item KIC 3645589 with only the $m=0$ rotational component observed (Fig.~\ref{fig-me-6});
\item the \cite{Kuszlewicz} star KIC 3531478 with the three $m=\{-1,0,1\}$ rotational components observed for which we are in agreement (Fig.~\ref{fig-Kuszlewicz-agreement-1});
\item the \cite{Kuszlewicz} star KIC 7504619 with the three $m=\{-1,0,1\}$ rotational components observed for which we are in agreement (Fig.~\ref{fig-Kuszlewicz-agreement-5});
\item the \cite{Kuszlewicz} star KIC 4731138 with only the two $m=\pm \, 1$ rotational components observed for which we are in agreement (Fig.~\ref{fig-Kuszlewicz-agreement-2});
\item the \cite{Kuszlewicz} star KIC 3634488 with only the two $m=\pm \, 1$ rotational components observed for which we are in marginal agreement (Fig.~\ref{fig-Kuszlewicz-agreement-4});
\item the \cite{Kuszlewicz} star KIC 4996676 with only the $m=0$ rotational component observed for which we are in agreement (Fig.~\ref{fig-Kuszlewicz-agreement-3});
\item the \cite{Kuszlewicz} star KIC 3446775 with only the $m=0$ rotational component observed for which we are in agreement (Fig.~\ref{fig-Kuszlewicz-agreement-6});
\item the \cite{Corsaro} NGC 6791 star KIC 2437325 with only the $m=0$ rotational component observed for which we are in agreement (Fig.~\ref{fig-Corsaro-1});
\item the \cite{Corsaro} NGC 6819 star KIC 5111718 with only the two $m=\pm \, 1$ rotational components observed for which we are in agreement (Fig.~\ref{fig-Corsaro-2});
\item the \cite{Corsaro} field star KIC 6117517 with only the $m=0$ rotational component observed for which we are in agreement (Fig.~\ref{fig-Corsaro-3});
\item the \cite{Corsaro} field star KIC 7060732 with the three $m=\{-1,0,1\}$ rotational component observed for which we are in marginal agreement (Fig.~\ref{fig-Corsaro-4});
\item the \cite{Corsaro} field star KIC 8475025 with only the two $m=\pm \, 1$ rotational components observed for which we are in agreement (Fig.~\ref{fig-Corsaro-5}).
\end{itemize}
All these stars lie on the red giant branch.

\clearpage

\clearpage
\onecolumn

\begin{longtable}{ccccc}
\caption{\label{table:Kuszlewicz} Properties of the 72 RGB stars from \cite{Kuszlewicz} sample.}\\
\hline\hline
KIC & Number of & $i$ ($^\circ$) & $\sigma \ind{i, +}$ ($^\circ$) & $\sigma \ind{i, -}$ ($^\circ$)\\
& rotational & & & \\
& components & & & \\
\hline
\endfirsthead
\caption{continued.}\\
\hline\hline
KIC & Number of & $i$ ($^\circ$) & $\sigma \ind{i, +}$ ($^\circ$) & $\sigma \ind{i, -}$ ($^\circ$)\\
& rotational & & & \\
& components & & & \\
\hline
\endhead
\hline
\endfoot
2166709 & 3 & 58.0 & 10.7 & 10.7\\
3111383 & 1 & 31.7 & 13.8 & 31.7\\
3446775 & 1 & 31.8 & 13.8 & 31.8\\
3531478 & 3 & 41.2 & 9.3 & 9.3\\
3634488 & 2 & 86.2 & 3.8 & 9.7\\
4042882 & 3 & 48.6 & 10.0 & 10.0\\
4139632 & 2 & 79.2 & 10.8 & 2.8\\
4445966 & 2 & 81.1 & 8.9 & 3.2\\
4459359 & 2 & 76.0 & 14.0 & 5.6\\
4638467 & 3 & 55.1 & 11.1 & 11.1\\
4646477 & 1 & 26.1 & 11.2 & 26.1\\
4731138 & 2 & 79.8 & 10.2 & 3.1\\
4738693 & 2 & 76.8 & 13.2 & 3.8\\
4996676 & 1 & 30.1 & 13.0 & 30.1\\
5033397 & 2 & 78.8 & 11.2 & 3.9\\
5115688 & 1 & 30.4 & 13.2 & 30.4\\
5119742 & 3 & 48.9 & 9.23 & 9.23\\
5198982 & 3 & 56.4 & 7.7 & 7.7\\
5305291 & 2 & 74.9 & 15.1 & 5.6\\
5428405 & 2 & 78.1 & 11.9 & 4.0\\
5553307 & 2 & 81.9 & 8.1 & 3.1\\
5623097 & 3 & 48.4 & 10.4 & 10.4\\
5649129 & 2 & 78.8 & 11.2 & 4.0\\
5731852 & 2 & 74.5 & 15.5 & 5.0\\
5773365 & 2 & 77.4 & 12.6 & 3.6\\
5879486 & 2 & 75.3 & 14.7 & 4.3\\
5961985 & 2 & 75.7 & 14.3 & 5.8\\
6208018 & 2 & 71.2 & 18.9 & 5.9\\
6222530 & 2 & 64.6 & 12.8 & 12.8\\
6307132 & 3 & 50.5 & 10.3 & 10.3\\
6352407 & 3 & 59.5 & 19.1 & 19.1\\
6776494 & 3 & 55.4 & 10.9 & 10.9\\
6783217 & 2 & 81.5 & 8.5 & 2.8\\
6924074 & 3 & 58.3 & 7.4 & 7.4\\
6952783 & 2 & 76.6 & 13.4 & 4.4\\
7046554 & 2 & 79.3 & 10.7 & 3.9\\
7468195 & 3 & 60.2 & 8.3 & 8.3\\
7504619 & 3 & 56.8 & 16.6 & 16.6\\
7584122 & 3 & 55.8 & 8.4 & 8.4\\
7595722 & 2 & 77.9 & 12.1 & 4.2\\
7693845 & 2 & 76.5 & 13.5 & 3.5\\
7769544 & 3 & 52.7 & 16.0 & 16.0\\
7898594 & 2 & 75.5 & 14.5 & 3.8\\
8098454 & 3 & 67.5 & 8.6 & 8.6\\
8107355 & 1 & 28.2 & 12.1 & 28.2\\
8145017 & 3 & 46.9 & 10.6 & 10.6\\
8192753 & 3 & 48.3 & 10.2 & 10.2\\
8645227 & 1 & 18.2 & 7.7 & 18.2\\
8827367 & 3 & 54.6 & 9.0 & 9.0\\
8893299 & 2 & 69.1 & 20.9 & 9.9\\
9145781 & 2 & 83.6 & 6.4 & 2.6\\
9157260 & 3 & 67.0 & 7.9 & 7.9\\
9219983 & 1 & 42.9 & 9.5 & 42.9\\
9335457 & 3 & 59.2 & 7.8 & 7.8\\
9418101 & 2 & 74.2 & 15.8 & 7.6\\
9814077 & 2 & 80.0 & 10.0 & 2.7\\
9893437 & 2 & 76.9 & 13.1 & 4.3\\
9896174 & 2 & 74.1 & 15.9 & 4.6\\
9956184 & 2 & 71.3 & 18.8 & 6.4\\
10198496 & 3 & 65.3 & 8.9 & 8.9\\
10199289 & 2 & 80.3 & 9.7 & 4.0\\
10353556 & 3 & 43.1 & 10.0 & 10.0\\
10482211 & 2 & 77.9 & 12.1 & 5.3\\
10581491 & 2 & 69.3 & 20.7 & 11.7\\
10675916 & 3 & 54.5 & 8.1 & 8.1\\
10734124 & 3 & 52.1 & 11.3 & 11.3\\
11015392 & 2 & 82.5 & 7.6 & 2.4\\
11038809 & 3 & 45.3 & 16.8 & 16.8\\
11043770 & 2 & 80.0 & 10.0 & 3.3\\
11852899 & 3 & 55.8 & 7.9 & 7.9\\
12115374 & 2 & 80.8 & 9.3 & 4.7\\
12203197 & 3 & 57.4 & 8.3 & 8.3\\
\end{longtable}

\begin{longtable}{cccccc}
\caption{\label{table:Corsaro} Properties of the 21 RGB stars from \cite{Corsaro} sample.}\\
\hline\hline
KIC & Number of & $i$ ($^\circ$) & $\sigma \ind{i, +}$ ($^\circ$) & $\sigma \ind{i, -}$ ($^\circ$) & Field or\\
& rotational & & & & cluster star\\
& components & & & &\\
\hline
\endfirsthead
\caption{continued.}\\
\hline\hline
KIC & Number of & $i$ ($^\circ$) & $\sigma \ind{i, +}$ ($^\circ$) & $\sigma \ind{i, -}$ ($^\circ$) & Field or\\
& rotational & & & & cluster star\\
& components & & & &\\
\hline
\endhead
\hline
\endfoot
2437325 & 1 & 39.1 & 17.4 & 39.1 & NGC 6791\\
2570244 & 1 & 36.3 & 16.0 & 36.3 & NGC 6791\\
3744043 & 3 & 47.1 & 16.4 & 16.4 & Field\\
5111718 & 2 & 68.1 & 21.9 & 8.4 & NGC 6819\\
5112072 & 3 & 63.3 & 13.0 & 13.0 & NGC 6819\\
5113441 & 1 & 37.7 & 16.7 & 37.7 & NGC 6819\\
6117517 & 1 & 13.0 & 5.4 & 13.0 & Field\\
6144777 & 3 & 63.6 & 12.5 & 12.5 & Field\\
7060732 & 3 & 65.7 & 11.1 & 11.1 & Field\\
7619745 & 2 & 78.1 & 11.9 & 4.8 & Field\\
8366239 & 2 & 73.1 & 8.4 & 8.4 & Field\\
8475025 & 2 & 80.2 & 9.8 & 5.1 & Field\\
8718745 & 1 & 14.7 & 6.2 & 14.7 & Field\\
9267654 & 2 & 72.4 & 17.6 & 6.0 & Field\\
9475697 & 1 & 20.2 & 8.5 & 20.2 & Field\\
9882316 & 1 & 17.6 & 7.4 & 17.6 & Field\\
10123207 & 1 & 13.5 & 5.6 & 13.5 & Field\\
11353313 & 2 & 71.0 & 9.7 & 9.7 & Field\\
11913545 & 2 & 75.4 & 14.6 & 3.7 & Field\\
11968334 & 3 & 54.0 & 14.3 & 14.3 & Field\\
12008916 & 2 & 70.6 & 19.4 & 6.5 & Field\\
\end{longtable}

\end{appendix}

\end{document}